\documentclass[a4paper,fleqn,usenatbib]{mnras}

\usepackage[T1]{fontenc}
\usepackage{ae,aecompl}

\usepackage{graphicx}
\usepackage{amsmath}
\usepackage{amssymb}
\usepackage{breakurl}
\usepackage{color}
\usepackage{lscape}
\usepackage{afterpage}
\usepackage[normalem]{ulem}

\usepackage[usenames,dvipsnames,svgnames,table]{xcolor}
\definecolor{corr}{HTML}{ff0000}
\definecolor{comm}{HTML}{276d00}

\usepackage{tikz}
\usetikzlibrary{arrows.meta,shapes,snakes}
\tikzset{%
  >={Latex[width=2mm,length=2mm]},
    noderect/.style = {rectangle, rounded corners, draw=black,
                           minimum width=2cm, minimum height=1cm,
                           text centered, font=\sffamily, fill=orange!30},
    noderesult/.style = {rectangle, rounded corners, draw=black,
                           minimum width=2cm, minimum height=1cm,
                           text centered, font=\sffamily, fill=green!30},
    nodecirc/.style = {draw, circle, draw=black, minimum width=1cm, text centered, font=\sffamily, fill=blue!30},
    nodediam/.style = {diamond, draw=black, minimum width=1cm, font=\sffamily, fill=red!30,
                        text centered},
}

\newcommand{\lsun}{\mbox{L$_{\sun}$}}
\newcommand{\msun}{\mbox{M$_{\sun}$}}

\title[Multi-wavelength modelling of AFGL~2591]{
    Multi-wavelength modelling of the circumstellar environment of the massive
    proto-star AFGL~2591 VLA 3
}

\author[F. A. Olguin et al.]{
    F. A. Olguin$^{1,2}$\thanks{E-mail:\,folguin@phys.nthu.edu.tw}, 
M. G. Hoare$^2$,
K. G. Johnston$^2$,
F. Motte$^{3,4}$,
H.-R. V. Chen$^{1}$, \newauthor
H. Beuther$^{5}$, 
J. C. Mottram$^{5}$, 
A. Ahmadi$^{5}$,
C. Gieser$^{5}$,
D. Semenov$^{5,6}$,
T. Peters$^{7}$, \newauthor
A. Palau$^{8}$,
P. D. Klaassen$^{9}$,
R. Kuiper$^{10}$,
\'A. S\'anchez-Monge$^{11}$,
Th. Henning$^{5}$\\
$^{1}$Institute of Astronomy, National Tsing Hua University, Taiwan\\
$^{2}$School of Physics \& Astronomy, E.C. Stoner Building, University of Leeds, 
Leeds LS2 9JT, UK\\
$^{3}$Institut de Plan\'etologie et d'Astrophysique de Grenoble, Univ. Grenoble 
Alpes -- CNRS-INSU, BP 53,\\ 38041 Grenoble Cedex 9, France\\
$^{4}$AIM Paris-Saclay, CEA/IRFU -- CNRS/INSU -- Univ. Paris Diderot, Service
    d'Astrophysique, CEA-Saclay,\\ 91191 Gif-sur-Yvette Cedex, France\\
$^{5}$Max Planck Institute for Astronomy, K\"onigstuhl 17, 69117 Heidelberg,
Germany\\
$^{6}$Department of Chemistry, Ludwig Maximilian University, Butenandtstr. 5-13, 81377 Munich, Germany\\
$^{7}$Max-Planck-Institut f\"ur Astrophysik, Karl-Schwarzschild-Str. 1, 85748, Garching, Germany\\
$^{8}$Instituto de Radioastronom\'ia y Astrof\'isica, Universidad Nacional Aut\'onoma de M\'exico, 58090, Morelia, Michoac\'an, M\'exico\\
$^{9}$UK Astronomy Technology Centre, Royal Observatory Edinburgh, Blackford Hill, Edinburgh EH9 3HJ, UK\\
$^{10}$Institute of Astronomy and Astrophysics, University of T\"ubingen, Auf der Morgenstelle 10, 72076, T\"ubingen, Germany\\
$^{11}$I. Physikalisches Institut, Universit\"at zu K\"oln, Z\"ulpicher Str. 77, 50937, K\"oln, Germany
}

\date{Accepted 2020 August 05. Received 2020 August 05; in original form 2019 August 19}
\pubyear{2020}

\begin{document}
 \label{firstpage}
 \pagerange{\pageref{firstpage}--\pageref{lastpage}}
 \maketitle
 
 \begin{abstract}
    We have studied the dust density, temperature and velocity distributions of the archetypal massive young stellar object (MYSO) AFGL~2591. 
    Given its high luminosity ($L=2\times10^5\,\lsun$) and distance ($d=3.3$\,kpc), AFGL~2591 has one of the highest $\sqrt{L}/d$ ratio, giving better resolved dust emission than any other MYSO.
	As such, this paper provides a template on how to use resolved multi-wavelength data and radiative transfer to obtain a well-constrained 2-D axi-symmetric analytic rotating infall model.
    We show for the first time that the resolved dust continuum emission from \textit{Herschel} 70\,\micron\ observations is extended along the outflow direction, whose origin is explained in part from warm dust in the outflow cavity walls.
    However, the model can only explain the kinematic features from CH$_3$CN observations with unrealistically low stellar masses ($<15$\,\msun), indicating that additional physical processes may be playing a role in slowing down the envelope rotation.
    As part of our 3-step continuum and line fitting, we have identified model parameters that can be further constrained by specific observations.
    High-resolution mm visibilities were fitted to obtain the disc mass (6\,\msun) and radius (2200\,au).
    A combination of SED and near-IR observations were used to estimate the luminosity and envelope mass together with the outflow cavity inclination and opening angles.
 \end{abstract}
 
 
\section{Introduction}
\label{sect:intro}

High-mass stars play an important role in the evolution of galaxies, but their formation is still not well understood. 
Do high-mass stars form as a scaled-up version of isolated low-mass star formation \citep[e.g.][]{2012ApJ...754...71K,2018A&A...616A.101K} or by some other means (e.g. competitive accretion, \citealt[][]{2009MNRAS.400.1775S}; fragmentation-induced starvation, \citealp{2010ApJ...725..134P}; global hierarchical collapse, \citealp{2019arXiv190311247V})? 
To understand how high-mass stars form we need to study the distribution of matter in star-forming regions, which is the result of the interaction of inflow and outflow processes driven by gravitational collapse, rotation, turbulence, magnetic fields and radiation. 
The imprints left by these processes in the circumstellar material can be well studied in early stages of massive young stellar objects (MYSOs), when the ionizing nature of the stellar radiation has not started to evaporate the core.

These MYSOs are radio weak \citep[radio fluxes of few mJy,][]{2002ASPC..267..137H}, bright in the IR and have luminosities $L \gtrsim 10^4$~\lsun\ \citep[e.g.][]{2011ApJ...730L..33M}. 
The radiation pressure on dust produced by such massive stars can be reduced by the high optical depths of an accretion disc \citep{2010ApJ...722.1556K}, and also through bipolar outflow cavities opened by the interaction of jets and winds with the infalling material \citep[e.g.][]{2005ApJ...618L..33K,2011ApJ...740..107C,2015ApJ...800...86K,2016ApJ...832...40K}.
Temperatures lower than their zero age main sequence (ZAMS) counterparts are also thought to be responsible for the lack of ionizing radiation given their luminosity \citep{2007ASSP....1...61H,2011MNRAS.416..972D}, which may be due to the forming star being swollen due to accretion \citep[][]{2010ApJ...721..478H,2013ApJ...772...61K,2013MNRAS.428.1537P,2016MNRAS.458.3299H,2018A&A...616A.101K}. 

MYSOs are located within giant molecular clouds at larger distances (typical distances of 3\,kpc) compared to nearby low-mass star-forming clouds, and usually in clustered environments. 
Their study is therefore observationally challenging due to their location, the amount of gas and dust in their parental clouds and their fast formation ($10^4-3\times10^5$\,yr timescales; \citealt{2011ApJ...730L..33M,2010A&A...515A..55R,2013A&A...558A.125D}).
Hence, to study these regions we must observe them at longer wavelengths and at high angular resolution.  
In the last decade an effort to map large regions of the sky at high angular resolution has been undertaken using ground- (e.g. UKIRT\footnote{United Kingdom Infrared Telescope}, JCMT\footnote{James Clerk Maxwell Telescope}, APEX\footnote{Atacama Pathfinder EXperiment}) and space-based telescopes (e.g. \textit{Spitzer}, \textit{Herschel}) from near-IR to sub-millimetre wavelengths.
These observations are also complemented by interferometric observations of specific sources at even higher resolutions, which allow the study of regions closer to the forming star \citep[e.g.][]{2013MNRAS.428..609M,2018A&A...617A.100B}.

The study by \citet{2015MNRAS.449.2784O} of Red \textit{MSX} Source (RMS) survey \citep{2013ApJS..208...11L} MYSOs as mapped by the \textit{Herschel} IR Galactic Plane Survey \citep[Hi-GAL;][]{2010PASP..122..314M}, shows that relatively isolated sources with high $\sqrt{L}/d$, with $L$ the source luminosity and $d$ its distance, may be resolved at 70\,\micron.
They also analysed three sources in the $l=30\degr$ and $59\degr$ fields by fitting their spectral energy distributions (SEDs) and 70\,\micron\ data with 1-D radiative transfer models assuming a power law density distribution. 
The density power law index they obtained was shallower (${\leq}1$) than expected for infalling material (index of 1.5). 
These results suggest that the far-IR emission may be dominated by warm dust from the outflow cavity walls rather than rotational flattening as suggested by earlier studies \citep[e.g.][]{2009A&A...494..157D}, as the mapped emission is larger than the expected centrifugal radius. 
These findings are in line with the results of 2-D axisymmetric radiative transfer modelling of mid-IR observations \citep[e.g.][]{2010A&A...515A..45D}.

In the light of these new data, a multi-wavelength study of MYSOs can provide further constraints to the dust/gas density, temperature and velocity distributions of their circumstellar matter. 
In this work we present a method to fit multi-wavelength data and assess the importance of specific observations to constrain these distributions.
We apply this method to the well-studied MYSO AFGL~2591 to explain a large range of observations and compare the results with those in the literature.

\subsection{The selected source: AFGL~2591}

AFGL~2591 is a well studied luminous ($L\sim10^5$\,\lsun) source located at $3.3\pm0.1$\,kpc \citep{2012A&A...539A..79R} in the Cygnus X sky region.
Several radio continuum sources have been identified in this region: four sources have been classified as \ion{H}{II} regions \citep[VLA 1, 2, 4, 5; see][]{2003ApJ...589..386T,2013A&A...551A..43J}, and one as a MYSO or young hot core \citep[VLA 3;][]{2003ApJ...589..386T,2019A&A...631A.142G}. An unknown source(s) associated with maser emission has also been detected \citep[VLA 3-N][]{2013MNRAS.430.1309T}.
The MYSO (AFGL~2591 VLA 3) will be referred to as AFGL~2591 as it is the source that dominates the SED from the near-IR to millimetre wavelengths \citep{2013A&A...551A..43J}.
As such, this proto-typical MYSO/young hot core is one of the sources with the highest $\sqrt{L}/d$ in the RMS survey sample.
It was covered by the \textit{Herschel}/HOBYS survey\footnote{
    The \textit{Herschel} imaging survey of OB Young Stellar objects (HOBYS) is a \textit{Herschel} key programme.
    See \url{http://hobys-herschel.cea.fr}} \citep{2010A&A...518L..77M}. 
The HOBYS data for the Cygnus X region were published by \citet{2016A&A...591A..40S}, however there is no published close-up study of this source at 70\,\micron\ to date.
The HOBYS survey was designed to map specific star-forming regions at lower scan speeds. 
Therefore, it is one of the best candidates to resolve the 70\,\micron\ emission since the observations are less subject to smearing effects compared to other \textit{Herschel} observations such as Hi-GAL. 

The presence of a jet within a large-scale outflow was inferred from early CO and HCO$^+$ molecular line observations \citep[e.g.][]{1984ApJ...286..302L,1995ApJ...451..225H}, with the bipolar outflow cavity oriented in the E-W direction.
The jet was also detected with radio interferometric observations at 3.6\,cm, and has a position angle of ${\sim}100\degr$ \citep{2013A&A...551A..43J}.
The blue-shifted outflow cavity can also be observed in scattered light in the $K$-band \citep[2\,\micron, e.g.][]{2003A&A...412..735P}, whilst shocked  H$_2$ bipolar emission is detected in the same band \citep{1992ApJ...391..710T,1992A&A...262..229P}. 
The blue-shifted cavity $K$-band emission presents several features (loops) formed probably by entrainment of material \citep[e.g.][]{2009MNRAS.400..629P}.
\citet{1995ApJ...451..225H} qualitatively constrained the full cavity opening angle to be $<90\degr$ and its inclination with respect to the line of sight ${\leqslant}45\degr$ from CO emission.
Based on geometrical considerations from Gaussian fits to molecular line emission maps (e.g. SO$_2$), \citet{2006A&A...447.1011V} constrained the inclination angle to between 26 and $38\degr$.
The near-IR polarization study by \citet{1991MNRAS.251..508M} found that a cone with an inclination of $35\degr$ can reproduce their data, however \citet{2013MNRAS.435.3419S} found an inclination angle of $15\degr$ by fitting the SED and then modifying this model in order to match the morphology of their \textit{HST} polarization data. 
\citet{1999ApJ...522..991V} fitted a 1-D spherically symmetric power law density distribution to CS isotopologues observations and then modified their best model to include an empty bipolar cavity, and found that an opening angle of $60\degr$ can better fit their data. 
However, the opening angle may be wider at the base of the cavity ($\theta_c{\sim}100$\degr) as implied by the spatial distribution and proper motion of water maser emission \citep{2012ApJ...745..191S}. 
The detailed radiative transfer modelling by \citet{2013A&A...551A..43J}, whose objective was to fit the SED and \textit{JHK} 2MASS slices along and perpendicular to the outflow direction, found that the inclination angle with respect to the line of sight is constrained to be between $30-65\degr$, and that the real cavity opening angle is not well constrained by their observations as it is degenerate with the inclination angle. 
The position angle of the Herbig-Haro objects observed in the near-IR varies between $258-261\degr$ \citep{1992A&A...262..229P}, whilst \citet{2003A&A...412..735P} adopted a value of $259\degr$ for the outflow cavity symmetry axis even though the position angle of the loops ranges between $263-265\degr$ as determined from their $K$-band speckle interferometric observations.

The presence of a disc in AFGL~2591 is still uncertain. 
Through 2-D axi-symmetric radiative transfer modelling of the SED and 2MASS observations, \citet{2013A&A...551A..43J} found that models with and without a disc can reproduce the observations that they modelled. 
However, HDO, H$_2^{18}$O and SO$_2$ interferometric observations at millimetre wavelengths point towards the presence of a sub-Keplerian disc-like structure and expanding material in the inner ${\sim}1000$\,au, and the continuum is extended nearly perpendicular to the jet/outflow direction \citep{2012A&A...543A..22W}. 
The partially resolved source identified in the $K$-band speckle interferometric visibilities of \citet{2003A&A...412..735P} seems to be tracing the inner rim of this disc, which has a size of ${\sim}120$\,au at 3.3\,kpc as derived from their modelling. 

The kinematics of the outflow and the inner region have been studied by several authors. 
The blue-shifted radial velocity of the jet is constrained between $200-500$\,km\,s$^{-1}$ as measured from the line wings of the Herbig-Haro objects \citep{1992A&A...262..229P} and $^{12}$CO \citep{1999ApJ...522..991V}. 
The C$^{18}$O observations of \citet{2013A&A...551A..43J} show evidence of Hubble-like expansion of the blue-shifted emission towards the outflow. 
Hubble-like expansion and rotation at disc scales (${<}5000$\,au) was also found from Plateau de Bure Interferometer (PdBI) observations of HDO and H$_2^{18}$O by \citet{2012A&A...543A..22W}. 
The rotation at these scales seems to be sub-Keplerian and anti-clockwise looking down from the blue-shifted cavity. 
\citet{2012A&A...543A..22W} argue that these water isotopologues are closer to the surface of a disc-like structure, thus the expansion motion is a result of the interaction of the stellar radiation/wind, and infall occurs in the inner region of the disc where these molecules are depleted. 
Similarly, their SO$_2$ observations trace material affected more by the wind than the rotation.

The change in the abundance for different molecules in AFGL~2591 has been addressed by several authors. 
\citet{2007A&A...475..549B} used the spherically symmetric abundance models of \citet{2005A&A...440..949S} to fit their CS and SO interferometric (sub-arcsec resolution) observations, and concluded that X-ray emission from the protostar is needed to produce a better fit.
\citet{2015A&A...574A..71K} studied the emission of 25 molecules observed with the Heterodyne Instrument for the Far-Infrared (HIFI) by \textit{Herschel} \citep{2010A&A...518L...1P}. 
They found that the emission from some molecules can be better described by abundances with a jump at 100 or 230\,K, which were then fitted by a theoretical chemical model. 
Their theoretical model fitted relatively well the abundance jumps in molecules like H$_2$O and NH$_3$, and predicted a chemical age of $10-50$\,kyr. 
The first jump at 100\,K is related to the evaporation of ices \citep{1998ARA&A..36..317V} whilst the second jump at temperatures $T{>}230$\,K is important for N-bearing molecules due to less formation of N$_2$ \citep{2001ApJ...553L..63B}.
Far UV and X-ray radiation, which penetrates a thin layer in the cavity walls, can also enhance the abundance of several diatomic molecules observed in the far-IR and (sub)mm as shown by the chemical modelling of \citet{2009ApJ...700..872B,2010ApJ...720.1432B}. 
The physical-chemical modelling of 14 molecules observed at 1\,mm with the Northern Extended Millimeter Array (NOEMA) by \citet{2019A&A...631A.142G} shows that many species have have abundances that change with radius (e.g. SO, SO$_2$), and derived a chemical age of ${\sim}20$\,kyr. 
They also obtained a flatter density distribution to explain the gas distribution (index of 1.0) than the one derived from dust emission at 1\,mm (index of 1.7).
These studies show that molecular line emission traces gas with different physical conditions, and imply the presence of different density structures from the changes in abundance as a function of temperature. 

In this work, we have combined high-resolution observations at key frequencies to constrain the dust density, temperature, and gas velocity through radiative transfer modelling. 
The data are presented in \S\ref{sect:data}.
The dust continuum and molecular line modelling procedures are presented in \S\ref{sect:modelling} and a detailed description of the procedures can be found in Appendices~\ref{ap:modelling}--\ref{ap:dataproc} (available online). 
The results of the modelling are presented in \S\ref{sect:results} and discussed in \S\ref{sect:discussion}. 
Finally, our conclusions are presented in \S\ref{sect:conclusions}.

\section{Data}
\label{sect:data}

\subsection{\textit{Herschel} 70\,\micron}
\label{sect:data:herschel}

\begin{figure}
 \centering
 \includegraphics[width=\columnwidth]{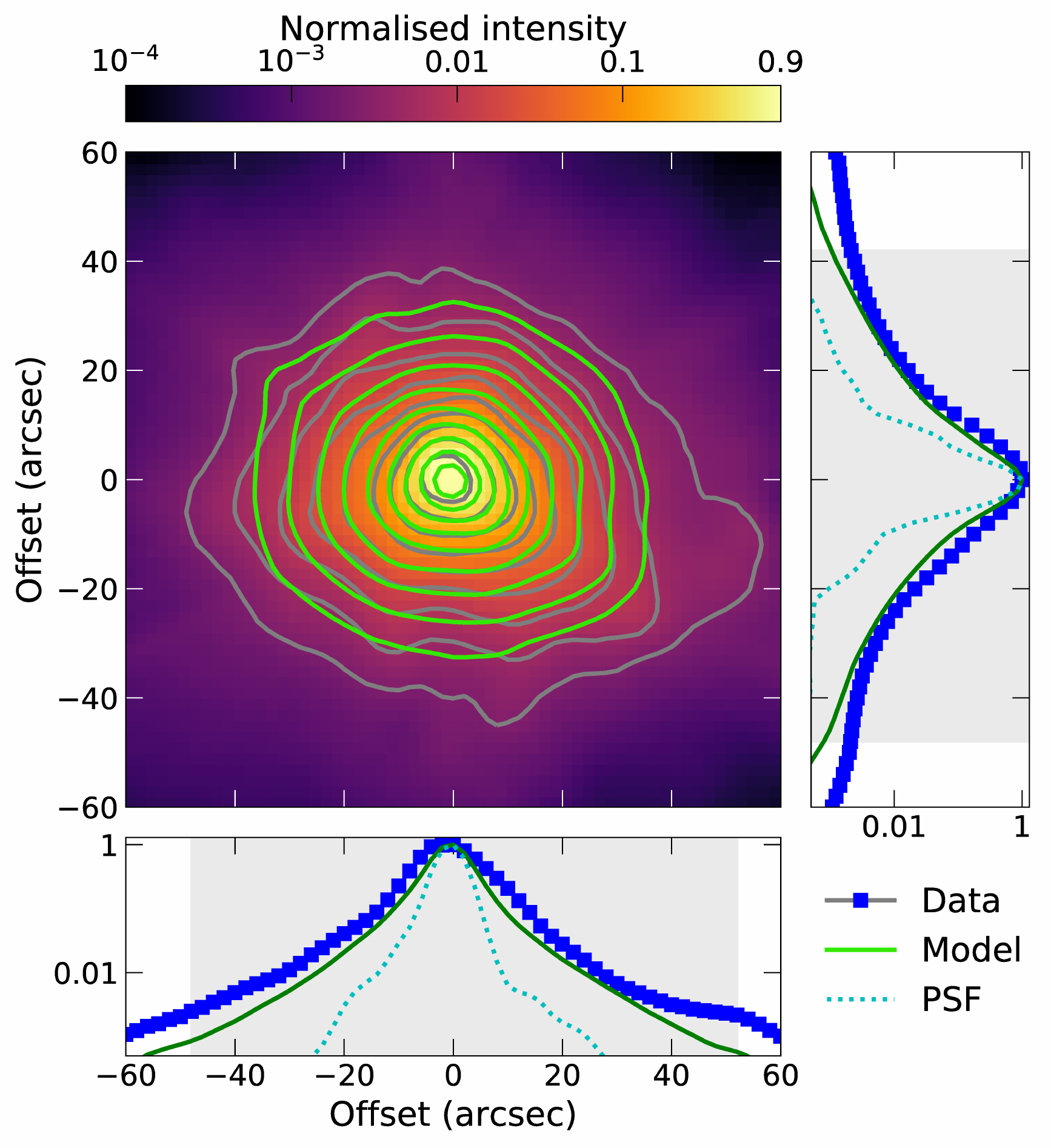}
 \caption{Observed (colour scale and grey contours) and model (green contours, see Section~\ref{sect:modelling}) 70\,\micron\ maps of AFGL~2591. 
     Contour levels, in intensity normalised to peak value (87.3\,Jy), are $5, 10, 20 ... 1280\times\sigma_{\rm rms}$ with $\sigma_{\rm rms}=50$\,mJy.
     Bottom and right panels show a slice through the peak in the horizontal and vertical directions, respectively. 
     The blue squares in the side panels correspond to the observations (errors are smaller than the point size), the green continuous line is the best-fitting model presented in Section~\ref{sect:results} and the dotted cyan line is the PSF. 
     The shaded region enclose the points higher than $5\sigma_{\rm rms}$.
}
\label{fig:cont:herschel}
\end{figure}

Resolved 70\,\micron\ data from \textit{Herschel} observed as part of the HOBYS survey were used.
The final images were obtained using procedures from the HOBYS team (F. Motte, private communication), and details of the reduction process can be found in \citet{2016A&A...591A..40S}.
These data  were taken with the Photodetector Array Camera and Spectrometer \citep[PACS][]{2010A&A...518L...2P} in the parallel mode of the telescope at a scan speed of 20\,arcsec\,s$^{-1}$. 
This speed produces a more symmetrical PSF ($6\farcs51\times5\farcs44$) with fewer artefacts than data scanned at 60\,arcsec\,s$^{-1}$, which has a $12\farcs58\times5\farcs85$ PSF \citep[][]{2012Lutz,2015MNRAS.449.2784O}.
A zoom on AFGL~2591 is shown in Fig.~\ref{fig:cont:herschel} and provides the highest resolution in the far-IR to date. 
Table~\ref{tab:herschel} lists the observed size of the 70~\micron\ emission as obtained from a 2-D Gaussian fit to the data, which shows that the source is well resolved by \textit{Herschel} and that the Gaussian position angle is also close to the E-W outflow direction. 
The former is also true for the horizontal and vertical slices shown in Fig.~\ref{fig:cont:herschel} side panels. 
The emission peak is shifted ${\sim}3.2$~arcsec to the west of the radio source position \citep{2003ApJ...589..386T}. 
However, the \textit{Herschel} $1\sigma$ astrometry error\footnote{\url{http://herschel.esac.esa.int/Docs/PACS/pdf/pacs_om.pdf}} is ${\sim}2$\,arcsec, 
hence their positions agree within $2\sigma$.
In what follows it will be assumed that the emission peak at 70\,\micron\ coincides with the radio one. 

\begin{table*}
 \centering
 \caption{70\,\micron\ angular sizes.}
 \label{tab:herschel}
 \begin{tabular}{lcccccc}
  \hline
     & \multicolumn{2}{c}{2-D Gaussian} & \multicolumn{2}{c}{Horizontal slice} & \multicolumn{2}{c}{Vertical slice}\\
    Type  & FWHM & PA & FWHM & FW1\% & FWHM & FW1\% \\
        & (arcsec) & (deg) & (arcsec)&(arcsec)&(arcsec)&(arcsec)\\
  \hline
Observed & $14.5\times12.9$ & $253.1\pm0.7$ & 12.2 & 58.0 & 11.3 & 47.7 \\
   Model & $10.7\times9.9$ & $244\pm0.8$    & 8.2  & 48.9 & 8.0  & 42.6 \\
     PSF & $5.9\times6.6^{a}$ & $56\pm1$    & 6.1  & 22.9 & 5.9  & 20.7 \\
  \hline
 \end{tabular}
 \begin{minipage}{\textwidth}
     \small{\textbf{Notes.} An error of 0.02\,arcsec is estimated for the semi-axes of the 2-D Gaussian fit. 
         FW1\% stands for full width at 1~per~cent the peak intensity.\\
     $^a$ Note this value is from the PSF binned to the observed pixel size and rotated to match the observations, hence it is different from the one presented in Table~\ref{tab:cont}.}
 \end{minipage}
\end{table*}

\subsection{NOEMA 1.3\,mm observations}
\label{sect:data:noema}

AFGL~2591 was observed in the 1.3\,mm spectral band with NOEMA between August 2014 and February 2016 as part of the CORE project \citep{2018A&A...617A.100B}. 
The observations were undertaken in three array configurations (A, B, and D) at 1.37\,mm with the wide-band correlator Widex and a narrow-band correlator \citep[for a complete overview of the spectral setup see][]{2018A&A...618A..46A}.
The data were reduced by the CORE team and continuum visibilities were extracted and then CLEANed \citep[see][]{2018A&A...617A.100B}. 
To facilitate the comparison between models and observations, we separate the analysis of the A+B visibilities, or extended configurations, and the D visibilities, or compact configuration.
For the purpose of this paper we use the continuum data from A+B and D configurations, and only the compact configuration for the CH$_3$CN molecular line data in order to analyse the kinematics of the inner envelope ($<10000$\,au scales).
A study of the CH$_3$CN chemistry using the combined A+B+D data can be found in \citet{2019A&A...631A.142G}. 
The study of the kinematics of the disc (if any) from the combined data set will be the focus of future publications (e.g. Ahmadi et al. in prep.).

\begin{figure*}
  \centering
\includegraphics[width=\textwidth]{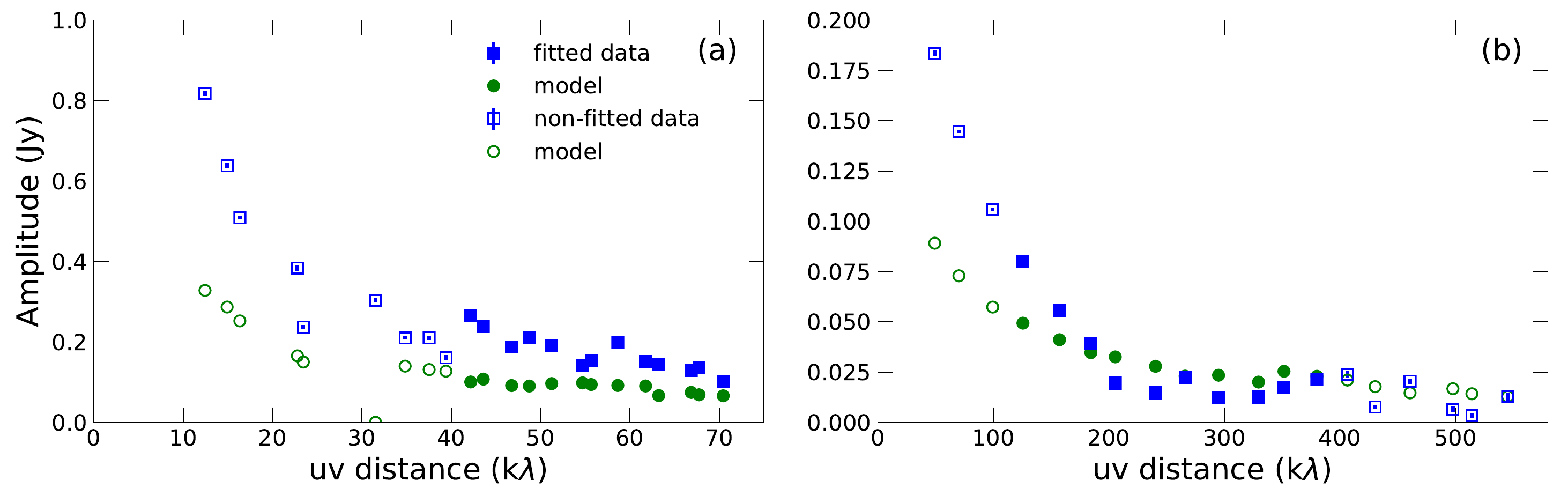}
  \caption{NOEMA 1.3\,mm observed (blue squares) and simulated (green circles, see Section~\ref{sect:modelling}) interferometry visibility profiles. 
(a) Compact (D) configuration visibility profile. 
(b) Extended (A+B) configurations visibility profile.
    The selection of the data to be fitted is explained in Appendix~\ref{ap:dataproc} (online).
    }
\label{fig:cont:mm:vis}
\end{figure*}

\begin{figure*}
  \centering
\includegraphics[width=0.98\textwidth]{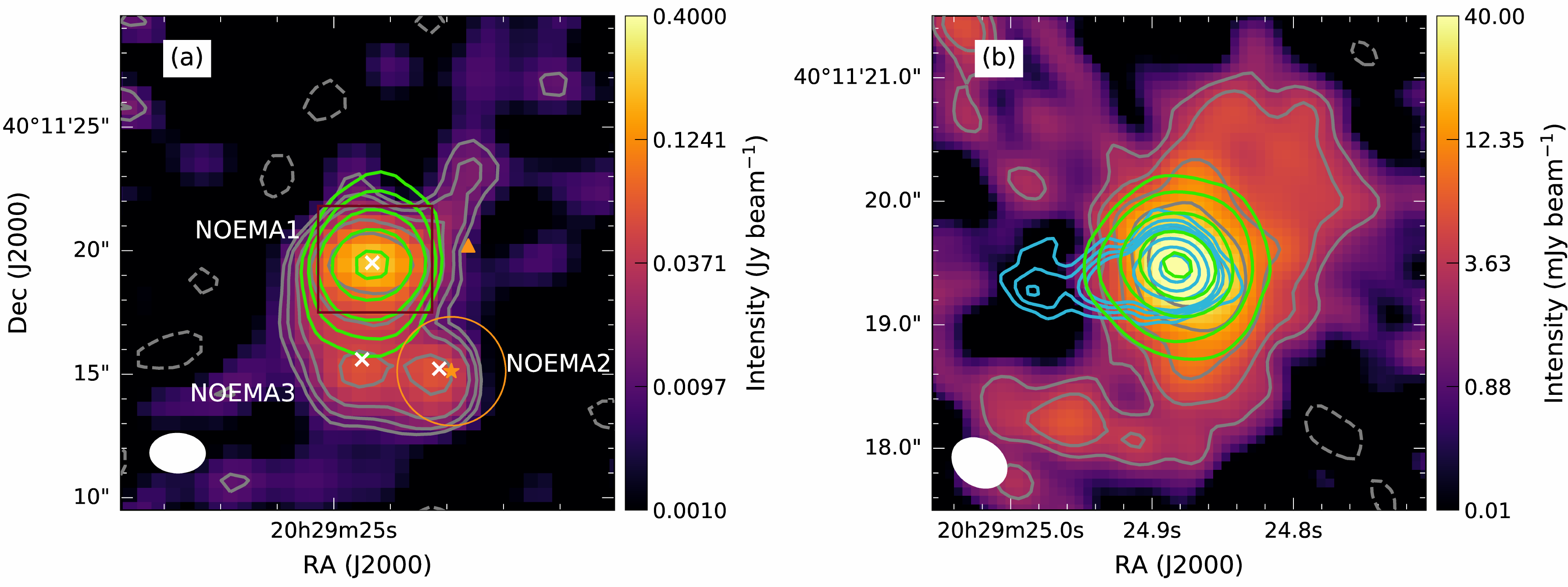}
\caption{NOEMA 1.3\,mm observed (colour scale and grey contours) and model (green contours, see Section~\ref{sect:modelling}) CLEAN images. 
    The compact configuration map is shown in (a) and the levels are $-2, 2, 3, 5, 10, 20\times\sigma_{\rm rms}$ with $\sigma_{\rm rms}=5.0$\,mJy\,beam$^{-1}$. 
    The white crosses mark the position of the continuum sources in Table~\ref{tab:noema}.
    The orange star and circle show the position and 3.6\,cm deconvolved size \citep[${\sim}4.4$\,arcsec;][]{2013A&A...551A..43J} of VLA~1.
    The orange triangle shows the position of VLA~2.
    The extended configurations map is shown in (b) and the levels are $-3, 3, 5, 10, 20, 40\times\sigma_{\rm rms}$ with $\sigma_{\rm rms}=0.8$\,mJy\,beam$^{-1}$. 
    For comparison, the VLA 3.6\,cm observations of \citet{2013A&A...551A..43J} are plotted in light blue contours at $3, 4, 5, 7, 10, 15, 20\times\sigma_{\rm rms}$ levels with $\sigma_{\rm rms}=30\,\mu$Jy\,beam$^{-1}$.
    The dark red square in (a) shows the area covered in (b).
    The observed beam is shown in the bottom left corner.
}
  \label{fig:cont:mm:clean}
\end{figure*}

The beam size of the compact configuration continuum observations is $2\farcs3\times1\farcs6\;{\rm PA}=88\fdg7$, whilst the beam size of the combined extended configurations is $0\farcs48\times0\farcs38\;{\rm PA}=56\degr$. 
Vector-averaged radial visibility profiles were extracted, where we used the standard error of the mean of the points in each uv-distance bin as measure of the errors. 
The profiles are shown in Fig.~\ref{fig:cont:mm:vis} whilst the CLEAN images are shown in Fig.~\ref{fig:cont:mm:clean}.
The NOEMA/CORE A+B observations recover more emission towards smaller baselines than previous PdBI 1.4\,mm A+B observations by \citet{2012A&A...543A..22W} after taking into consideration the difference in wavelengths, whilst achieving similar angular resolution.

The methyl cyanide CH$_3$CN $J=12-11$ line was covered by the narrow-band correlator with a spectral resolution of 0.312\,MHz (0.43\,km\,s$^{-1}$). 
A data cube covering the transition's $K$-ladder between $K=0-5$ was extracted. 
The beam size of these observations is $2\farcs6\times1\farcs8\;{\rm PA}=87\degr$.
The average noise measured in an annulus of sky along all channels is $\sigma_{\rm rms}=30$\,mJy\,beam$^{-1}$ per channel, with variations between $10-50$\,mJy\,beam$^{-1}$ at different channels. 

\begin{figure}
  \centering
  \includegraphics[width=\columnwidth]{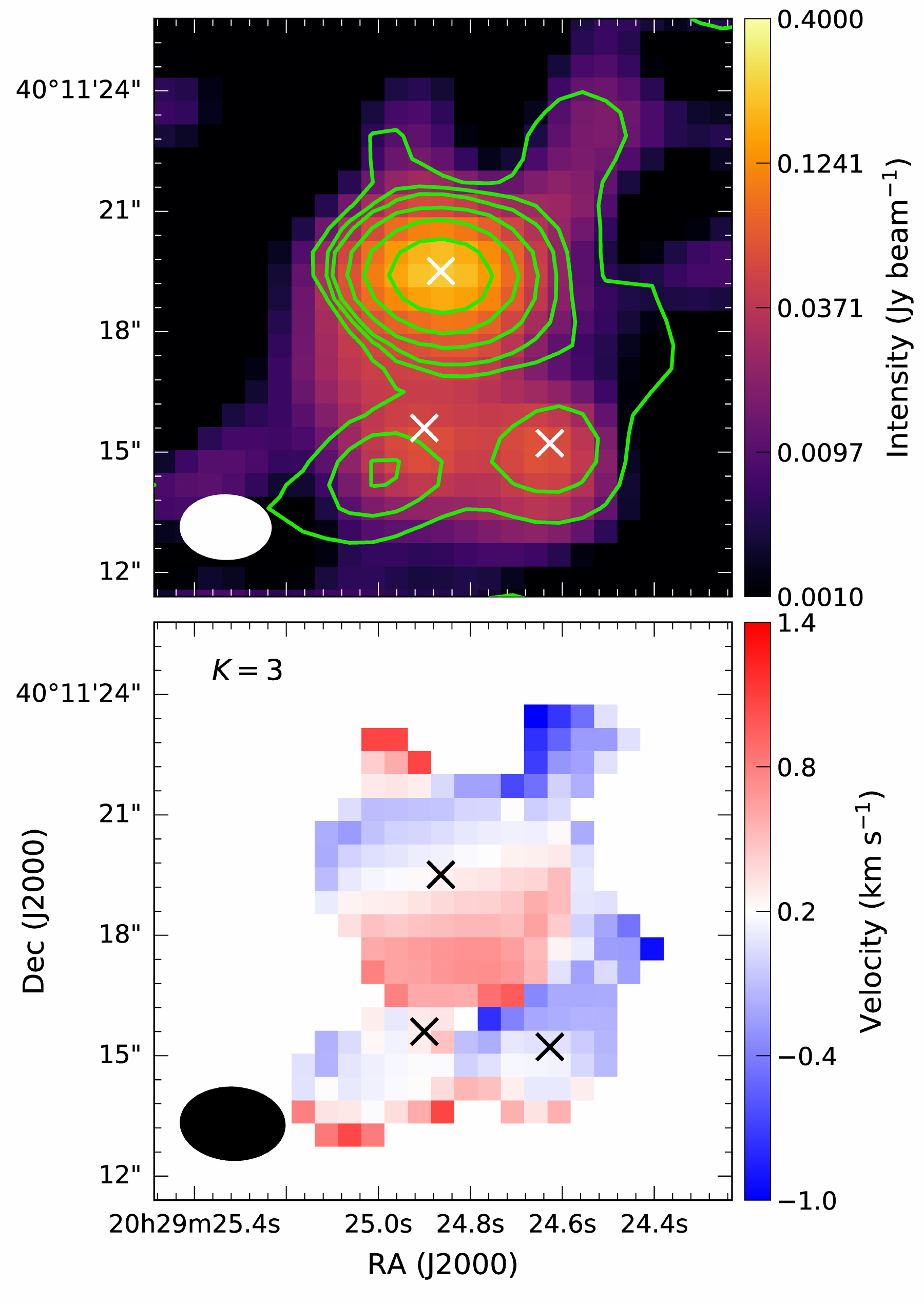} 
  \caption{NOEMA compact configuration zeroth (upper row) and first (lower row) moment maps of CH$_3$CN $J=12-11$ for the $K=3$ transition. 
        The continuum is plotted in colour scale in the zeroth moment maps.
        The zeroth moment contour levels are $1, 3, 5, 10, 20, 40\times\sigma_{\rm rms}$ with $\sigma_{\rm rms}=0.45$\,Jy\,beam$^{-1}$\,km\,s$^{-1}$.
        Zero velocity correspond to the systemic velocity in the LSR system ($v_{\rm LSR}=-5.5$\,km\,s$^{-1}$).
        The positions of the continuum sources are marked with a cross and the beam size is shown in the bottom left corner.
    }
  \label{fig:line:mom0_mom1}
\end{figure}

Zeroth- and first-order moment maps for the transition with $K=3$ are shown in Fig.~\ref{fig:line:mom0_mom1}.
The first moment maps were calculated considering all data with fluxes larger than $5\sigma_{\rm rms}=0.15$\,Jy\,beam$^{-1}$ per channel over a range of spectral channels covering the lines.
The local standard of rest (LSR) velocity measured by \citet[][$-5.5\pm0.5$\,km\,s$^{-1}$]{2014A&A...567A..53K} over 24 lines in \textit{Herschel}/HIFI observations was used to shift the velocities, so zero velocity corresponds to the line systemic velocity.

We identified 3 continuum sources from the compact configuration data named NOEMA~1--3 and their positions and peak intensities are summarised in Table~\ref{tab:noema:summary}.
The position of NOEMA~1 is consistent with the VLA~3 source, for consistency to previous works we will use the latter name.
Fig.~\ref{fig:cont:mm:clean}(a) reveals the presence of at least 2 sources south of VLA~3.
Thus, the NOEMA images (continuum and zeroth moment compact configuration maps) were fitted with three 2-D Gaussian components and the results are listed in Table~\ref{tab:noema}.
The observed size of VLA~3 decreases slightly with increasing $K$ transition, i.e. decreases with increasing upper energy of the line, and the source is only partially resolved by the molecular line observations.
The orientation of the source from the methyl cyanide emission seems to agree with the large-scale orientation of the near-IR images, but the beam is oriented in the same direction.
In fact, the deconvolved PAs shown in Table~\ref{tab:noema:deconv} have large uncertainties, thus the orientation is strongly affected by the beam.

\begin{table}
\centering
\caption{Summary of NOEMA compact configuration continuum sources.}
\label{tab:noema:summary}
\begin{tabular}{lcccc}
\hline
NOEMA  & VLA    & RA      & Dec     & Peak flux\\
Source & Source & (J2000) & (J2000) & (mJy beam$^{-1}$) \\
\hline
1	& 3		& 20 29 24.864 & 40 11 19.51 & 230\\
2	& 1		& 20 29 24.627 & 40 11 15.22 & 62\\
3	& --	& 20 29 24.900 & 40 11 15.60 & 61\\
\hline
\end{tabular}
\end{table}

\begin{figure}
  \centering
  \includegraphics[width=\columnwidth]{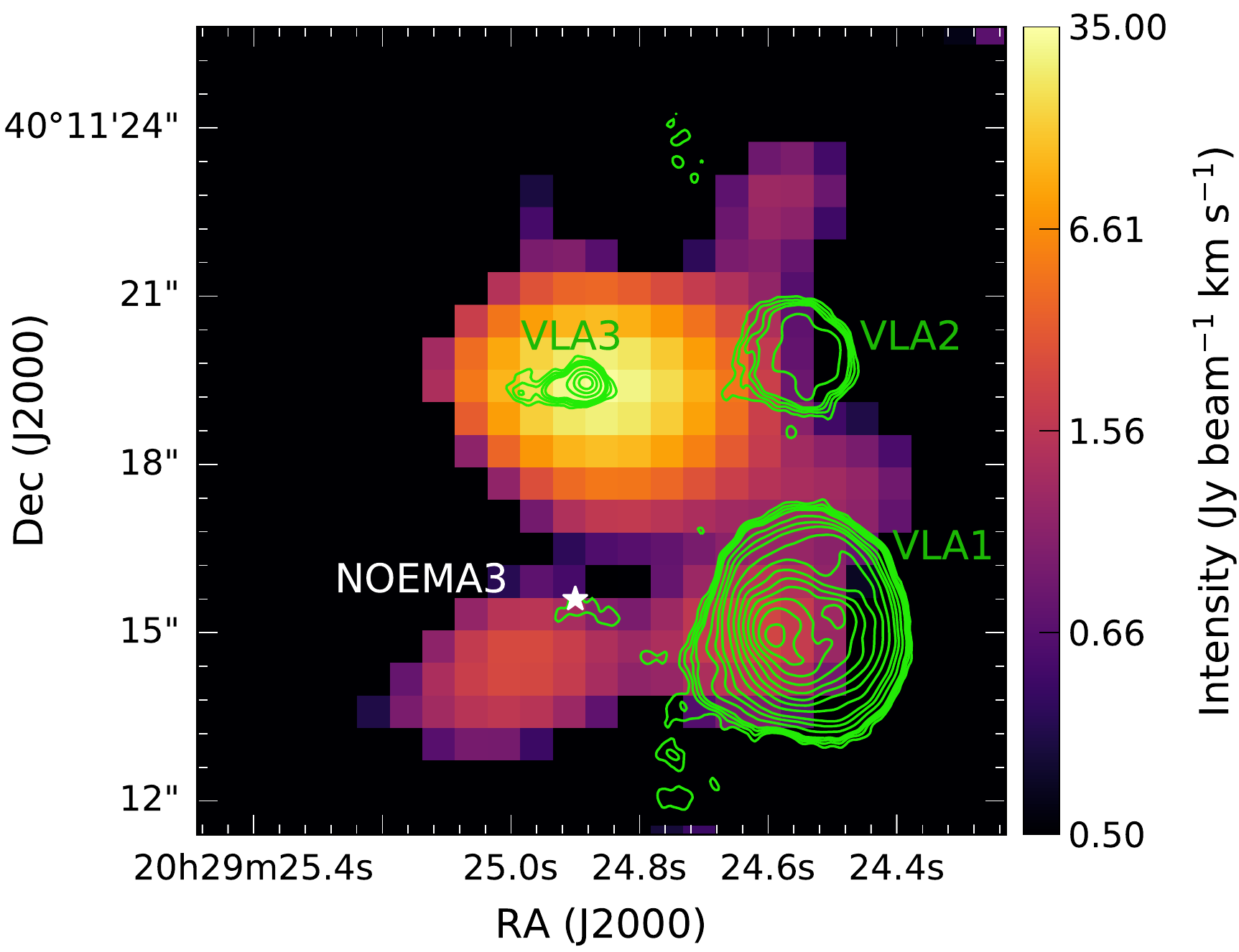} 
  \caption{VLA 3.6\,cm continuum emission (green contours) from \citet{2013A&A...551A..43J} superimposed to the NOEMA compact configuration CH$_3$CN $J=12-11$ $K=3$ zeroth moment map (colour scale). 
  Sources are labelled and the position of the continuum source NOEMA~3 is marked with a white star. 
  Following \citet{2013A&A...551A..43J}, contours are 3, 4, 5, 7, 10, 15, 20, 30--100$\times\sigma_{\rm rms}$ ($\sigma_{\rm rms}=30\,\mu$Jy\,beam$^{-1}$).
  }
  \label{fig:line:radio}
\end{figure}

Similarly, a single 2-D Gaussian was fitted to the extended configurations emission in  Fig.~\ref{fig:cont:mm:clean}(b).
The results listed in Table~\ref{tab:noema} show that the position angle of the emission is closer to perpendicular to the radio jet in Fig.~\ref{fig:line:radio} \citep[${\sim}100\degr$,][]{2013A&A...551A..43J} rather than oriented in the outflow direction.
Although the Gaussian fit is just an approximation, the orientation of the lower level contours in Fig.~\ref{fig:cont:mm:clean}(b) shows a similar orientation. 

\begin{landscape}
\begin{table}
\caption{Measured properties from 2-D Gaussian fit to the NOEMA 1.3\,mm continuum and CH$_3$CN zeroth-order moment emission.}
\label{tab:noema}
\begin{tabular}{lcccccccccc}
\hline
    $K^{a}$ & Frequency & $E_u^{a}$ & RA & Dec & Peak$^{b}$ & $T_{\rm B}^{c}$ &
    Flux$^{d}$  & Major & Minor & PA\\
     & (GHz) & (K) & (hh:mm:ss) & (\degr:':") &  & (K) &  & (arcsec) & (arcsec) & (deg)\\
\hline
    \multicolumn{10}{c}{NOEMA~1 (VLA~3)}\\
    \multicolumn{3}{l}{Continuum (compact)}  & 20 29 24.864 & 40 11 19.51 & 210$\pm$6 & -- & 436$\pm$12 & 3.28$\pm$0.04 & 2.37$\pm$0.06 & 267$\pm$2 \\ 
    \multicolumn{3}{l}{Continuum (extended)} & 20 29 24.875 & 40 11 19.46 &  50$\pm$3 & -- & 180$\pm$11 & 0.90$\pm$0.05 & 0.72$\pm$0.03 & 21$\pm$9 \\ 
    0 & 220.74726 & 68.87  & 20 29 24.864 & 40 11 19.41 & 32.7$\pm$0.6 & 40.9 & 39.3$\pm$0.7 & 2.87$\pm$0.03 & 2.04$\pm$0.05 & 87.1$\pm$0.9 \\
    1 & 220.74301 & 76.01  & 20 29 24.861 & 40 11 19.38 & 26.4$\pm$0.5 & 37.9 & 31.6$\pm$0.6 & 2.85$\pm$0.03 & 2.04$\pm$0.05 & 86.8$\pm$1.0 \\
    2 & 220.73026 & 97.44  & 20 29 24.861 & 40 11 19.39 & 24.5$\pm$0.6 & 34.5 & 28.9$\pm$0.7 & 2.82$\pm$0.04 & 2.03$\pm$0.06 & 86.1$\pm$1.3\\
    3 & 220.70902 & 133.16 & 20 29 24.861 & 40 11 19.39 & 32.1$\pm$0.6 & 53.9 & 37.8$\pm$0.7 & 2.81$\pm$0.04 & 2.03$\pm$0.05 & 84.7$\pm$1.1 \\
    4 & 220.67929 & 183.15 & 20 29 24.861 & 40 11 19.36 & 16.2$\pm$0.4 & 22.3 & 18.3$\pm$0.5 & 2.81$\pm$0.05 & 1.96$\pm$0.07 & 84.6$\pm$1.3 \\
    5 & 220.64108 & 247.40 & 20 29 24.856 & 40 11 19.36 & 12.4$\pm$0.3 & 15.3 & 13.7$\pm$0.4 & 2.86$\pm$0.05 & 1.93$\pm$0.07 & 87.7$\pm$1.3 \\
    \multicolumn{10}{c}{NOEMA~2}\\
    \multicolumn{3}{l}{Continuum (compact)} & 20 29 24.627 & 40 11 15.22 & 48$\pm$6 & -- & 76$\pm$9 & 2.9$\pm$0.3 & 2.0$\pm$0.2 & 18$\pm$7 \\ 
    0 & 220.74726 & 68.87  & 20 29 24.62 & 40 11 15.5 & 2.3$\pm$0.3 & 2.7 & 3.8$\pm$0.5 & 3.2$\pm$0.3 & 2.6$\pm$0.3 & 137$\pm$14 \\ 
    1 & 220.74301 & 76.01  & 20 29 24.62 & 40 11 15.5 & 1.9$\pm$0.3 & 3.3 & 3.3$\pm$0.4 & 3.1$\pm$0.3 & 2.7$\pm$0.3 & 140$\pm$15 \\
    2 & 220.73026 & 97.44  & 20 29 24.62 & 40 11 15.6 & 1.6$\pm$0.3 & 2.3 & 3.4$\pm$0.5 & 3.5$\pm$0.4 & 3.0$\pm$0.3 & 151$\pm$17 \\
    3 & 220.70902 & 133.16 & 20 29 24.63 & 40 11 15.4 & 2.1$\pm$0.3 & 2.9 & 4.1$\pm$0.6 & 3.4$\pm$0.4 & 2.8$\pm$0.3 & 136$\pm$15 \\
    4 & 220.67929 & 183.15 & 20 29 24.63 & 40 11 15.5 & 1.0$\pm$0.2 & 1.3 & 2.1$\pm$0.3 & 3.7$\pm$0.4 & 2.8$\pm$0.3 & 2.7$\pm$9.5 \\
    5 & 220.64108 & 247.40 & 20 29 24.63 & 40 11 15.4 & 1.2$\pm$0.1 & 1.1 & 2.0$\pm$0.2 & 3.2$\pm$0.2 & 2.6$\pm$0.3 & 110$\pm$9 \\
    \multicolumn{10}{c}{NOEMA~3}\\
    \multicolumn{3}{l}{Continuum (compact)} & 20 29 24.90 & 40 11 15.6 & 57$\pm$6 & -- & 250$\pm$25 & 4.9$\pm$0.2 & 3.3$\pm$0.2 & 49$\pm$5 \\
    0 & 220.74726 & 68.87  & 20 29 24.99 & 40 11 14.5 & 2.6$\pm$0.3 & 2.8 & 3.2$\pm$0.3 & 3.2$\pm$0.2 & 1.9$\pm$0.3 & 114$\pm$4 \\
    1 & 220.74301 & 76.01  & 20 29 24.99 & 40 11 14.5 & 2.1$\pm$0.2 & 3.1 & 2.8$\pm$0.3 & 3.2$\pm$0.2 & 2.0$\pm$0.2 & 113$\pm$4 \\
    2 & 220.73026 & 97.44  & 20 29 24.99 & 40 11 14.5 & 2.1$\pm$0.3 & 2.9 & 2.3$\pm$0.3 & 2.9$\pm$0.3 & 1.8$\pm$0.3 & 117$\pm$6 \\
    3 & 220.70902 & 133.16 & 20 29 24.98 & 40 11 14.5 & 2.5$\pm$0.3 & 3.4 & 3.2$\pm$0.4 & 3.3$\pm$0.2 & 1.9$\pm$0.3 & 110$\pm$4 \\
    4 & 220.67929 & 183.15 & 20 29 24.98 & 40 11 14.4 & 1.5$\pm$0.3 & 1.4 & 1.7$\pm$0.3 & 3.0$\pm$0.3 & 1.8$\pm$0.4 & 111$\pm$6 \\
    5 & 220.64108 & 247.40 & 20 29 24.97 & 40 22 14.6 & 1.0$\pm$0.2 & 1.3 & 1.2$\pm$0.2 & 3.2$\pm$0.3 & 1.8$\pm$0.4 & 110$\pm$6 \\
\hline
\end{tabular}
\\
    \begin{minipage}{\textwidth}
        \textbf{Notes.} The 2-D Gaussian components were fitted simultaneously to the continuum images, whilst for the methyl cyanide zeroth moment maps each source was fitted separately within a box isolating each source.\\
        $^a$ From Leiden Atomic and Molecular Database (LAMDA).\\
        $^b$ The units of peak fluxes are mJy\,beam$^{-1}$ for the continuum observations and Jy\,beam$^{-1}$\,km\,s$^{-1}$ for the molecular lines.\\
        $^c$ Brightness temperature at the line peak at the position of the zeroth moment peak.\\
        $^d$ The units of flux are mJy for the continuum observations and Jy\,km\,s$^{-1}$ for the molecular lines. 
    \end{minipage}
\end{table}
\end{landscape}

\begin{table}
\caption{Deconvolved 2-D Gaussian fit sizes for NOEMA~1 (VLA~3) from NOEMA compact configuration CH$_3$CN emission at different $K$ transition.}
\centering
\label{tab:noema:deconv}
\begin{tabular}{cccc}
\hline
    $K$ & Major & Minor & PA\\
     & (arcsec) & (arcsec) & (deg) \\
     \hline
    0 & 1.14$\pm$0.08 & 0.86$\pm$0.12 & 87$\pm$175\\ 
    1 & 1.09$\pm$0.09 & 0.88$\pm$0.13 & 83$\pm$173\\
    2 & 1.00$\pm$0.13 & 0.84$\pm$0.19 & 73$\pm$79 \\
    3 & 1.01$\pm$0.13 & 0.80$\pm$0.18 & 58$\pm$41 \\
    4 & 1.00$\pm$0.17 & 0.61$\pm$0.37 & 69$\pm$31 \\
    5 & 1.12$\pm$0.14 & 0.55$\pm$0.43 & 90$\pm$176\\
     \hline
\end{tabular}
\end{table}

\begin{figure}
  \centering
  \includegraphics[width=\columnwidth]{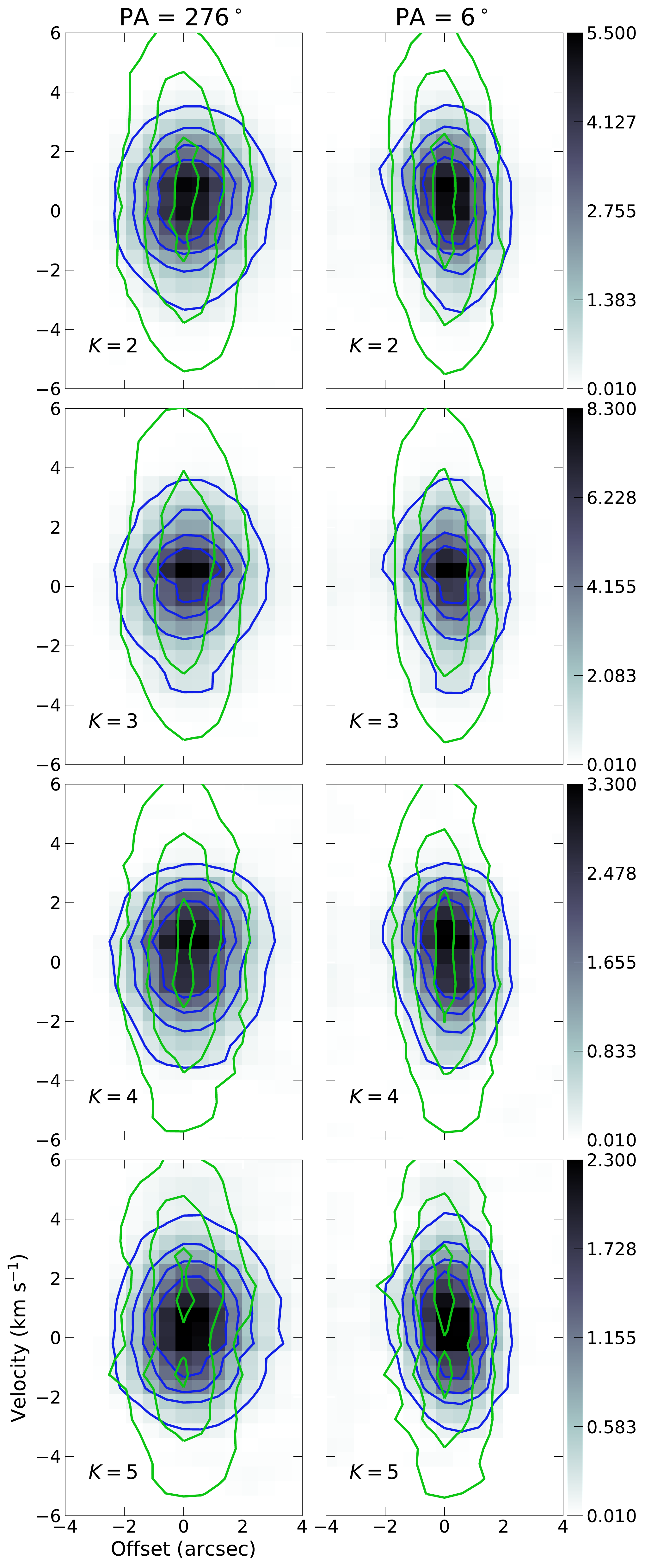} 
    \caption{Observed position-velocity (pv) maps (grey scale and blue contours) and best-fitting model pv maps (green contours, see Section~\ref{sect:modelling}) of the CH$_3$CN $J=12-11$ $K=2-5$ line emission for a position angle of 276\degr\ (left column), i.e. along the rotation axis, and 6\degr\ (right column). 
    Contours levels are 10--80~per~cent in 20~per~cent steps of 5, 8, 3 and 2\,Jy\,beam$^{-1}$ for each $K$ line, respectively.
    }
  \label{fig:line:pv}
\end{figure}

To extract the kinematic information from the observations of VLA~3, we calculated position-velocity (pv) maps for CH$_3$CN lines with $K=2-5$ and are shown in Fig.~\ref{fig:line:pv}.
The slices were obtained at 6\degr\ (along the putative disc major axis from \citealp{2012A&A...543A..22W}) and 276\degr\ (along the outflow direction; see online Appendix~\ref{ap:dataproc} for a discussion) with a slice width of 1.8\,arcsec (3 times the pixel size and close to the beam width).

The position of NOEMA~2 is close to VLA~1 and its total flux (${\sim}0.1$\,mJy) is consistent with the value predicted from the VLA~1 radio spectral index \citep[$\alpha=0$,][]{2013A&A...551A..43J}. 
In Fig.~\ref{fig:line:radio}, it can be seen that the methyl cyanide emission of NOEMA~2 coincides with emission from the eastern hemisphere of VLA~1. 
This is consistent with VLA~1 being a slightly cometary \ion{H}{II} region with the ionising gas expanding towards the less dense gas in the west.

The source immediately south of VLA~3, NOEMA~3, has an intensity peak which is shifted ${\sim}4$\,arcsec towards the south east from the continuum peak of VLA~3. 
This source is not associated with 3.6\,cm radio continuum emission above a level of $3\sigma=90$\,$\mu$Jy~beam$^{-1}$ (see Fig.~\ref{fig:line:radio}) and has not been detected in the IR, thus its nature is as yet unknown. 
Fig.~\ref{fig:line:mom0_mom1} also shows that the velocity structures of the three sources can be separated from each other. 

\subsection{Other continuum multi-wavelength data}
\label{sect:data:multi}

Multi-wavelength observations were used to constrain the dust density and temperature distributions. 
Since these images were previously processed, only the sky level was subtracted when necessary. 
Table~\ref{tab:cont} presents a summary of the observations which were used to extract spatial information for the fitting and are described below.

\begin{table}
 \centering
 \caption{Summary of continuum observations used to extract spatial information.}
 \label{tab:cont}
 \begin{tabular}{cccc}
  \hline
  $\lambda$ & Instrument & Resolution & Type$^\rmn{a}$\\
  (\micron)&&&\\
  \hline
  1.2 & UKIRT/WFCAM & 0\farcs9 & I \\
  1.6 & UKIRT/WFCAM & 0\farcs9 & I \\
  2.1 & UKIRT/WFCAM & 0\farcs9 & I \\
  2.1 & SAO/6~m & 0.17\arcsec & V \\
     70 & \textit{Herschel}/PACS & $6\farcs51\times5\farcs44$ & I\\
  450 & JCMT/SCUBA & 9\arcsec & P \\
  850 & JCMT/SCUBA & 14\arcsec & P \\
  1300 & NOEMA (D) & $2\farcs3 \times 1\farcs6$ PA $88\fdg7$& V \\
  1300 & NOEMA (A+B) & $0\farcs48 \times 0\farcs38$ PA $56\degr$ & V\\
  \hline
 \end{tabular}
 \begin{minipage}{\columnwidth}
 $^\rmn{a}$ Type of data product used to compare observations and models: I for images, V for visibility profiles and P for azimuthally averaged radial profiles.
 \end{minipage}
\end{table}

\subsubsection{Near-IR imaging}
\label{sect:data:multi:nir}

\begin{figure}
  \centering
  \includegraphics[scale=0.5]{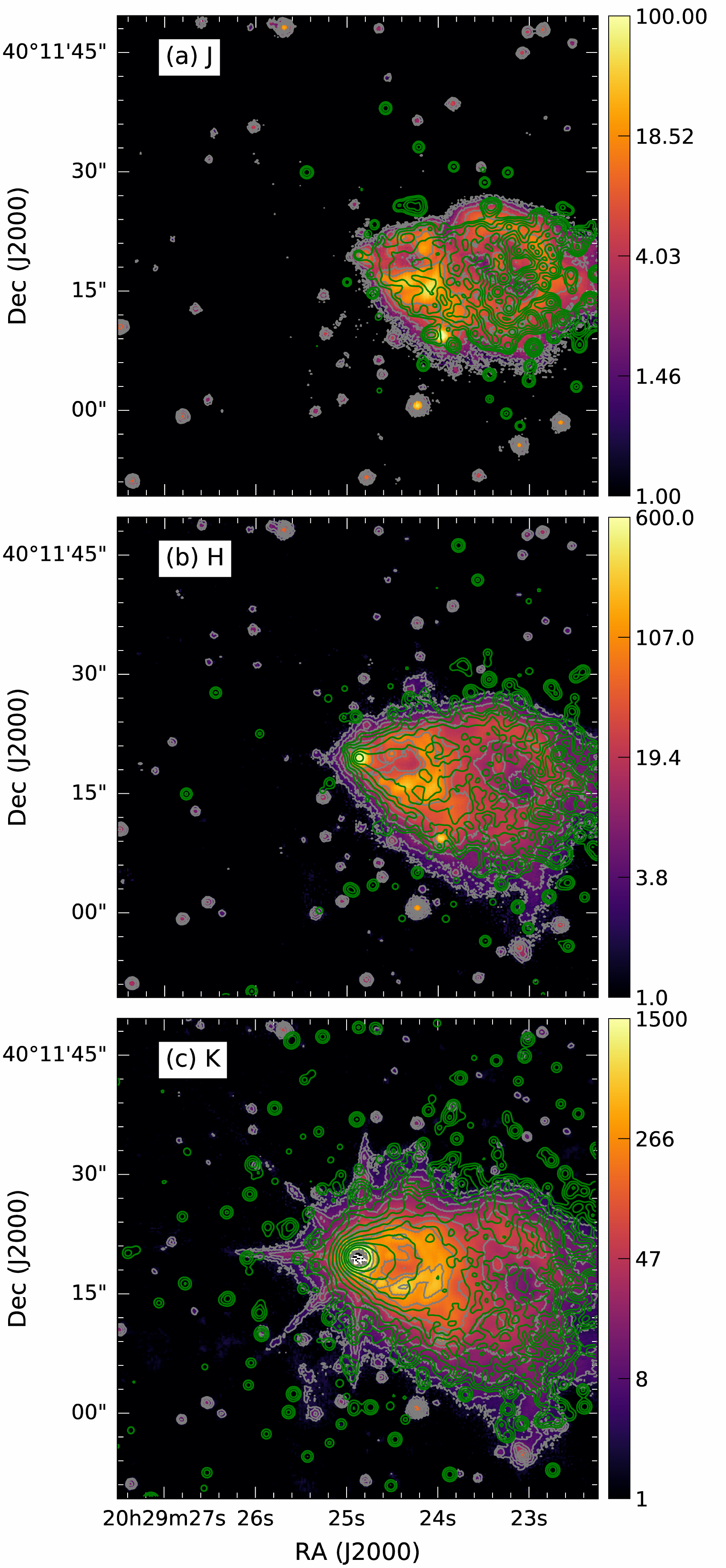}
  \caption{Observed (colour map and grey contours) and model (green contours, see Section~\ref{sect:modelling}) near-IR maps of AFGL~2591. 
      The $J$-band data and model are shown in (a) and contour levels are $3, 5, 10, 20, 40\times\sigma_{\rm rms}$ with $\sigma_{\rm rms}=0.3$\,MJy\,sr$^{-1}$. 
      (b) shows the $H$-band data and model in contours at levels $5, 10, 20-160\times\sigma_{\rm rms}$ with $\sigma_{\rm rms}=0.4$\,MJy\,sr$^{-1}$.
      (c) shows the $K$-band data and model in contours at levels $5, 10, 20-2560\times\sigma_{\rm rms}$ with $\sigma_{\rm rms}=0.6$\,MJy\,sr$^{-1}$. 
      The colour maps are in units of MJy\,sr$^{-1}$.
      }
  \label{fig:cont:nir}
\end{figure}

Near-IR images from the UKIRT Wide Field Camera \citep[WFCAM, ][]{2007A&A...467..777C} at 1.2, 1.6 and 2.1\,\micron, $J$, $H$ and $K$ bands respectively, were obtained from the UKIRT Infrared Deep Sky Survey \citep[UKIDSS;][]{2008MNRAS.391..136L} and are presented in Fig.~\ref{fig:cont:nir}. 
These data have higher resolution (${\sim}1$\,arcsec) than the 2MASS data used by \citet{2013A&A...551A..43J}, which had a resolution of ${\sim}2.5$\,arcsec.
The images were flux-calibrated and error maps were obtained by using Poisson statistics on the un-calibrated data converted to counts. 
A Moffat function was fitted to several saturated and unsaturated nearby point-like sources to produce a PSF image. 
This function has been proven reliable in reproducing the PSF wings \citep[e.g.][]{2011MNRAS.414.2055M}. 
The core of saturated sources were masked in order to fit better the PSF wings. 
The FWHM of the fitted Moffat function are $0.9\pm0.1$, $0.9\pm0.2$ and $0.9\pm0.2$\,arcsec and the atmospheric scattering coefficients are $3.0\pm0.4$, $2.7\pm0.4$ and $2.7\pm0.7$ for the $J$, $H$ and $K$-bands, respectively.
These widths are consistent with the observed seeing FWHM of 0.8\,arcsec, as recorded in the header of the images.

\begin{figure}
  \centering
\includegraphics[width=\columnwidth]{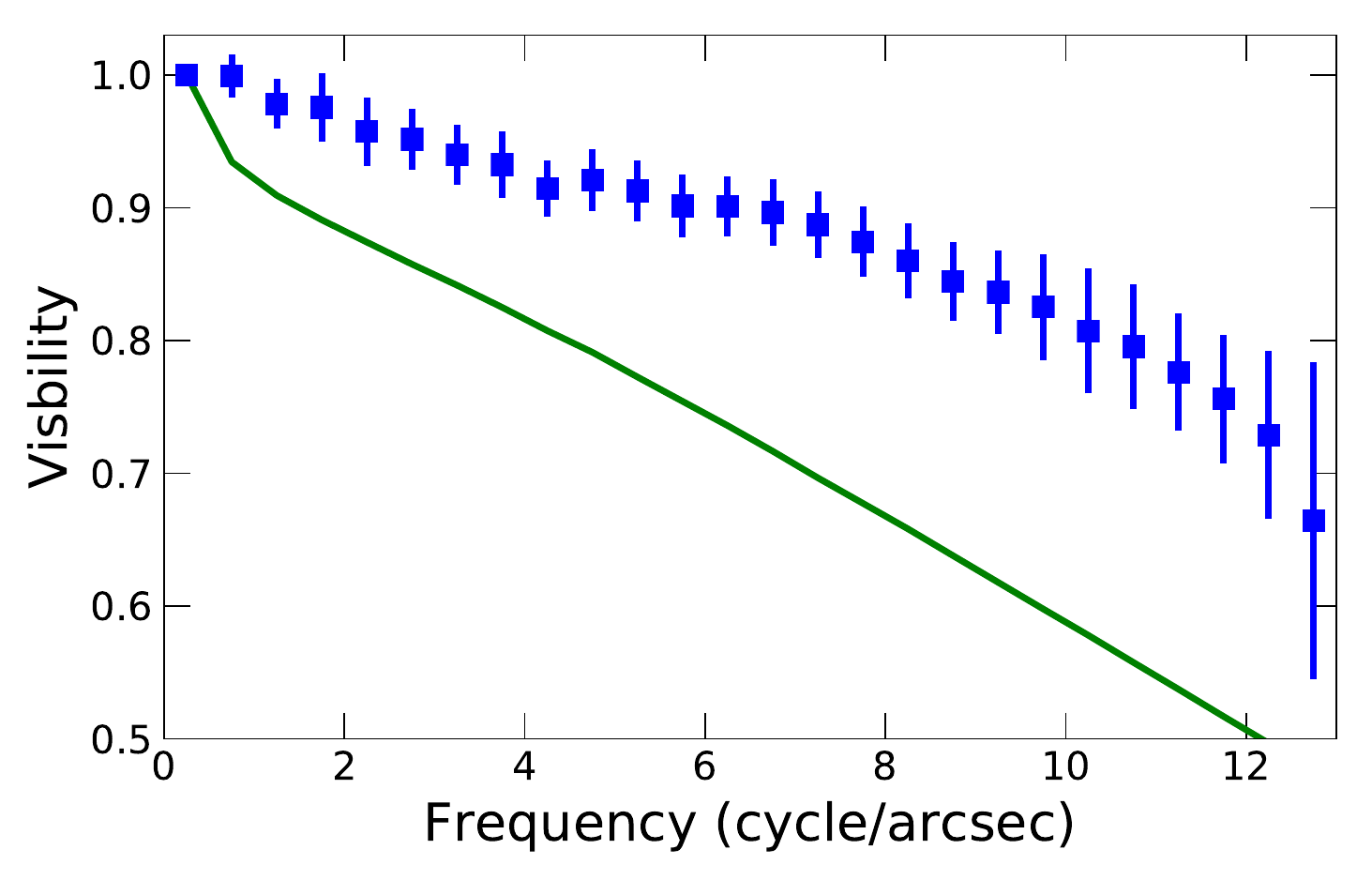}
  \caption{$K$-band speckle interferometry visibilities radial profile. 
      Observed points are represented by blue squares and the best-fitting model by a continuous green line. 
	  The model clearly underestimates the observations (see Section~\ref{sect:results:cont}).
  }
  \label{fig:cont:nir:vis}
\end{figure}

In order to study the regions close to the MYSO, which are saturated in the UKIRT images, we used Special Astrophysical Observatory (SAO) 6\,m telescope $K$-band speckle interferometric visibilities from \citet{2003A&A...412..735P}. 
The visibilities were averaged over annuli with constant width and the standard deviation of the data in each annulus was used as a measure of the errors. 
Fig.~\ref{fig:cont:nir:vis} shows the visibility radial profile. 
It is worth  noticing that the visibilities are reasonably symmetric which results in relatively small errors \citep[see][their fig.~2]{2003A&A...412..735P}.

\subsubsection{(Sub)mm}
\label{sect:data:multi:submm}

\begin{figure*}
  \centering
\includegraphics[width=\textwidth]{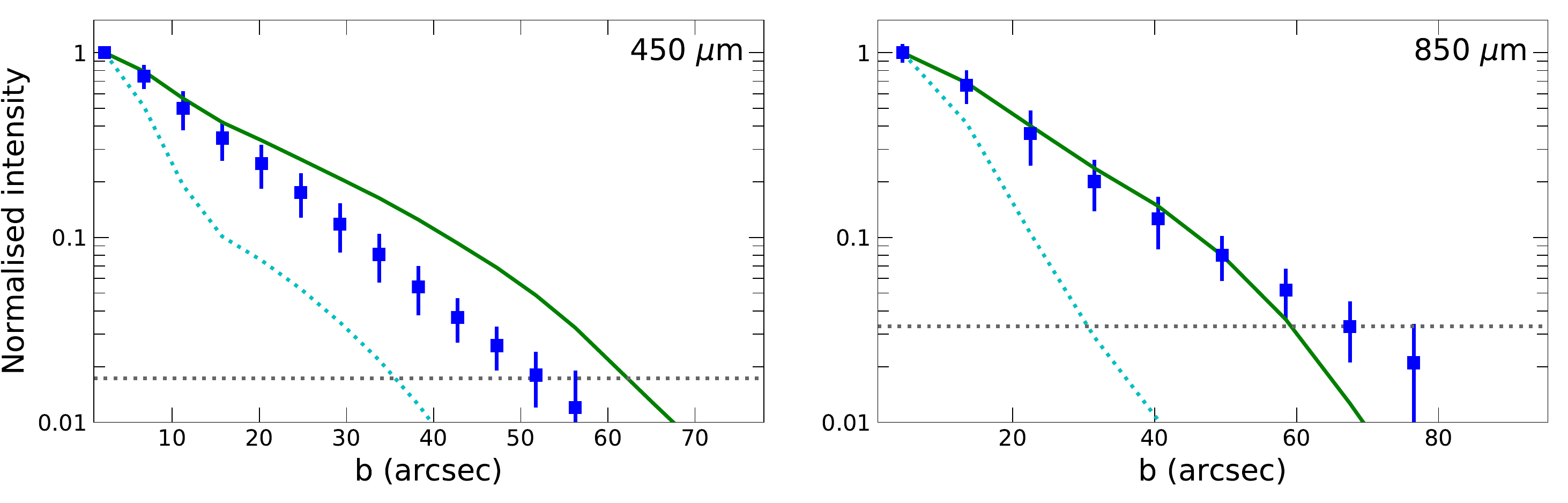}
  \caption{Normalised radial intensity profiles at 450 and 850\,\micron. 
      Observed profiles are represented with blue squares and the best-fitting model by a green continuous line (see Section~\ref{sect:modelling}). 
      The horizontal dotted line represents the $3\sigma$ level above zero, where $\sigma$ is the noise level. 
      Observed points above this level were compared with the model ones. 
      The cyan dotted line shows the PSF profile.
    }
  \label{fig:cont:submm:prof}
\end{figure*}

In the submm, images were obtained from the JCMT Legacy Catalogue \citep{2008ApJS..175..277D} which were taken with SCUBA\footnote{Submillimetre Common-User Bolometer Array} at 450 and 850\,\micron. 
Following \citet{2008ApJS..175..277D}, we prepared a PSF composed of two Gaussians which reproduce the main and secondary lobes. 
The main lobes have a FWHM of 9 and 14\,arcsec at 450 and 850\,\micron\ respectively, whilst 2-D Gaussians fitted to observed emission towards AFGL~2591 have a FWHM of $37\farcs1\times28\farcs5\;{\rm PA}=247\degr\pm1\degr$ and $47\farcs6\times38\farcs7\;{\rm PA}=233\degr\pm1\degr$ respectively. 
Hence the source is elongated with position angles of the 2-D Gaussian major axis consistent with the outflow cavity direction, as in the 70\,\micron\ data. 
However, since the real PSF is known to be more complicated and changes between nights \citep[e.g.][]{2000A&A...357..637H}, we fit azimuthally averaged radial intensity profiles (hereafter radial intensity profiles). 
The radial intensity profiles shown in Fig.~\ref{fig:cont:submm:prof} were calculated as the average flux of concentric annuli regions of constant width and their errors are the standard deviation of the enclosed fluxes. 

\subsubsection{Spectral energy distribution}
\label{sect:data:multi:sed}

\begin{figure}
  \centering
  \includegraphics[width=\columnwidth]{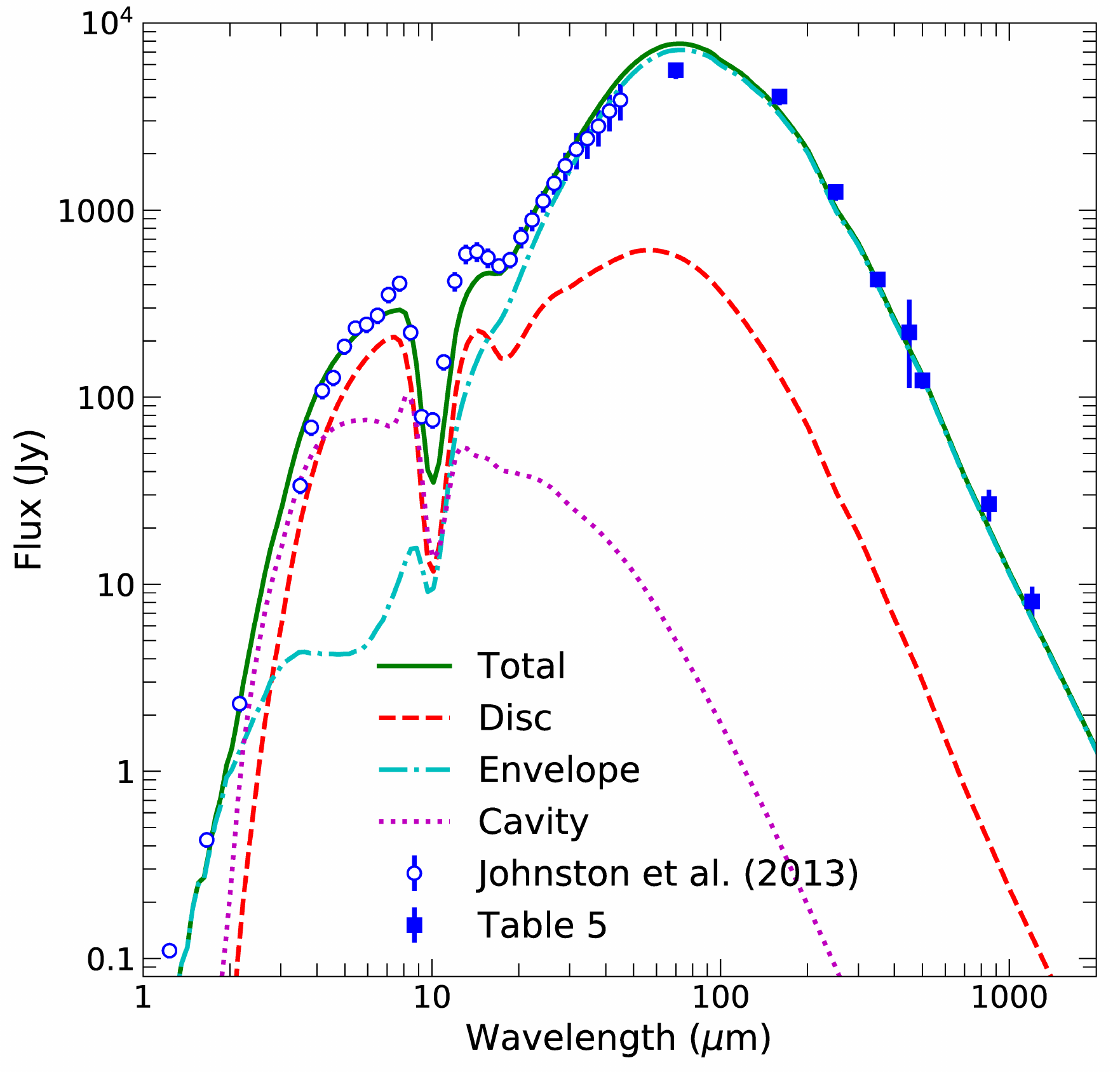}
  \caption{SED of AFGL~2591 and the best-fitting model.
      The emission from each of the different components included in the model are also shown (see legend).
      }
  \label{fig:cont:sed}
\end{figure}

The spectral energy distribution (SED) consists of data points from \citet{2013A&A...551A..43J} for $\lambda <60$\,\micron\ and the data points in Table~\ref{tab:sed} for longer wavelengths. 
The whole SED is shown in Fig.~\ref{fig:cont:sed}. 
The fluxes for the \textit{Herschel} bands, observed by PACS and the Spectral and Photometric Imaging Receiver \citep[SPIRE, ][]{2010A&A...518L...3G}, were obtained from the sum of all pixels with intensity higher than three times the image noise. 
This filters out the filamentary emission observed in SPIRE data whilst the fluxes are within the calibration uncertainties of the ones calculated with an aperture covering the whole region. 
The flux calibration uncertainties for PACS and SPIRE in the literature range from ${\sim}5$ to 15~per~cent 
\citep[e.g.][and \textit{Herschel} online documentation\footnote{SPIRE: \url{http://herschel.esac.esa.int/twiki/bin/view/Public/SpireCalibrationWeb}\\PACS: \url{http://herschel.esac.esa.int/twiki/bin/view/Public/PacsCalibrationWeb}}]{2013ApJ...767..126S, 2014A&A...562A.138R}. 
We adopted an uncertainty of 10~per~cent for both. 

\begin{table}
 \centering
 \caption{SED data points for $\lambda\geqslant60$~\micron.}
 \label{tab:sed}
 \begin{tabular}{cccc}
  \hline
  Wavelength & Flux &  Instrument & Ref.\\
  (\micron) & (Jy) &  &\\
  \hline
  70   & $5600\pm560$ & \textit{Herschel}/PACS  & (1)\\
  160  & $4055\pm406$ & \textit{Herschel}/PACS  & (1)\\
  250  & $1250\pm125$  & \textit{Herschel}/SPIRE & (1)\\
  350  & $426\pm43$   & \textit{Herschel}/SPIRE & (1)\\
  450  & $222\pm110$  & JCMT/SCUBA              & (2)\\
  500  & $123\pm12$    & \textit{Herschel}/SPIRE & (1)\\
  850  & $26.8\pm5.2$ & JCMT/SCUBA              & (2)\\
  1200 & $8.1\pm0.8$  & IRAM 30m/MAMBO          & (3)\\
  \hline
 \end{tabular}
 \begin{minipage}{\columnwidth}
     (1) This work. 
     Data from PACS and SPIRE were obtained from the HOBYS observations, with SPIRE data from high gain dedicated observations.\\
     (2) \citet{2008ApJS..175..277D}\\
     (3) \citet{2011ApJ...727..114R} based on \citet{2007A&A...476.1243M}
 \end{minipage}
\end{table}

\section{Data modelling}
\label{sect:modelling}

\subsection{Multi-wavelength data modelling}
\label{sect:modelling:multi}

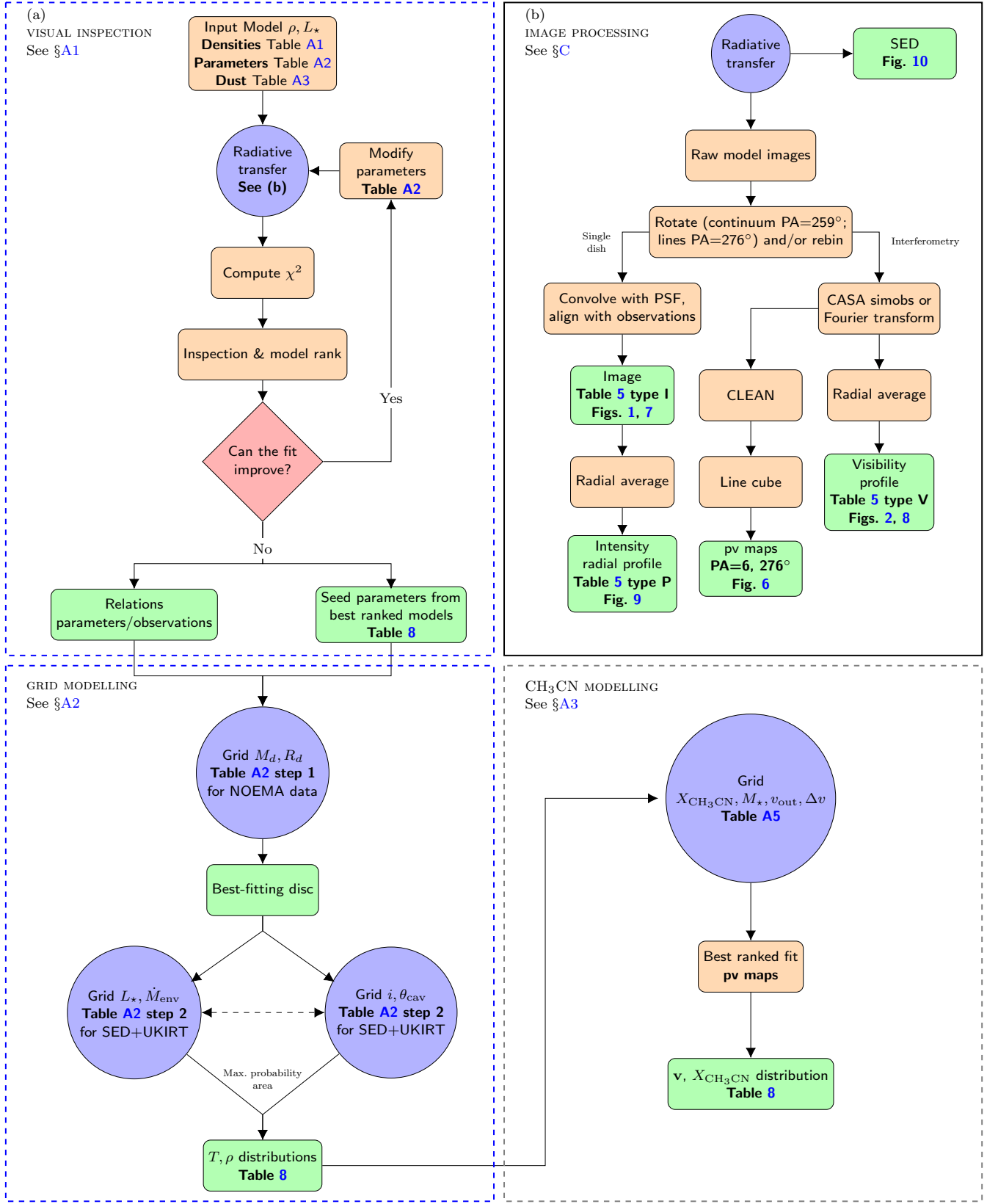
\begin{figure*}
\resizebox{\textwidth}{!}{
\begin{tikzpicture}[node distance=1.5cm,
    every node/.style={fill=white}, align=center]
    \node (start)             [noderect]                                      {Input Model $\rho, L_\star$\\\textbf{Densities} Table~\ref{tab:densities}\\\textbf{Parameters} Table~\ref{tab:cont:parameters}\\\textbf{Dust} Table~\ref{tab:dust}};
    \node (rtrun1)            [nodecirc, below of=start, yshift=-0.8cm]       {Radiative\\transfer\\\textbf{See (b)}};
  \node (compare1)          [noderect, below of=rtrun1, yshift=-0.5cm]      {Compute $\chi^2$};
  \node (rank1)             [noderect, below of=compare1, yshift=-0.1cm]                   {Inspection \& model rank};
  \node (decision1)         [nodediam, below of=rank1, yshift=-0.6cm]       {Can the fit\\improve?};
    \node (modify1)           [noderect, right of=rtrun1, xshift=1.0cm]       {Modify\\parameters\\\textbf{Table~\ref{tab:cont:parameters}}};
  \node (result1)           [noderesult, below of=decision1, xshift=-2.5cm, yshift=-1.5cm]   {Relations\\parameters/observations};
    \node (result2)           [noderesult, below of=decision1, xshift=2.5cm, yshift=-1.5cm]   {Seed parameters from\\best ranked models\\\textbf{Table~\ref{tab:cont:best}}};
  \node (rtrun2)            [nodecirc, below of=result1, xshift=2.5cm, yshift=-1.6cm]                    {Grid $M_d, R_d$\\\textbf{Table~\ref{tab:cont:parameters} step 1}\\for NOEMA data};
  \node (result3)           [noderesult, below of=rtrun2, yshift=-0.8cm]      {Best-fitting disc};
  \node (rtrun3)            [nodecirc, below of=result3, xshift=-2.5cm, yshift=-0.9cm, minimum size=2.5cm]     {Grid $L_\star, \dot{M}_{\rm env}$\\\textbf{Table~\ref{tab:cont:parameters} step 2}\\for SED+UKIRT};
  \node (rtrun4)            [nodecirc, below of=result3, xshift=2.5cm, yshift=-0.9cm, minimum size=2.5cm]      {Grid $i, \theta_{\rm cav}$\\\textbf{Table~\ref{tab:cont:parameters} step 2}\\for SED+UKIRT};
    \node (result4)           [noderesult, below of=rtrun3, xshift=2.5cm, yshift=-1.5cm]      {$T, \rho$ distributions\\\textbf{Table~\ref{tab:cont:best}}};
  
  \node (images1)            [nodecirc, right of=start, xshift=8cm]       {Radiative \\ transfer};
    \node (images2)            [noderesult, right of=images1, xshift=1.5cm]       {SED\\\textbf{Fig.~\ref{fig:cont:sed}}};
  \node (images3)            [noderect, below of=images1, yshift=-0.5cm]       {Raw model images};
  \node (images4)            [noderect, below of=images3]       {Rotate (continuum PA=259\degr; \\lines PA=276\degr) and/or rebin};
  \node (images5)            [noderect, below of=images4, xshift=-2.5cm]       {Convolve with PSF,\\align with observations};
  \node (images6)            [noderect, below of=images4, xshift=2.5cm]       {CASA simobs or\\Fourier transform};
    \node (images7)            [noderesult, below of=images5, yshift=-0.2cm]       {Image\\\textbf{Table~\ref{tab:cont} type I}\\\textbf{Figs.~\ref{fig:cont:herschel}, \ref{fig:cont:nir}}};
  \node (images8)            [noderect, below of=images7, yshift=-0.2cm]       {Radial average};
  \node (images9)            [noderect, below of=images6, xshift=-2.5cm, yshift=-0.2cm]       {CLEAN};
  \node (images10)            [noderect, below of=images6, yshift=-0.2cm]       {Radial average};
    \node (images11)            [noderesult, below of=images8, yshift=-0.3cm]       {Intensity\\radial profile\\\textbf{Table~\ref{tab:cont} type P}\\\textbf{Fig.~\ref{fig:cont:submm:prof}}};
  \node (images12)           [noderect, below of=images9, yshift=-0.2cm]       {Line cube};
  \node (images13)           [noderesult, below of=images12, yshift=-0.2cm]       {pv maps\\\textbf{PA=6, 276\degr}\\\textbf{Fig.~\ref{fig:line:pv}}};
    \node (images14)           [noderesult, below of=images10, yshift=-0.4cm]    {Visibility\\profile\\\textbf{Table~\ref{tab:cont} type V}\\\textbf{Figs.~\ref{fig:cont:mm:vis}, \ref{fig:cont:nir:vis}}};
  
    \node (rtrun5)            [nodecirc, right of=rtrun2, xshift=8cm, yshift=-0.5cm] {Grid\\$X_{\rm CH_3CN}, M_\star, v_{\rm out}, \Delta v$\\\textbf{Table~\ref{tab:line:grid}}};
    \node (rank2)           [noderect, below of=rtrun5, yshift=-1.8cm]      {Best ranked fit\\\textbf{pv maps}};
    \node (result5)           [noderesult, below of=rank2, yshift=-0.8cm]      {$\bf v$, $X_{\rm CH_3CN}$ distribution\\\textbf{Table~\ref{tab:cont:best}}};
  
  \draw[->]             (start) -- (rtrun1);
  \draw[->]            (rtrun1) -- (compare1);
  \draw[->]          (compare1) -- (rank1);
  \draw[->]             (rank1) -- (decision1);
  \draw[->]           (modify1) -- (rtrun1);
  \draw[->]       (decision1) -| node[yshift=1.25cm, text width=3cm]
                                   {Yes}
                                   (modify1);
 \draw[->]      (decision1) -- +(0.0,-2.0) -- +(-2.5,-2.0) -- +(-2.5,-2.5) node[yshift=0.8cm,xshift=2.5cm]{No} (result1);
 \draw[->]      (decision1) -- +(0.0,-2.0) -- +(+2.5,-2.0) -- +(+2.5,-2.5) node[yshift=0.8cm,xshift=-2.5cm]{No} (result2);
 \draw[->]      (result1) -- +(0.0,-1.2) -- +(2.5,-1.2) -- +(+2.5,-1.8) (rtrun2);
 \draw[->]      (result2) -- +(0.0,-1.2) -- +(-2.5,-1.2) -- +(-2.5,-1.8) (rtrun2);
  \draw[->]           (rtrun2) -- (result3);
 \draw[->]      (result3) -- +(0.0,-0.8) -- +(+1.4,-1.8) (rtrun3);
 \draw[->]      (result3) -- +(0.0,-0.8) -- +(-1.4,-1.8) (rtrun4);
 \draw[<->, dashed]      (rtrun3) -- (rtrun4);
 \draw[->]      (rtrun3) -- +(+2.5,-2.0) -- +(+2.5,-2.5) node[yshift=1.2cm, scale=0.7, opacity=0, text opacity=1]{Max. probability\\area} (result4);
 \draw[->]      (rtrun4) -- +(-2.5,-2.0) -- +(-2.5,-2.5) (result4);
 \draw[->]      (result4) -| +(5.5,0.0) -- +(5.5,7.2) -- +(7.7,7.2) (rtrun5);
 
 \draw[->]      (images1) -- (images2);
 \draw[->]      (images1) -- (images3);
 \draw[->]      (images3) -- (images4);
 \draw[->]      (images4) -| node[xshift=-0.5cm, yshift=-0.2cm, scale=0.7]{Single\\dish} (images5);
 \draw[->]      (images4) -| node[xshift=0.9cm, yshift=-0.2cm, scale=0.7]{Interferometry} (images6);
 \draw[->]      (images5) -- (images7);
 \draw[->]      (images7) -- (images8);
 \draw[->]      (images8) -- (images11);
 \draw[->]      (images6) -| (images9);
 \draw[->]      (images9) -- (images12);
 \draw[->]      (images12) -- (images13);
 \draw[->]      (images6) -- (images10);
 \draw[->]      (images10) -- (images14);
 
  \draw[->]           (rtrun5) -- (rank2);
  \draw[->]           (rank2) -- (result5);
  
 \draw [color=blue,thick,dashed](-5cm,1cm) rectangle (4.5cm,-11.8cm);
 \node at (-0.5,1) [above=-6.0mm, right=-4.2cm, align=left] {(a)\\\textsc{visual inspection}\\See \S\ref{ap:modelling:cont:vis}};
 \draw [color=blue,thick,dashed](-5cm,-12.cm) rectangle (4.5cm,-22.5cm);
 \node at (-0.5,-12.1) [above=-5mm, right=-4.2cm, align=left] {\textsc{grid modelling}\\See \S\ref{ap:modelling:cont:grid}};
 \draw [color=black,thick](4.7cm,1cm) rectangle (14.0cm,-11.8cm);
 \node at (9.2,1) [above=-6.0mm, right=-4.2cm, align=left] {(b)\\\textsc{image processing}\\See \S\ref{ap:dataproc}};
 \draw [color=gray,thick,dashed](4.7cm,-12.cm) rectangle (14.0cm,-22.5cm);
 \node at (9.2,-12.1) [above=-5mm, right=-4.2cm, align=left] {\textsc{CH$_3$CN modelling}\\See \S\ref{ap:modelling:line:mollie}};
 
\end{tikzpicture}
}

\caption{Radiative transfer modelling procedure.
(a) Overall procedure.
The blue dashed boxes show the 2 steps followed to fit the dust continuum data, whilst the grey dashed box shows the procedure used to fit the CH$_3$CN molecular line emission observations.
Purple circles involve performing radiative transfer steps.
(b) Image processing procedure performed at the end of each radiative transfer computation.
Green boxes represent final products of each step.
Detail descriptions of the procedures and additional tables are available in the online Appendices.
}\label{fig:modelling}
\end{figure*}

We performed a radiative transfer modelling of the multi-wavelength dust continuum and CH$_3$CN observations to obtain density, temperature and velocity distributions.
The fit was performed in 3 steps and is described in Fig.~\ref{fig:modelling}(a).
Each step involved the production of synthetic observations as described in Fig.~\ref{fig:modelling}(b) to obtain the model data products (images, intensity radial profiles, etc.) which were then compared with the observations.
To produce synthetic data, we used the 3-D radiative transfer codes \textsc{\lowercase{Hyperion}} \citep{2011A&A...536A..79R} for dust continuum and \textsc{\lowercase{Mollie}} \citep{2010ApJ...716.1315K} for line emission.
The modelling steps included one visual inspection for the continuum fitting and model grid fitting for some continuum observations and the CH$_3$CN observation.
Details and the particularities of the modelling and synthetic observations are described in Appendices~\ref{ap:modelling}--\ref{ap:dataproc} (online).

\subsection{1-D LTE and non-LTE CH$_3$CN modelling}
\label{sect:modelling:cassis}

We fitted the methyl cyanide 12--11 $K$-ladder with the program \textsc{\lowercase{cassis}}\footnote{Based on analysis carried out with the \textsc{\lowercase{CASSIS}} software and the Cologne Database for Molecular Spectroscopy (CDMS; \url{https://cdms.astro.uni-koeln.de}) molecular database. \textsc{\lowercase{CASSIS}} has been developed by IRAP-UPS/CNRS (\url{http://cassis.irap.omp.eu}).} to study some physical properties of the gas (e.g. temperature).
This program uses a Markov chain Monte Carlo (MCMC) which samples the parameter space in order to minimise the $\chi^2$ between the observed and synthetic spectra. 
We used one isothermal density component model to fit the observations and their parameter ranges are listed in Table~\ref{tab:cassispars}. 
\textsc{\lowercase{CASSIS}} calculates spectra under the LTE approximation, or the non-LTE approximation by using the radiative transfer code \textsc{\lowercase{RADEX}} \citep{2007A&A...468..627V}. 
Two density component models were experimented upon, but similar results were obtained.

\begin{table}
\caption{Parameter constraints of the modelling with \textsc{\lowercase{CASSIS}}.}
\label{tab:cassispars}
\begin{tabular}{lc}
\hline
    Parameter & Limits\\
\hline
    Column density ($N_{\rm CH_3CN}$) & $10^{13}-10^{18}$\,cm$^{-2}$ \\
	H$_2$ density ($n_{\rm H_2}$)$^{a}$ & $10^{3}-10^{9}$\,cm$^{-3}$ \\ 
    Excitation temperature ($T_{\rm ex}$) & 10--500\,K \\ 
	Size & 0--1.2\,arcsec \\
	Line width ($\Delta v$) & 1--7\,km\,s$^{-1}$ \\
	Offset to LSR velocity ($\Delta v_{\rm LSR}$)$^{b}$ & -2--2\,km\,s$^{-1}$\\
\hline
\end{tabular}
    \begin{minipage}{\columnwidth}
        $^a$ Only for the non-LTE case.\\
        $^b$ $v_{\rm LSR}=-5.5$\,km\,s$^{-1}$.
    \end{minipage}
\end{table}

In order to limit the region to analyse, only spectra with at least one line with a flux higher than 0.25\,Jy\,beam$^{-1}$ (${>}5\sigma$) were considered in the fit. 
Additionally, \textsc{\lowercase{CASSIS}} requires an estimate of each line rms and the flux calibration error. The first was estimated from boxes around the lines and vary between 0.04--0.06\,K, where the higher the K line the higher the error. 
The relative flux calibration error was fixed to 20\,per\,cent \citep{2018A&A...617A.100B}.

The size parameter in Table~\ref{tab:cassispars} is the source size and is used to calculate the beam dilution factor, which is defined in \textsc{\lowercase{CASSIS}} as
\begin{equation}
    \Omega = \frac{\theta^2}{\theta^2 + \theta_{\rm B}^2}\,,
\end{equation}
where $\theta$ is the source size and $\theta_{\rm B}$ is the beam size derived from the configuration files specific for each telescope, which include its diameter/largest baseline. 
In this case, the baseline used was calculated to match the resolution of the observations. 
The brightness temperature is then calculated by \textsc{\lowercase{CASSIS}} as
\begin{equation}\label{eq:tbright}
    T_b = \Omega J_\nu(T_{\rm ex})(1-e^{-\tau}) + f({\rm CMB, dust})\,,
\end{equation} 
where $J_\nu(T_{\rm ex})=h\nu k^{-1} (e^{h\nu/kT_{\rm ex}} - 1)^{-1}$, $T_{\rm ex}$ is the excitation temperature, $\tau$ is the line optical depth and $f({\rm CMB, dust})$ accounts for the contribution from the cosmic microwave background and dust emission.

\section{Results}
\label{sect:results}

\subsection{Continuum modelling}
\label{sect:results:cont}

\begin{figure}
  \centering
\includegraphics[width=\columnwidth]{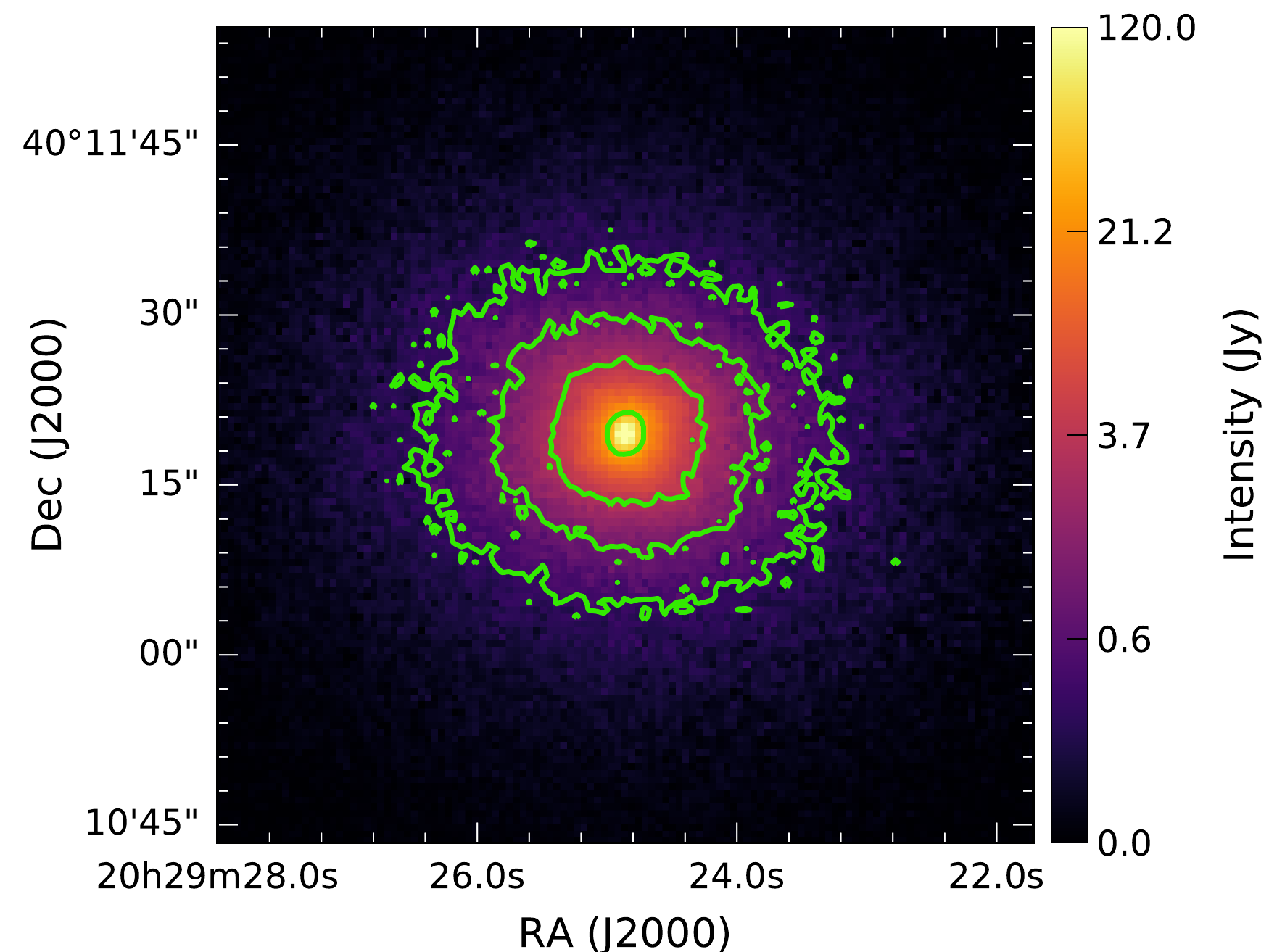}
\caption{Unprocessed model image of the best-fitting model at 70\,\micron. 
Contour levels are 0.1, 0.3, 1 and 10~per~cent of the peak intensity.
Note the elongation along the cavity axis oriented at PA$=259\degr$ as the distance to the source increases.
    }
  \label{fig:cont:herschel:raw}
\end{figure}

The parameters of the overall best-fitting model, i.e. after the continuum grid fitting step, are listed in Table~\ref{tab:cont:best}.
The extended emission observed at 70\,\micron\ is relatively well fitted by the best-fitting model up to ${\sim}40$\,arcsec from the centre as shown in Fig.~\ref{fig:cont:herschel}. 
In general, the model matches the elongation in the data. 
At 1~per~cent of the peak intensity, where the model has a better fit, the horizontal slice, which cuts through the cavity, is more elongated than the vertical one in the model and the observations as listed in Table~\ref{tab:herschel}. 
This cannot be explained by any asymmetry in the 70\,\micron\ PSF, which is relatively symmetric as shown in the same table.
Fig.~\ref{fig:cont:herschel:raw} shows the 70\,\micron\ image before convolution with the PSF, where the contribution of the cavity to the elongation can be observed. 
The emission is underestimated close to the source and along the cavity axis (PA$=259\degr$, see Fig.~\ref{fig:cont:herschel}). 
As a result, the angular sizes of the model at the FWHM from slices and 2-D Gaussian fit listed in Table~\ref{tab:herschel} are similar between each direction.

The parameters of the best-fitting model during the visual inspection step are listed in Table~\ref{tab:cont:best} and the ranking and reduced $\chi^2$ values for each observation are listed in Table~\ref{tab:cont:chi2}.
The best-fitting model has a type A density distribution (\citealp{1976ApJ...210..377U} envelope and a flared disc).
The envelope radius was well-fitted during this step and has a value roughly a factor 1.4 smaller than the initial visual inspection model and has a similar size to the model with disc of \citet[][]{2013A&A...551A..43J}.
The disc scale height is smaller by roughly one order of magnitude than those in \citet[][]{2013A&A...551A..43J}, but this parameter cannot be constrained by our observations, thus it was less explored. 
The shape of the cavity changed from the initial exponent of 1.5 to 2.0, and its reference density decreased a ${\sim}15$~per~cent from its initial value and its density exponent did not change.

\begin{table*}
 \centering
    \caption{Parameters of the continuum and CH$_3$CN line best-fitting model, which consists of an Ulrich envelope and a flared disc.}
\label{tab:cont:best}
 \begin{tabular}{lccc}
  \hline
     Parameter & Visual & Grid & Constrained by\\
               & inspection &   & \\
  \hline
     \multicolumn{4}{c}{Continuum}\\
     Distance (kpc)                                   & \multicolumn{2}{c}{3.3$\pm$0.1}   & Fixed \\
     Position angle (PA)                              & \multicolumn{2}{c}{259\degr}      & Fixed \\
     Inclination angle ($i$)                          & 40\degr  & $25\degr\pm5\degr$     & \textbf{SED + near-IR}\\
     Stellar temperature ($T_\star$/K)                & 20000    & 18000                  & --\\
     Stellar luminosity ($L_\star/10^5$\,\lsun)       & 1.6      & $1.6\pm0.2$            & \textbf{SED}\\
     Inner radius ($R_\rmn{in}$/au)                   & 36       & 35                     & SED + Speckle\\
     Outer radius ($R_\rmn{out}$/au)                  & 195000   & 195000                 & \textbf{Submm}\\
     Envelope infall rate ($\dot{M}_\rmn{env}/10^{-3}\,\msun\,{\rm yr}^{-1}$) & 1.1 & $2.0\pm0.2$ & \textbf{SED + (sub)mm}\\
     Envelope dust                                    & \multicolumn{2}{c}{OHM92}         & -- \\
     Centrifugal radius ($R_\rmn{c}$/au)              & 200      & $2200\pm200$           & \textbf{NOEMA ext.} \\
     Disc mass ($M_\rmn{d}$/\msun)                    & 1.0      & $6\pm1$                & \textbf{NOEMA ext.}\\
     Disc scale height at $r=100$ au ($h_0$/au)       & \multicolumn{2}{c}{0.6}           & --\\
     Disc flaring exponent ($\alpha$)                 & 1.88     & 2.25                   & --\\
     Disc vertical density exponent ($\beta$)         & 1.13     & 1.25                   & --\\
     Disc dust                                        & \multicolumn{2}{c}{OHM92}                  & --\\
     Cavity opening angle$^{a}$ ($2\theta_{\rm cav}$) & 60\degr  & $60\degr\pm5\degr$     & \textbf{Near-IR + SED}\\
     Cavity shape exponent ($b_\rmn{cav}$)            & \multicolumn{2}{c}{2.0}           &  \textbf{Near-IR + SED}\\
     Cavity density exponent ($p_{\rm cav}$)          & \multicolumn{2}{c}{2.0}           &  \textbf{Near-IR + SED}\\
     Cavity reference density$^{a}$ ($\rho_{\rm cav}/10^{-21}$\,g\,cm$^{-3}$) & \multicolumn{2}{c}{1.4} & \textbf{Near-IR + SED}\\
     Cavity dust                                      & \multicolumn{2}{c}{KMH}           & --\\
     \multicolumn{4}{c}{Extra parameters for line emission}\\
     Position angle (PA)                              & -- & 276\degr                     & Fixed \\
     Stellar mass ($M_\star/\msun$)                   & -- & 35                           & Fixed \\
     Abundance for $T>100$~K ($X_1$)                  & -- & $2\times 10^{-8}$            & -- \\
     Turbulent width ($\Delta v_{\rm nth}/{\rm km\,s}^-1$) & -- & 2                       & -- \\
     Outflow velocity$^a$ ($v_0/{\rm km\,s}^-1$)      & -- & 0, 1$^{b}$                   & -- \\
  \hline
 \end{tabular}
    \begin{minipage}{\textwidth}
        \textbf{Notes.} Parameters in bold under the last column are the 10 free parameters that can be constrained by the observations. 
        The inner radius depends on the stellar luminosity.\\
        $^a$ Defined at a height $z=10^4$~au.\\
        $^b$ 2 models had the same best overall ranking value.
    \end{minipage}
\end{table*}

\begin{table*}
    \centering
    \caption{Overall ranking and reduced $\chi^2$ values for the continuum 
	best-fitting models of each type.}
    \label{tab:cont:chi2}
    \begin{tabular}{lccccccccccccc}
        \hline
        Type & Rank$^a$ & 70~\micron\ & SED & $J$ & $H$ & \multicolumn{2}{c}{$K$} &
        450~\micron\ & 850~\micron\ & \multicolumn{2}{c}{1.3~mm} & $\chi^2$ & $\langle \chi^2 \rangle$\\
            & & & & & & UKIRT & Speckle & & & extended & compact &&\\
        \hline
        \multicolumn{14}{c}{Overall best-fitting model}\\
        A & --   & 211 & 7.0  & 26 & 64  & 316 & 32  & 13 & 0.40 & 408  & 557  & -- & -- \\
        \multicolumn{14}{c}{Best-fitting model by density type}\\
        A & 662  & 114 & 9.7  & 23 & 79  & 311 & 49  & 20 & 0.56 & 993  & 852  & 137 & 164 \\
        B & 854  & 160 & 7.3  & 25 & 81  & 305 & 35  & 17 & 0.51 & 1040 & 1142 & 137 & 194 \\
        C & 1019 & 305 & 44.1 & 21 & 107 & 523 & 469 & 3  & 0.10 & 557  & 300  & 217 & 175 \\
        \multicolumn{14}{c}{Best-fitting model by $\chi^2$}\\
        A & 1134 & 100 & 8.9  & 26 & 91  & 223 & 44  & 45 & 3.60 & 926  & 1570 & 113 & 160 \\
        \multicolumn{14}{c}{Best-fitting model by $\langle \chi^2 \rangle$}\\
        C & 1278 & 274 & 54.2 & 23 & 155 & 624 & 25  & 0.7& 1.16 & 156  & 195  & 267 & 125 \\
        \hline
    \end{tabular}
    \begin{minipage}{\textwidth}
        $^a$ Average weighted ranking as described in Appendix~\ref{ap:modelling:cont:vis:proc} (online).
    \end{minipage}
\end{table*}

Similarly, in Table~\ref{tab:cont:best} and \ref{tab:cont:chi2} we list the results for the overall best-fitting model after the grid fitting step.
The disc mass and radius increased with respect to the visual inspection best-fitting model.
The former is roughly a third of the disc mass in the control model of \citet[][]{2013A&A...551A..43J}, but the disc radii are similar.
This difference is explained by the difference in the disc scale height.
The cavity opening angle is equal to the one from the visual inspection best-fitting model, and the inclination angles differ by 15\degr.
The luminosities are the same between the best-fitting models at each step, whilst the envelope infall rate is ${\sim}1.8$ times higher in the overall best-fitting model.

The model SED agrees reasonably well with the observations as shown in Fig.~\ref{fig:cont:sed}. 
The total mass of the model, $M=2.1\times10^3$\,\msun, is enough to fit the 10\,\micron\ silicate feature relatively well at an inclination of 25\degr\ and closely fit the submm points. 
This is 900\,\msun\ more massive than the model with disc of \citet[][]{2013A&A...551A..43J}, but 600\,\msun\ less massive than their envelope only model. 
Models with WD01 dust can also fit in the submm with lower masses, however they underestimate the fluxes between 3--8\,\micron\ by a factor ${\sim}10$.

The near-IR maps in Fig.~\ref{fig:cont:nir}(a)--(c) show that the near-IR band images are well reproduced. 
However, Fig.~\ref{fig:cont:nir:vis} shows that the small scales closer to the star position probed by the speckle observation are not well reproduced. 
In general, most models with a cavity density of the same order as the best-fitting model do not fit this observation. 

At 450\,\micron, the best-fitting model overestimates the emission at larger scales as shown in Fig.~\ref{fig:cont:submm:prof}. 
Type C models with WD01 dust are able to reproduce this profile but not the near-IR observations. 
At 850\,\micron, most of the models reproduce the radial profile relatively well. 
These two fits are unaffected by the inclination angle. 
\citet{2014ApJ...785...42P} successfully fitted these profiles with a spherically symmetric model and obtained a density power law index of 1.8, which is slightly lower than our type C models but steeper than the Ulrich envelope.
However, an analysis of the observed images shows that the observed source is more elongated along the cavity at both wavelengths, but this is not seen in our models.

The 1.3\,mm extended configuration visibilities are relatively well fitted as shown in Fig.~\ref{fig:cont:mm:vis}(b). 
Models with WD01 dust or larger and more massive discs are able to match the data for baselines smaller than ${\sim}150$~k$\lambda$, but these baselines were not included in the fitting process. 
For longer baselines, the intensity is determined mainly by the mass of the disc, $6$\,\msun\ in the best-fitting model.
The best-fitting model underestimates the observed 1.3\,mm compact configuration visibilities (Fig.~\ref{fig:cont:mm:vis}a) at large spatial scales (smaller uv distances). 
This is observed in most models. 
However, those models with WD01 dust give an improvement of the fit, but they do not fit the short wavelengths ($\lambda<12$~\micron) of the SED. 

At ${\sim}10^4$\,au scales, as mapped by the 1.3\,mm compact configuration observations in Fig.~\ref{fig:cont:mm:clean}(a), the model FWHM is $3\farcs2\times3\farcs0$ PA$=325\degr \pm 2\degr$ as obtained from a 2-D Gaussian fit. 
This is slightly more extended than the observed size in Table~\ref{tab:noema} and has a different orientation.
The presence of at least one source to the south of VLA~3, as shown in Fig.~\ref{fig:cont:mm:clean}(a), may also contaminate the submm observations making them look more elongated than they really are. A more complete modelling of VLA~3 including the contribution of NOEMA~1 is beyond the scope of this paper.

Fig.~\ref{fig:cont:mm:clean}(b) shows the degree of agreement between the observations and model at ${\sim}100$\,au scales from the 1.3\,mm extended configuration observations. 
It shows that the model emission is more compact than the observed one, with a FWHM of $0\farcs7\times0\farcs6\,\rmn{PA}=247\degr\pm10\degr$ from a 2-D Gaussian fit (cf. Table~\ref{tab:noema}).
The PA of the model is closer to the orientation of the outflow cavity than to the orientation of the disc, which can be explained by the inclination angle.
The extended emission in the observations also show features which cannot be explained by the model due to its symmetry.

\subsubsection{Other density distributions}
\label{sect:results:cont:other}

Table~\ref{tab:cont:chi2} lists the $\chi^2$ values for the best-fitting models for the different types of model densities (A, B and C) for each observation in the visual inspection fitting.
The values of the parameters of the best-fitting models are listed in Table~\ref{ap:tab:cont:best} (online).
The best-fitting type B model has the same stellar properties as the best-fitting type A model (from grid fitting). 
The envelope of the best-fitting type B model has the same radius and has an infall rate $0.5\times10^{-3}\,\msun\,{\rm yr}^{-1}$ lower than the best-fitting type A model.
The inclination angle is 10\degr\ lower than the best-fitting type A model, which may explain the better fit to the SED than the best-fitting type A from visual inspection.
The distribution and shape of the cavity is the same as the best-fitting type A model.
This model has an accretion luminosity of ${\sim}20$\,\lsun. 

The best-fitting type C model has the same inclination angle as the best-fitting type A model. 
It fits relatively well the submm and mm observations, but is worse at fitting the near-IR observations and the SED because its envelope has a WD01 dust.
The envelope has a mass of $1.0\times10^3$\,\msun\ and the density in the cavity is lower ($\rho_{0,\rmn{cav}}=3.5\times10^{-24}$\,g\,cm$^{-3}$) but decreases at the same rate as the best-fitting type A model. 
Its disc is more compact than the best-fitting type A model one from visual inspection, and slightly more massive (1.2\,\msun). 
In general, this type of model does not fit the 70\,\micron\ image well. 

\subsection{CH$_3$CN line modelling}
\label{sect:results:line}

\subsubsection{1-D \textsc{\lowercase{CASSIS}} modelling}
\label{sect:results:line:cassis}

\begin{figure*}
  \centering
  \includegraphics[width=\textwidth]{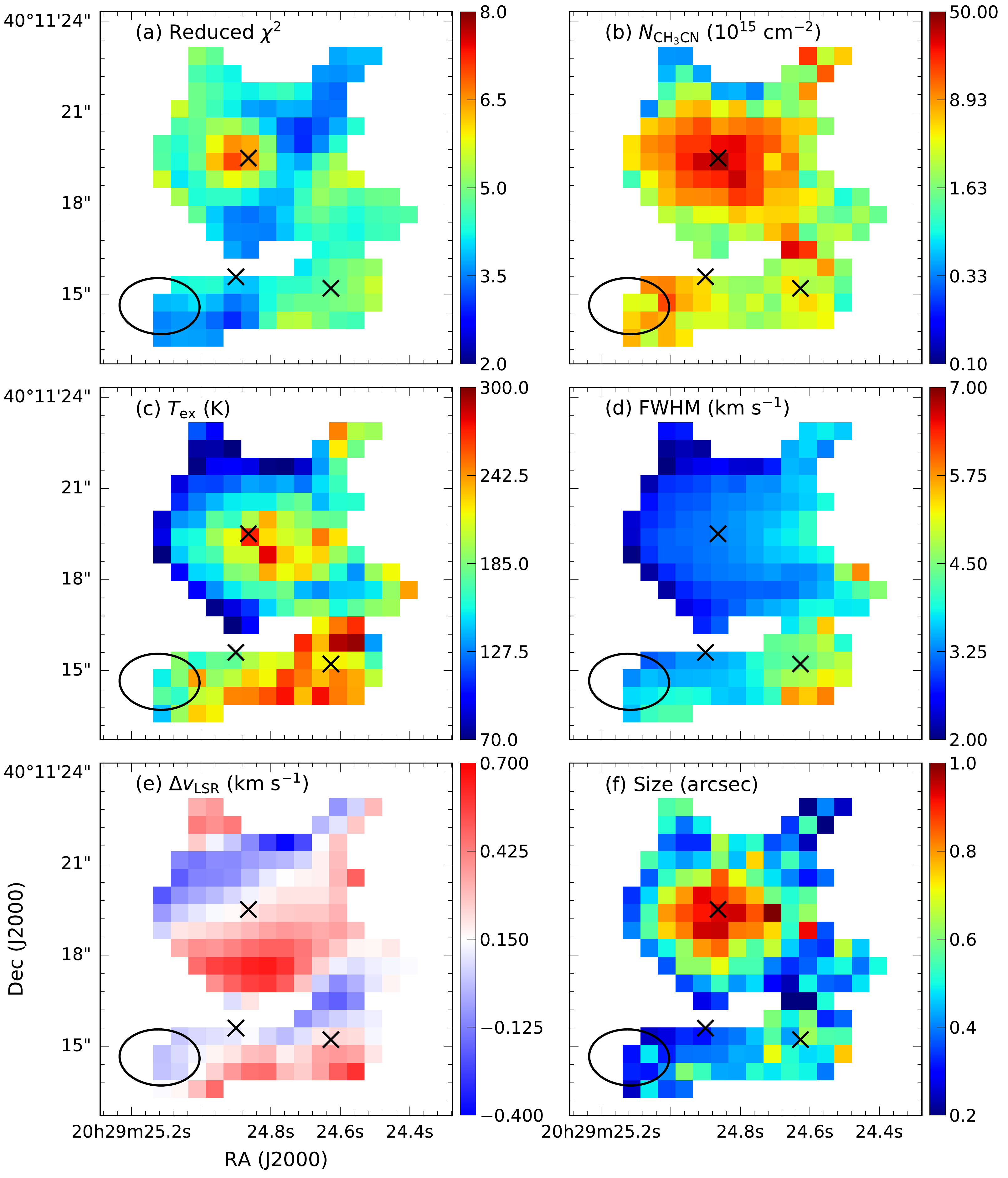} 
  \caption{Results for each parameter from the 1-D LTE modelling of the NOEMA compact configuration CH$_3$CN $J=12-11$ transition observations. 
    The crosses mark the position of the continuum sources in Table~\ref{tab:noema}.
    The beam is shown in the lower left corner (black ellipse).
    }
  \label{fig:line:cassis}
\end{figure*}

We obtained similar fits to the data and results under the LTE and non-LTE approximations. 
The results for the LTE modelling are shown in Fig.~\ref{fig:line:cassis}.
The peak column density for VLA~3 is $4.2\times10^{16}$\,cm$^{-2}$ in the LTE model, and is located at the position of the source. 
The average column density inside a region with radius 2\,arcsec is $1.5\times10^{16}$\,cm$^{-2}$. 
An approximate value of the ${\rm H}_2$ column density can be obtained from the continuum map in the optically thin limit \citep[e.g.][]{2009ApJ...698..488M}:
\begin{equation}\label{eq:columnfromdust}
    N_{{\rm H}_2} = \frac{S_\nu R_{\rm gd}}{\Omega_{\rm B} B_\nu(T_{\rm d})
    \kappa_\nu \mu m_{\rm H}}\,,
\end{equation}
with $S_\nu$ the observed dust continuum flux and $\Omega_{\rm B}$ the solid angle of the beam.
If eq.~\ref{eq:columnfromdust} is solved with a dust temperature $T_{\rm d}=190$\,K (see below), a dust opacity $\kappa_{\rm 1.3mm} =1$\,cm$^2$\,g$^{-1}$, a mean molecular weight $\mu=2.3$ and a gas-to-dust ratio $R_{\rm gd}=100$, a peak ${\rm H}_2$ column density value of $2.0\times10^{23}$\,cm$^{-2}$ and an average of $9.9\times10^{22}$\,cm$^{-2}$ within the same 2\,arcsec region are obtained. 
Hence the methyl cyanide abundance is $2.1\times10^{-7}$ for the peak value and $1.5\times10^{-7}$ for the average.

The excitation temperature for the LTE model in VLA~3 peaks at the source position and has a value of 271\,K. 
In a region of radius 2\,arcsec the minimum temperature is 120\,K and the average is 190\,K.

The FWHM of the line gets wider along the blue-shifted outflow direction. 
This may be due to turbulent motion of the gas in the outflow. 
The red-shifted outflow direction does not show the same trend, probably because the emission is optically thick towards that side.
However, the fit is worse towards the red-shifted outflow direction, as shown by the reduced $\chi^2$ maps, hence this result should be taken with caution.
The line velocity maps of VLA~3 are consistent with the first moment maps in Fig.~\ref{fig:line:mom0_mom1}.

The sizes obtained from the modelling are smaller than the beam size of the observations and the deconvolved angular size derived from continuum observations (${\sim}2.0$\,arcsec from Table~\ref{tab:noema}). 
The less beam diluted areas are located towards the denser regions.
As the flux is proportional to the beam dilution factor $\Omega$ and proportional to the column density in the optically thin regime, the size and the column density parameters may be degenerate.
However, if the core was spherical, the depth along the line of sight would increase towards the centre of the core.

The regions NOEMA~2 and 3 show a different temperature and velocity structure. 
As expected for an \ion{H}{II} region (VLA~1), NOEMA~2 has wider lines due to turbulent gas expansion and is hotter than the other two regions.
NOEMA~2 also has a lower column density in comparison to the other two regions, which is consistent with it being optically thin at mm/radio wavelengths as found by e.g. \citet[][]{2003ApJ...589..386T} at 3.6\,cm. 
NOEMA~3 peaks in temperature and column density at the same position. 
Its peak temperature is 240\,K and its column density is $1.6\times10^{16}$\,cm$^{-2}$, hence is slightly colder and less dense than VLA~3.
There appears to be a similar velocity gradient in NOEMA~2 and 3 to that of VLA~3, thus there may be a common rotational motion for all the sources present.

\subsubsection{3-D modelling}
\label{sect:results:line:mollie}

\begin{figure}
  \centering
  \includegraphics[width=\columnwidth]{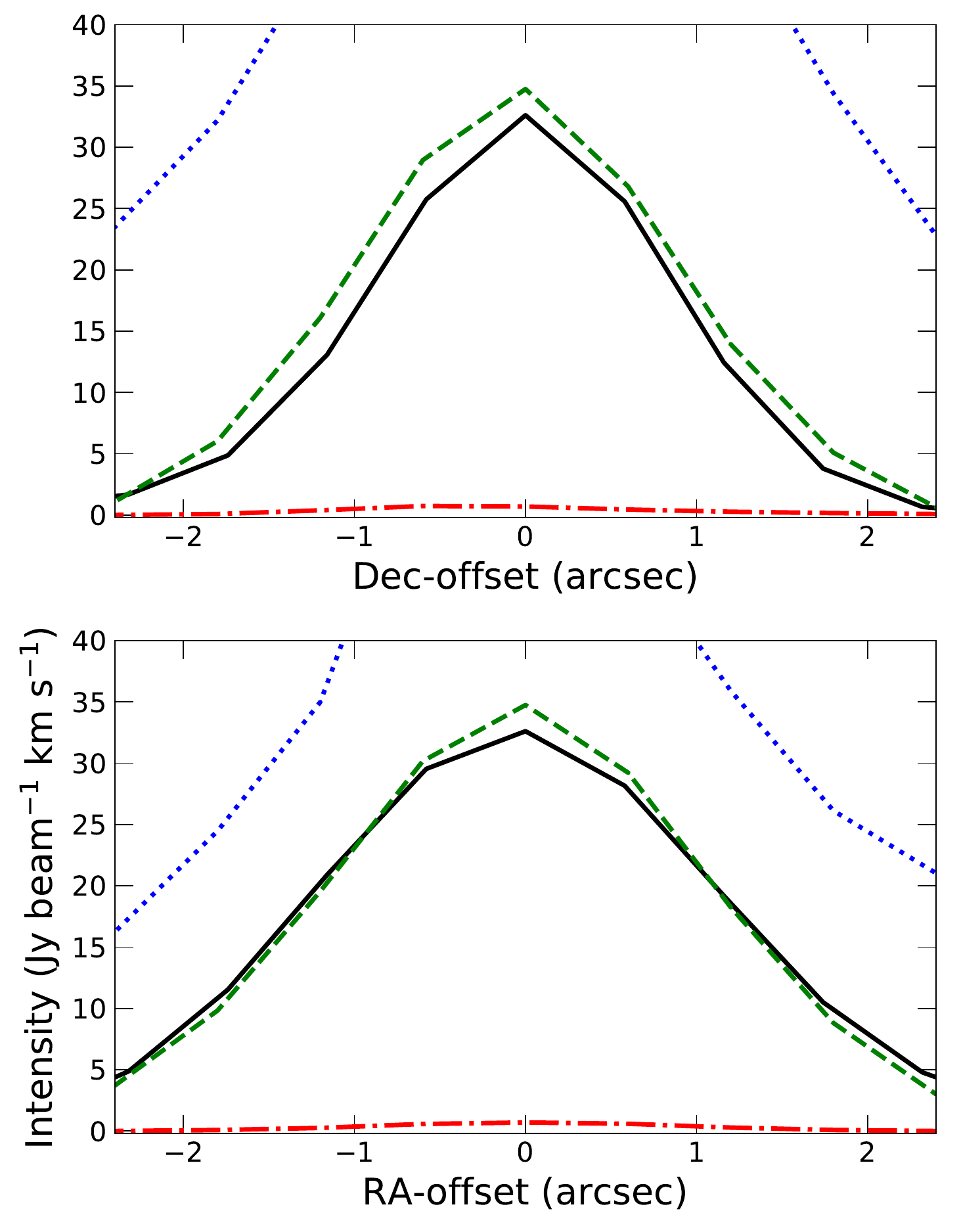} 
  \caption{Observed (black line) and models zeroth moment map slices through the peak along the vertical (upper) and horizontal (lower) axes for $K=3$. 
      The dashed green line corresponds to the best-fitting model with zero outflow reference velocity. 
      The dotted red line corresponds to models with the same parameters as the best-fitting one but with the temperature step at $T_0=200$\,K. 
      The dash-dotted blue line shows a model with the same parameters as the best fitting model but with a constant abundance.
    }
  \label{fig:line:mom0:slc}
\end{figure}

There are 2 best-fitting models (models with the same overall ranking), but the only difference is the outflow reference velocity.
Table~\ref{tab:cont:best} lists the parameters of the best-fitting models for the CH$_3$CN $J=12-11$ $K=2$ to $K=5$ molecular line emission for a model with a stellar mass of 35~\msun.
Unless otherwise stated, we will refer to these models as the best-fitting ones as the stellar mass matches the mass derived from the source luminosity.
The best-fitting models have outflow reference velocities $v_0=0$ and 1\,km\,s$^{-1}$.
However, this parameter may not be well constrained because the observations do not show any clear evidence of outflow motion in the first moment map, e.g. online Fig.~\ref{fig:line:mom1:k3}. 
Hence for simplicity we plot the results for the model with zero outflow reference velocity.
Fig.~\ref{fig:line:pv} shows the pv maps of the best-fitting model for each line and cut direction. 
The model reproduces relatively well the angular extent of the pv maps, but not the spectral extent.
The zeroth moment slices in Fig.~\ref{fig:line:mom0:slc} also show that the angular extent is well fitted.

\begin{figure}
  \centering
  \includegraphics[width=\columnwidth]{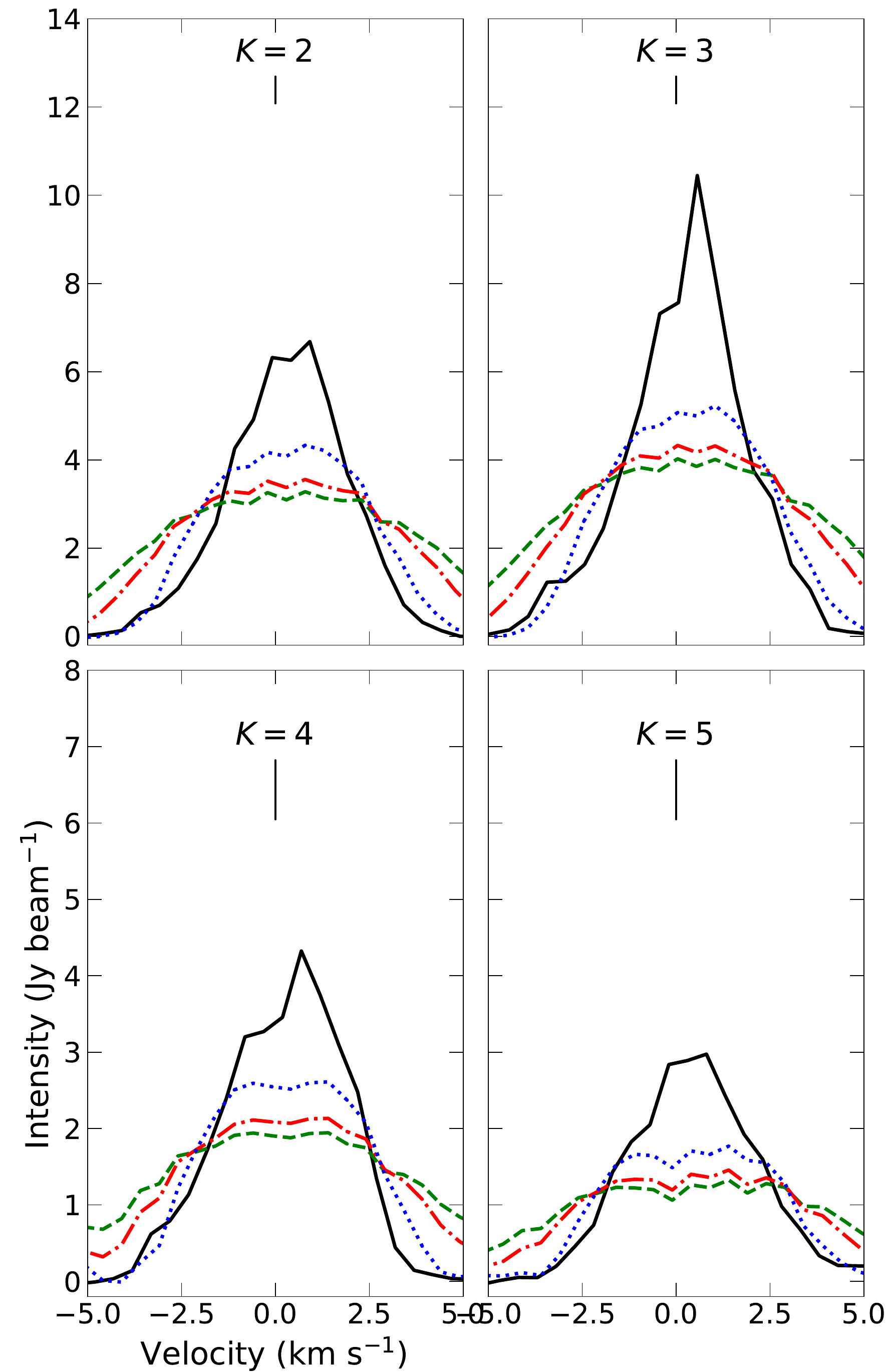} 
    \caption{Observed (black line) and best-fitting model peak spectra (dashed green line) for (a) $K=2$, (b) $K=3$, (c) $K=4$ and (d) $K=5$. 
    The dash-dotted red lines correspond to the best-fitting model with $M_\star=25$\,\msun.
    The dotted blue lines correspond to the best-fitting model with $M_\star=15$\,\msun.
    }
  \label{fig:line:best:peak}
\end{figure}

Fig.~\ref{fig:line:best:peak} shows the spectra at the peak of the zeroth moment map, which is equivalent to the spectra at zero offset in the pv maps, for each $K$ level.
The best-fitting models do not match the width and the peak of the observed spectral line, mainly because the model infall and rotation velocities are high enough to produce double peaked spectra.
Fig.~\ref{fig:line:best:peak} shows that models with lower stellar masses are needed to match them, but these models are not physically realistic given the stellar luminosity.
In general, the abundance scales the line intensity.
The best-fitting models prefer low abundances because larger values start to overestimate the values at the line wings.
On the other hand, the observed lines are skewed towards higher velocities whilst the model lines are more symmetric.

The best-fitting models have a turbulent width of 2.0\,km\,s$^{-1}$.
Although this is in the upper end of our model grid, the lines are already wide due to the large rotation/infall velocities.
The third model in the tally has a turbulent width of 1.5\,km\,s$^{-1}$.

\section{Discussion}
\label{sect:discussion}

\subsection{70\,\micron\ morphology}
\label{sect:discussion:herschel}

\begin{figure}
  \centering
  \includegraphics[width=\columnwidth]{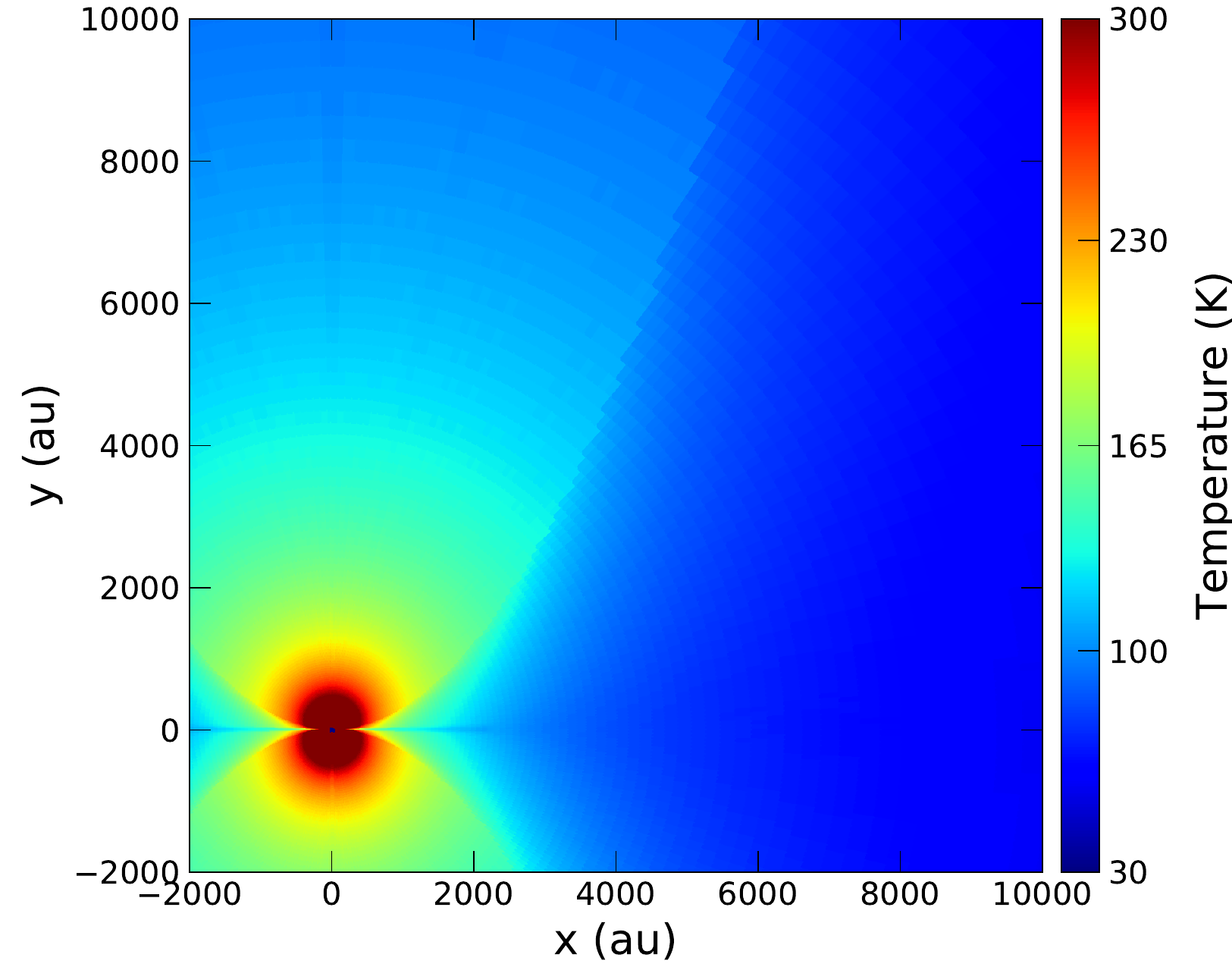} 
  \caption{Temperature distribution in the mid-plane of the inner 10000\,au region as obtained for the best-fitting model.
    }
  \label{fig:cont:temp}
\end{figure}

Our detailed radiative transfer modelling has shown that the bipolar outflow cavities play an important role in reproducing the 70\,\micron\ observations. 
The modelling has allowed the fit of the extended emission at both this wavelength and the SED. 
The observed and model 70\,\micron\ images are elongated along the cavity axis.
The results in Figs.~\ref{fig:cont:sed} and \ref{fig:cont:herschel:raw} imply that the warm/hot dust in the envelope cavity walls bring an important contribution to the emission at 70\,\micron. 
Furthermore, the temperature distribution in Fig.~\ref{fig:cont:temp} shows that cavity walls are being heated by the stellar radiation escaping through the outflow cavity more than other regions of the envelope. 
The figure also shows that the dust in the cavity is hotter than the walls, however there is less dust in the cavity. 
Hence the contribution of the envelope cavity wall is more important at 70\,\micron, which is also shown in the SED.

In the models of \citet[][]{2013ApJ...766...86Z} the 70\,\micron\ emission is extended along the outflow direction. 
The origin of the extended emission in their models is due to dust in the outflow at distances larger than the core radius, otherwise the emission is dominated by the cavity walls of the envelope.
They used a wide cavity angle (${\sim}100$\degr) thus at an inclination of 25\degr\ the observer is looking directly to the source, hence the emission of the cavity walls in their models is relatively symmetrical. 
In comparison, in our best-fitting model the line of sight passes through a section of the envelope.

\subsection{Rotation of the envelope}
\label{sect:discussion:vel}

Our analysis of methyl cyanide interferometric observations shows that they are tracing rotating gas at intermediate scales (${\sim}5000$\,au) in the envelope of AFGL~2591. 
The velocities in the line of sight and the radiative transfer modelling imply that rotation dominates rather than infall or outflow motions.
The direction of the rotation is consistent with the observations of smaller scales made by \citet{2012A&A...543A..22W}. 
They also found that the linear velocity gradient in the HDO line pv map is 16.2\,km\,s$^{-1}$\,arcsec$^{-1}$ at a ${\rm PA}=0\degr$. 
Similarly, we calculated a linear velocity gradient from the pv maps at ${\rm PA}=6\degr$ for each $K$ line and obtained an average value of $13.3\pm0.5$\,km\,s$^{-1}$\,arcsec$^{-1}$, which is consistent with slower rotation of the envelope in comparison to the disc.
Thus the small-scale rotating motion observed in other molecules can be connected to the large scale motion traced by methyl cyanide. 

We estimated a stellar mass of 35\,\msun\ from the source luminosity.
This mass is high enough for the models to produce double-peaked lines which do not fit the observations.
The width of the line in these models are dominated by the large velocity gradient produced by a larger stellar mass rather than the non-thermal velocity width.
We explored models with lower masses, and determined that for the Ulrich envelope with Keplerian disc model to fit the lines, stellar masses below 15\,\msun\ are needed (see Fig.~\ref{fig:line:best:peak}) which is unrealistic given the luminosity of the source (under this simple scenario where a single proto-star would form).
Hence the rotation and/or infall motions of the inner envelope are slower than the ones predicted from the Ulrich envelope with a Keplerian disc.
A binary or multiple system would provide less luminosity per total stellar mass, which would not fit SED.
The slower rotation may be indicative of material being funneled through the envelope to the disc and/or the result of magnetic braking.
Models at lower inclination angles (face-on), i.e. dominated by infalling motions rather than rotation, are discarded by the dust continuum models.

\citet{2007A&A...475..549B} noted that their observations of HCN molecular line emission at the continuum peak were skewed towards the red as in the observations presented here for CH$_3$CN, whilst other lines (e.g. SO) were skewed towards blue. 
They interpreted the blue-skewed lines as produced by material in the inner region absorbing the red-shifted emission. 
Their observations however trace a smaller region close to the star, thus this can be explained by an optically thick disc. 
On the other hand, the CH$_3$CN radiative transfer modelling lines do not show any asymmetry close to the line centre. 
Although our radiative transfer modelling seems to favour models without outflows, this is not well constrained since the observations do not show a clear evidence of tracing outflow material. 
The width of the lines seems to be different in the outflow cavity than in the envelope. 
The wider lines are located in the outflow cavity and can be explained as turbulent motions of gas in the outflow.
The emission from the region where the red-shifted outflow is located do not show the same trend. 
It can be argued that the emission is optically thick towards the red-shifted outflow, which is consistent with the inclination of the outflow axis with respect to the line of sight being closer to face-on.

\subsection{Circumstellar matter distribution}
\label{sect:discussion:props}

The physical properties derived from our modelling are consistent with the current picture of AFGL~2591, but its density distribution is better constrained.
\citet[][]{2012A&A...543A..22W} estimated a disc dust mass between 1--3\,\msun\ from the total flux at 1.4\,mm from PdBI observations, and a disc radius of ${\sim}1300$\,au (at 3.3\,kpc) from a Gaussian fit to the 1.4\,mm visibilities. 
We have shown that to fit the 1.3\,mm extended configuration visibilities (Fig.~\ref{fig:cont:mm:vis}b) a more massive and bigger disc is necessary.
However, such a disc does not match the spatial extent of the 1.3\,mm emission well (Fig.~\ref{fig:cont:mm:clean}b). 
This can be explained by a closer to edge-on disc, which would look more compact in the outflow direction where the extent of the emission differ the most.
Such a closer to edge-on disc can be the result of precession changing the PA and inclination angle at the smaller scales, but as argued in the previous section, the inner envelope favours an inclination closer to face-on.
\citet{2013A&A...551A..43J} obtained disc radii, which were also fixed to the centrifugal radius, of $3.5\times10^{4}$ and $2.7\times10^{3}$\,au for their envelope with disc and control models, respectively. 
They also could not constrain the disc mass and radius as they did not fit the mm interferometric data.

The inclination with respect to the line of sight is consistent with the values in the literature \citep[30--45\degr, e.g.][]{1995ApJ...451..225H,2006A&A...447.1011V} which are generally found based on geometrical considerations, and it is constrained between 25--35\degr. 
Larger values like those found by \citet{2013A&A...551A..43J} are ruled out by the modelling since at 40\degr\ the red-shifted outflow cavity starts to emit in the near-IR. 
This discrepancy may be explained by the data and methods used. 
Although the red-shifted cavity is observed in the models at higher inclinations, the intensity is much lower than the peak of the blue-shifted cavity. 
Hence the red-shifted cavity emission will be negligible due to the lower surface brightness sensitivity of the 2MASS observations. 
\citet{2013A&A...551A..43J} fitted the SED and 2MASS image profiles, thus their opening angle is not well constrained (cf. their fig.~4).
Another reason the models rule out higher inclinations is that they require a larger envelope mass to compensate for the lower flux resulting from the higher inclination, and higher inclinations produce a deeper silicate feature which does not fit the mid-IR spectrum.                                

The inner radius is ${\sim}3$ times smaller than the one found by \citet{2003A&A...412..735P} and \citet{2013A&A...551A..43J}. 
Although the sublimation radius was allowed to change in order to obtain a maximum dust temperature of 1600\,K in the inner radius, the sublimation radii of the models did not increase noticeably. 
Smaller inner radii, near the sublimation radius, are also not strongly ruled out by \citet{2013A&A...551A..43J}.
In order to analyse the effect of a larger inner radius on the near-IR observations, a modified version of the best-fitting model was calculated with an inner radius twice as large. 
This model did not improve the fit to the speckle visibilities ($\chi^2=49$). 
In addition, the fit to the $JHK$-bands observations is worse than the best-fitting model (reduced $\chi^2$ differences of about 46, 53 and 99, respectively). 

Fig.~\ref{fig:cont:temp} shows that the peak temperature at 1000\,au scales reaches values between 200-300\,K inside the outflow cavities, whilst in the denser cavity walls the peak temperature decreases to 200-250\,K.
\citet[][]{2014ApJ...785...42P} obtained a temperature of 250\,K at 1000\,au scales from their simultaneous modelling of the SED for $\lambda>70\,\micron$ and SCUBA 450\,\micron\ and 850\,\micron\ radial profiles.
\citet{2019A&A...631A.142G} obtained a temperature distribution from 1-D LTE fitting of CH$_3$CN and H$_2$CO transitions of the combined NOEMA/CORE observations.
From their equations, we obtain a value of ${\sim}220$\,K at 1000\,au from the 1-D LTE modelling.
All these values are consistent with the temperature distribution from our dust emission modelling.

The parabolic shape of the cavity can help to explain the wider cavity opening angle close to the star as observed in maser emission \citep{2012ApJ...745..191S}. 
The density in the cavity is also consistent with the one expected for an outflow \citep[e.g.][]{1999ApJ...526L.109M} and the opening angle is consistent with previous results (e.g. CS line modelling by \citealp{1999ApJ...522..991V}, maser emission from \citealp{2012ApJ...745..191S}). 
Due to the lack of data, previous modelling of MYSOs used a constant small density (or empty) in the cavity \citep[e.g.][]{2010A&A...515A..45D}. 
This is not preferred in the modelling, as a constant density tends to produce more extended emission in the near-IR than observed. 
For a fixed cavity density distribution exponent, the density is determined mainly by the near-IR observations and the near/mid-IR regime of the SED. 
If the gas in the cavity is expanding isotropically, the outflow rate $\dot{M}_{\rm out}$ can be described as
\begin{equation}\label{eq:outflowrate}
    \dot{M}_{\rm out} = 4\pi r^2 v_{\rm out} \rho_{\rm cav}\,,
\end{equation}
with $v_{\rm out}$ the outflow velocity and $\rho_{\rm cav}$ the cavity mass density at radius $r$. 
The outflow rate is roughly 10~per~cent the accretion rate $\dot{M}_{\rm acc}$ \citep[e.g.][]{2000prpl.conf..867R,2018A&A...620A.182K}.
If the accretion rate is $\dot{M}_{\rm acc}=\dot{M}_{\rm env}$, an outflow velocity $v_{\rm out}=260$\,km\,s$^{-1}$ is obtained by solving eq.~\ref{eq:outflowrate} with the values in Table~\ref{tab:cont:best}. 
This velocity is within the range assumed by \citet{2003A&A...412..735P} for the expansion of the loops (100--500\,km\,s$^{-1}$) based on the $^{12}$CO line wings in the observations of \citet{1999ApJ...522..991V}. 
For the radiative transfer model cavity density, the outflow velocity should be in the range 200--320\,km\,s$^{-1}$ for infall rates within 20~per~cent of the best-fitting model.

In general, the 450\,\micron\ and 850\,\micron\ images from the modelling are not elongated along the cavity axis.
For the best-fitting model, the major-to-minor axis ratio of 2-D Gaussians fitted to the model images is ${\sim}1$ at both wavelengths, instead of ${\sim}1.3$ as observed (cf. Section~\ref{sect:data:multi:submm}). 
Elongation along the cavity axis was predicted by \citet{2010A&A...515A..45D} for 350\,\micron\ observations of the MYSO W33A, but this is not observed in the modelling presented here at pc scales. 
However, their source has a larger inclination angle (60\degr\ for their best-fitting model) and their cavity has a constant density distribution of $\rho\approx 2\times 10^{-20}$\,g\,cm$^{-3}$. 
Our results indicate that the elongation may be the result from earlier stages in the core formation and/or later evolution rather than produced by the cavities.
To a lesser extent and as stated in Section~\ref{sect:data:multi:submm}, changes in the PSF may also help to explain why the models are not elongated.

Similarly to other molecules, the methyl cyanide abundance needed to be defined as a piecewise function in order to fit the extension of the emission and the ratios between different $K$ lines. 
Fig.~\ref{fig:line:mom0:slc} shows that a constant abundance model produces much more extended emission. 
Moving the temperature step from $T_0=100$\,K to 200\,K underestimates the peak zeroth moment, thus the $T=100-200$\,K range is the most relevant one in order to fit the peak of the zeroth moment emission.
The inclusion of a step at 230--300\,K was not needed during the line radiative transfer modelling. 
This jump should increase the abundance of N-bearing molecules in the inner hotter regions \citep[e.g.][]{2001ApJ...546..324R}. 
Such temperatures are reached in regions much closer to the star which are not resolved by the methyl cyanide observations presented here. 
The gas temperature from the chemical modelling of the high-resolution NOEMA/CORE A+B+D data by \citet{2019A&A...631A.142G} also does not go above ${\sim}220$\,K at the peak of the emission.
Further refining of the radiative transfer grid is also needed in order to further explore the changes in abundance.

\subsection{Multi-wavelength modelling}
\label{sect:discussion:model}

Here we discuss several aspects of the modelling procedure.
First we discuss the sensitivity of the model parameters to our observations and the selection of the best-fitting model given the large amount of data available for this work.
Then we suggest improvements to the modelling with a focus in an observational aspect to constrain the less sensitive parameters and a theoretical aspect to improve the physical model, and thus obtain a better interpretation of the data for this and other MYSOs.

\subsubsection{Sensitivity of model parameters}
\label{sect:discussion:model:params}

In total there are 15 free parameters defining each density distribution and heating source for model types A--C without including the dust models, the stellar mass and stellar radius.
Additionally, there are 3 free parameters describing the kinematics of the gas.
The inner radius depends on other parameters, and a few of them are not expected to be well-constrained by the modelling based on the observations available. 
These include the parameters defining the disc flaring and density structure (3 parameters), and the stellar temperature.
The latter converged to different values for the three types of models presented in \citet[][]{2013A&A...551A..43J}, but their inner radius, which was not well-constrained, was a multiple of the sublimation radius which in turn depends on the stellar temperature.
Therefore, there are 10 free parameters which can be constrained by the continuum observations (see column ``constrained by'' in Table~\ref{tab:cont:best}).

\begin{figure*}
  \centering
  \includegraphics[width=\textwidth]{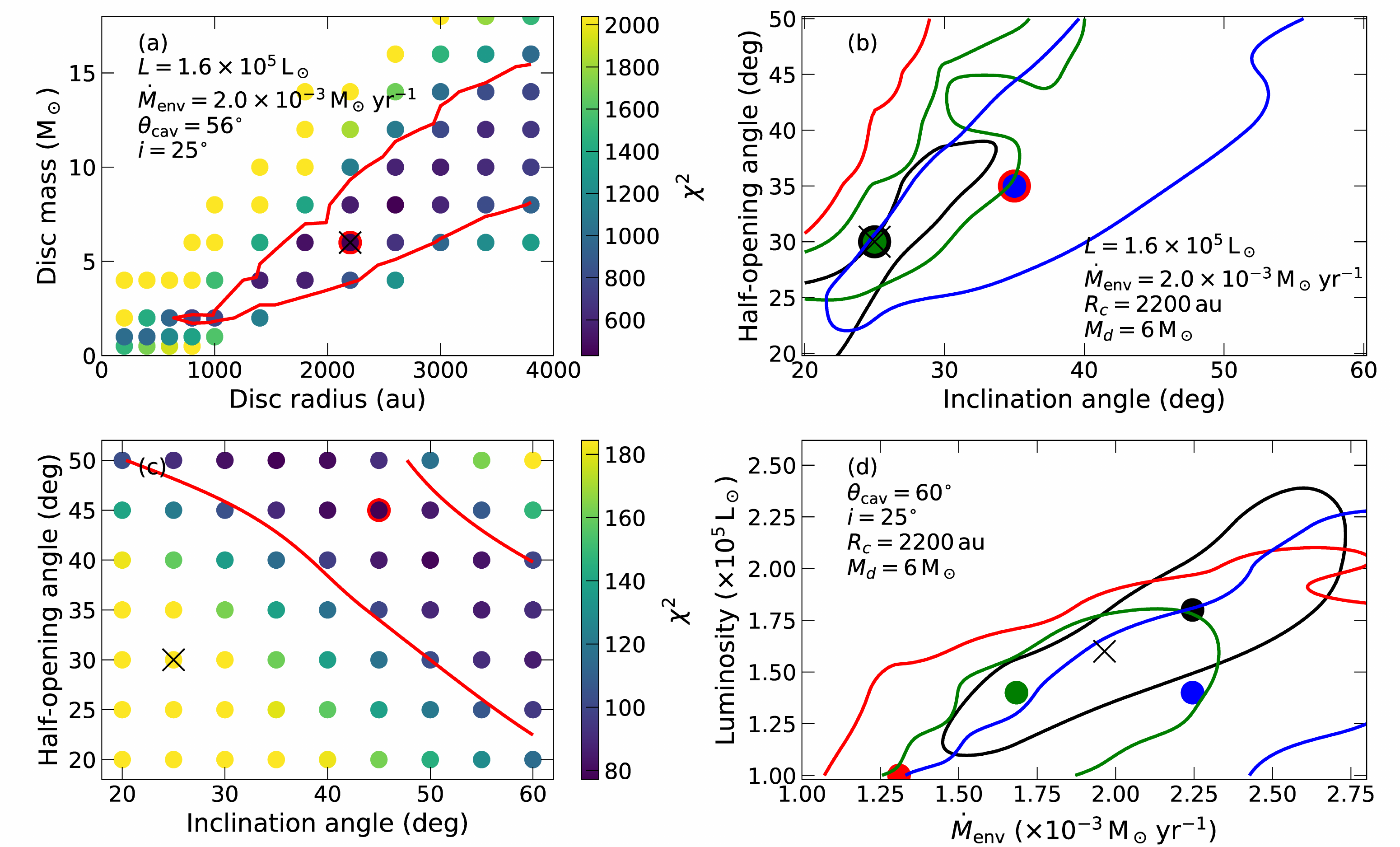}
  \caption{$\chi^2$ maps in parameter space. 
(a) $\chi^2$ of 1.3\,mm extended configuration interferometric visibilities in the disc mass/radius plane. 
The red contour correspond to a 50$\sigma_{\chi^2}$ level with $\sigma_{\chi^2}=10$. 
The dot with red border marks the minimum value. 
(b) Minimum $\chi^2$ value (dots) and 5$\sigma_{\chi^2}$ (see online Table~\ref{tab:cont:chi2std}) contour level for the SED (black), and $J$ (red), $H$ (green) and $K$ (blue) observations in the cavity half-opening/inclination angle plane. 
(c) \textit{Herschel} 70\,\micron\ $\chi^2$ map in the half-cavity opening/inclination angle plane. 
The dot with red border marks the minimum value and red contour corresponds to the 5$\sigma_{\chi^2}$ level with $\sigma_{\chi^2}=1$.
(d) Same as (b) but for the luminosity/infall rate plane. 
The black cross marks the values adopted for the best-fitting model. 
      }
  \label{fig:cont:chi2}
\end{figure*}

We calculated grids of models for parameters that can be well constrained by the observations available: inclination angle, cavity opening angle, envelope infall rate, disc mass and disc/centrifugal radius.
These grids allow us to estimate the uncertainties of the varied parameters.
Fig.~\ref{fig:cont:chi2} shows $\chi^2$ maps as a function of tentatively correlated parameters for specific observations.

Fig.~\ref{fig:cont:chi2}(a) shows the $\chi^2$ values for 1.3\,mm extended configuration visibilities in a grid of models with varying disc mass and radius, and the other parameters equal to the best-fitting model ones.
From the figure, we estimate an error of half the parameter resolution in this region of the grid, i.e. errors of 200\,au in disc radius and 1\,\msun\ in disc mass. 
We explored the correlation between these parameters and other observations, and found that in general IR observations (including 70\,\micron) and the SED prefer smaller disc radii.
On the other hand, sub(mm) observations prefer models with bigger discs and higher masses.
However none of these observations constrained these parameters as well as the 1.3\,mm extended configuration visibilities.

The inclination angle and cavity opening angle plane in Fig.~\ref{fig:cont:chi2}(b) shows that the minima of the SED and $H$-band observations coincide, whilst the $J$- and $K$-bands also coincide but at higher angles.
However, the contour levels show that the latter have flatter $\chi^2$ distributions than the former, and Fig.~\ref{fig:cont:nir} shows that good fits are obtained for the $J$- and $K$-bands at the best-fitting angles.
The error in this case is dominated by the SED, because its $\chi^2$ distribution is more steeper than the other observations as pointed by its contour.
We estimate an error of 5\degr\ for both angles.
For models with constant envelope infall rates, the SED is best fitted by models with higher values for both angles as the luminosity increases.
A similar behaviour is observed in the best fits of the $JHK$-bands, but due to the Monte Carlo noise the positions of the minima are more erratic.
On the other hand, if the luminosity remains fixed the inclination angle of the best-fitting model to the SED decreases as the envelope infall rate increases, and there is not a clear trend in the values of the cavity opening angle.
This is consistent with a deeper silicate feature at higher envelope infall rates (higher envelope masses) which is compensated by a lower inclination angle.
The same is observed in the minima for the $JHK$-bands observations.

We did not use the 70\,\micron\ to constrain the inclination and cavity opening angles.
The 70\,\micron\ distribution in Fig.~\ref{fig:cont:chi2}(c) shows that this observation is fitted by larger values for both angles.
This is probably pushed by the higher fluxes in the blue-shifted cavity as observed in the horizontal slice in Fig.~\ref{fig:cont:herschel}. 
The inclination angle affects mainly the extension of the emission along the outflow cavity whilst the change in opening angle affects both directions but the change is more subtle.
However, the increase in these angles does not increase the emission closer to the source as in the horizontal slice.
In addition, the density distribution cannot account for e.g. inhomogeneities in the cavity, which may improve the fit.
Submm observations are best fitted by models with lower opening angles, except for the 1.3\,mm extended configuration visibilities which prefer slightly larger values.
The latter are best fitted by slightly larger inclination angles (mainly in the 30--40\degr\ range) than the best-fitting model one. 
Even though we did not calculate a new grid of disc mass and radius with the best-fitting values for the inclination and cavity opening angles, Fig.~\ref{fig:cont:mm:vis} shows that a relatively good fit is obtained.

In Fig.~\ref{fig:cont:chi2}(d) we show a cut in the stellar luminosity and envelope infall rate for the best-fitting inclination and cavity opening angles.
In the figure, the position of the best-fitting model does not coincide with any of the individual observations, but it is in the intersection of the contours of the observations.
Its position is at the same distance to the SED and the $H$- and $K$-band observations, hence it provides relatively good fits to them.
The distribution of $\chi^2$ values for the $J$-band image is flatter than the other observations.
From these grids, we estimate an error in the envelope infall rate of $0.2\times10^{-3}\,\msun\,{\rm yr}^{-1}$, i.e. roughly 10~per~cent of the best-fitting model, and an error of $0.2\times10^5$\,\lsun\ for the stellar luminosity.
The minima for the best-fitted models to the 70\,\micron\ and 1.3\,mm compact configuration data are generally located towards the higher end of luminosity and infall rate in the grid, whilst the $K$-band speckle and the SCUBA radial profiles prefer values in the lower end of the grid of these parameters.
The minima of the 1.3\,mm extended configuration prefer higher luminosities and lower infall rates than the best-fitting model.

\begin{figure}
  \centering
  \includegraphics[width=\columnwidth]{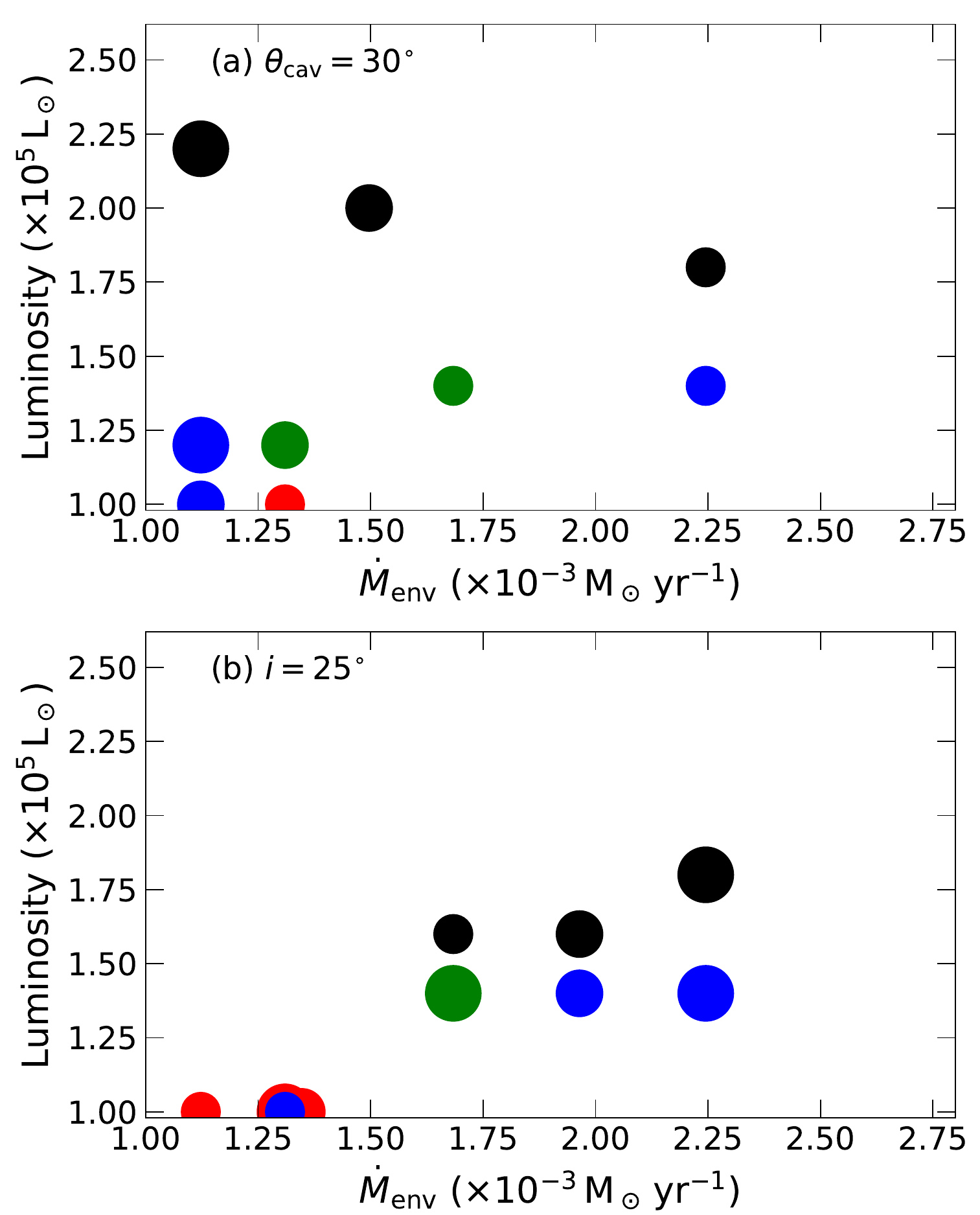}
  \caption{Position of the minimum $\chi^2$ in the luminosity-envelope infall rate plane for different (a) inclination angles and (b) cavity opening angles.
      The colour of the circles represent the SED (black), and $J$ (red), $H$ (green) and $K$ (blue) observations.
      The sizes of the circles increase with inclination angles (25, 35, 45\degr) and cavity opening angles (25, 28, 30\degr).
      }
  \label{fig:cont:chi2evol}
\end{figure}

Fig.~\ref{fig:cont:chi2evol} shows the position of the minima for grids with changing cavity opening angles and inclinations.
For a fixed cavity opening angle (Fig.~\ref{fig:cont:chi2evol}a), in the best-fitting models of the SED the luminosity increases and the infall rate decreases as the inclination angle increases.
This is expected because as the inclination angle increases the silicate feature is increasingly deeper, 
whilst a larger opening angle overestimates the fluxes between ${\sim}2-8$\,\micron\ because this range is the most affected by the cavity emission (see Fig.~\ref{fig:cont:sed}).
Hence, the envelope infall rate must decrease and the luminosity increase to compensate for the loss of flux in the envelope emission.
On the other hand, in the best-fitting models of the $JHK$-band observations the luminosity and the infall rate decrease as the inclination increases.
For a fixed inclination angle (Fig.~\ref{fig:cont:chi2evol}b), in the best-fitting models of the SED and the $JHK$-band observations the infall rate increases as the opening angle increases, and in the case of the SED the luminosity slowly increases.

As part of the visual inspection step, we explored models with different cavity properties.
Between models with cavity shape exponents $b_{\rm cav}=2$ and 1.5, the SED and the near-IR observations are best fitted by the former.
Models with lower cavity shape exponents the minimum $\chi^2$ has a larger inclination angle by 5\degr\ between the shape exponents explored.
For larger cavity shape exponents, the line of sight crosses a wider section of the envelope for a fixed inclination and cavity opening angle, hence the difference in minimum inclination angle.
Models with a steeper density distribution tend to fit the SED best, reaching a factor of ${\sim}2$ difference between the $\chi^2$ of a constant and a $p_{\rm cav}=2$ density distribution.
Similarly, the $J$- and $H$-band observations also favour steeper density distributions, whilst the $K$-band observation prefer intermediate exponents.
None the less, the difference in $\chi^2$ is not as sharp as for the SED.
A decrease in the cavity reference density of a factor of ${\sim}2$ or larger increases the $\chi^2$ of the SED by a similar factor, but the fit is less sensitive to changes of $0.2-0.4\times10^{-21}$\,g\,cm$^{-3}$.
To fit the near-IR observations including the speckle interferometry, different combinations of the cavity parameters can produce noticeable changes as these determine the amount scattered emission.
Some combinations produce local minima in $\chi^2$ values for the UKIRT observation fit, with comparable values between these minima.
Given that the emission is not homogeneous, different values may be trying to fit different sections of the cavity.
In general, cavity reference densities of 2 or more orders of magnitude smaller than the best fitting values for the SED are required to fit well the speckle visibilities, with increasingly larger density values for flatter density distributions.
All in all, it cannot be ruled out that a different combination of the 5 parameters determining the aspect of the cavity may produce better results as these are probably degenerate, and we cannot break these degeneracies because the models cannot account for the inhomogeneities observed.

As expected, models with higher submm dust opacities, such as WD01, require a lower accretion rate, i.e. lower envelope mass, to match the submm points, but still within ${\sim}20$~per~cent of the best-fitting model value. 
Another parameter that determines the envelope mass is its radius which is constrained mainly by the 850\,\micron\ intensity radial profile. 
Values between ${\sim}1.5-2.5\times10^5$\,au fit this profile at least up to 40\,arcsec. 
The simulated observation did not take into consideration chopping during the real observations, which can make the observed radius smaller.

Although models with different flaring, scale heights and vertical density exponents were calculated (cf. online Appendix~\ref{ap:modelling:cont:vis:params}), none of them improved the overall fit. 
Since most of the observations are not sensitive enough to these parameters, an accurate validity range cannot be given.

Stellar parameters are difficult to constrain since the proto-stellar emission cannot be observed directly in the near-IR. 
By taking into account the ratios of dust extinction/emission and scattering from our best-fitting model, we tried to constrain the stellar temperature by using the highest resolution near-IR point source data. 
However, the continuum is still dominated by photons scattered or emitted by dust, so this did not allow us to constrain the stellar temperature due to the low amount of direct stellar photons predicted by the model. 

On the other hand, the line fitting is affected mainly by the abundance and line width for a fixed stellar mass.
This was taken into account during the developing of the model grids.
We estimate that the data can be sensitive to a finer grid of abundance values, with steps of e.g. $0.1\times10^{-8}$.
The line width can also be better explored by a finer grid, but with lines whose width is not dominated by the gas kinematics.

\subsubsection{Best-fitting model selection}
\label{sect:discussion:best}

In our analysis we have used two methods to select the best-fitting models: one based on the ranking of $\chi^2$ values of each observation to select a seed model from the visual fit to dust continuum observations (Appendix~\ref{ap:modelling:cont:vis:proc}, online) and to select the line best-fitting model (Appendix~\ref{ap:modelling:line:mollie}, online), and another method where we develop a grid of models to fit specific dust continuum observations and parameters (Appendix~\ref{ap:modelling:cont:grid}, online).
The former ranking method has been used by a few authors rather than using a e.g. $\chi^2$ goodness of fit statistic. 
\citet{2002A&A...389..908J}, who combined SED and submm intensity profiles, argue that the $\chi^2$ values cannot be combined because the observations are not completely independent, thus it is not statistically correct to combine them. 
They also argue that intensity profiles and SEDs constrain different parameters of the density distribution. 
\citet{2005A&A...434..257W} add that the $\chi^2$ values are in different numerical scales, thus they should be compared in an ordinal way. 
It can also be added that e.g. radial profiles can have a number of data points comparable to the SED, thus certain wavelengths will carry more weight in a total $\chi^2$. 
In order to explore this further, we show in Table~\ref{tab:cont:chi2} the continuum best-fitting models for the visual inspection step selected by using a total reduced $\chi^2$ and an average reduced $\chi^2$. 

The best-fitting model as selected by the total reduced $\chi^2$ has a hotter star (36000\,K) with the same luminosity, a larger envelope ($R_\rmn{out}=270000$\,au) and a slight higher accretion rate ($\dot{M}_\rmn{env}=1.4\times10^{-3}\,\msun\,\rmn{yr}^{-1}$) than the best-fitting model as selected by ranking during the visual inspection step.
Their discs are the same except for the flaring and vertical density exponents, and their cavities have the same density distribution.
As was shown in Section~\ref{sect:results:cont} and in Fig.~\ref{fig:cont:chi2} higher inclination and opening angles are preferred by the 70\,\micron\ observations but not by the $J$ and $H$-band ones, and the SED.
It can also be seen in Table~\ref{tab:cont:chi2} that better $\chi^2$ values than in the best-fitting model by ranking are obtained in those observations with a higher number of points, namely the UKIRT $K$-band and 70\,\micron\ observations. 
Hence, this method for selecting the best-fitting model is dominated by the observations with a larger number of points.

The best-fitting model selected by the average $\chi^2$ is a type C model equal to the best-fitting type C model selected by ranking (see online Table~\ref{ap:tab:cont:best}), except for the envelope radius (200000\,au).
As argued, this type of models do not fit the near-IR observations and the SED.
In general, this method gives more weight to observations which have in general a high $\chi^2$ value, namely the 1.3\,mm NOEMA observations. 
It also reflects the difference in $\chi^2$ scale between the different observations. 
If all the observations were well fitted ($\chi^2\sim1$) then this best-fitting model should be the same as the ranking one.

In general the method used in the visual inspection fitting is independent of the intrinsic $\chi^2$ scale of each observations and the number of points of each one.
Overall the best-fitting model selected by ranking provides a good fit for key observations and its total and average $\chi^2$ values are relatively close to the best-fitting ones selected by these two $\chi^2$.
The disadvantage of the ranking is that it reduces the difference between models with a large difference in $\chi^2$ or increases it for models with similar $\chi^2$.  

On the other hand, in the grid modelling step of Fig.~\ref{fig:modelling} (see also Appendix~\ref{ap:modelling:cont:grid}, online) we determined the best-fitting model from the intersection of confidence regions in parameter space built from $\chi^2$ values for different dust continuum observations. 
This procedure is commonly used to constrain parameters in cosmological models.
The advantage of this method is that it does not depend on the intrinsic scale of the $\chi^2$ values and does not reduce the differences in $\chi^2$ value between models.
However, prior knowledge of the effect of each parameter in the fit to each observation is needed, and it requires more computing time to develop a grid.
Hence combining this method with a prior filter from our visual inspection fitting improved the $\chi^2$ values of key observations.

\subsubsection{Dust size distribution and grain growth}
\label{sect:discussion:dust}

\begin{figure}
  \centering
\includegraphics[width=\columnwidth]{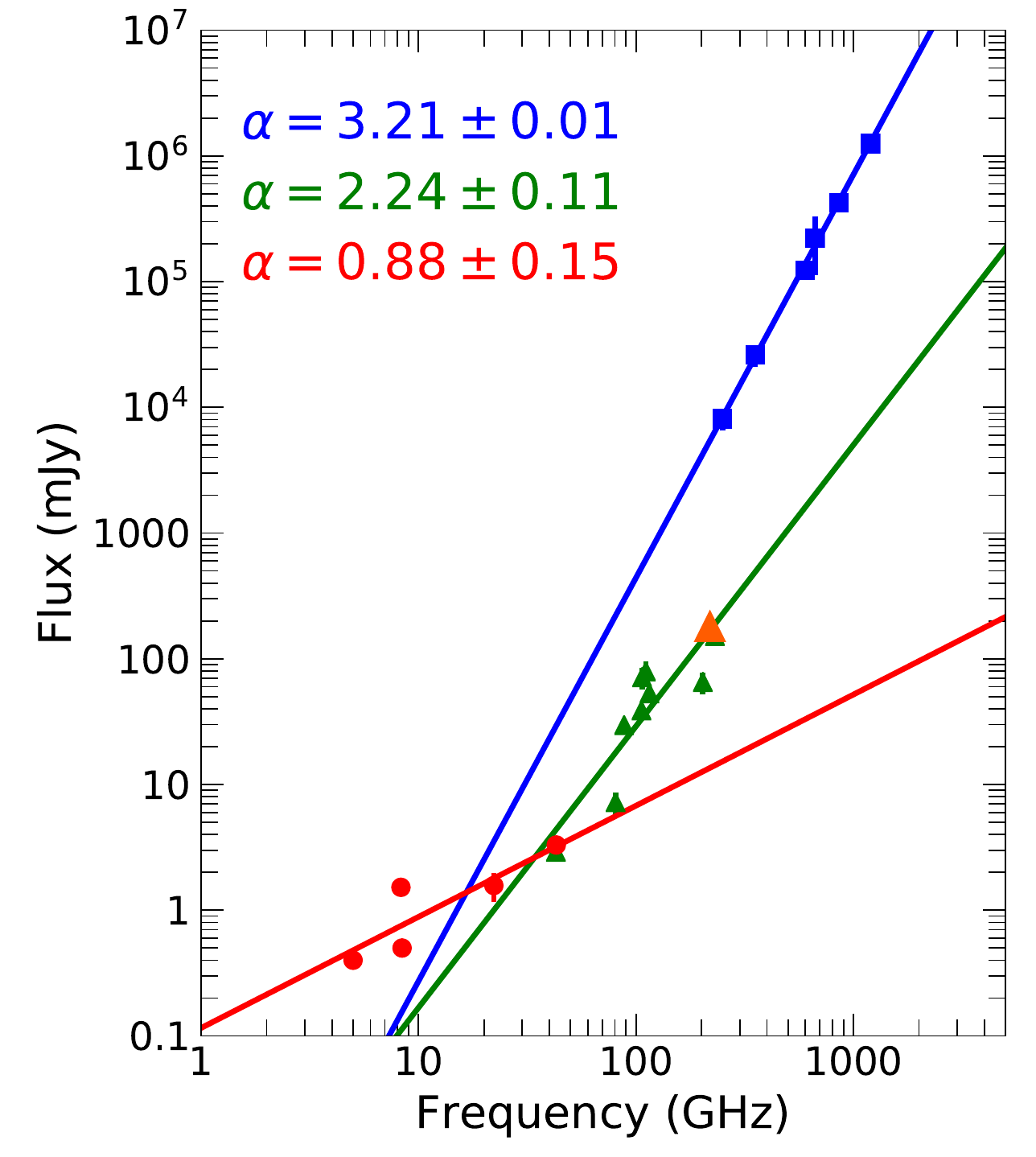}
  \caption{SED of the cm/submm regime as observed at different scales. 
      Blue squares fluxes correspond to single-dish observations and are presented in Table~\ref{tab:sed}. 
	  The orange triangle corresponds to the 1.3\,mm extended configurations flux in Table~\ref{tab:noema}, and the green triangles come from mm interferometric observations from \citet{1999ApJ...522..991V}, \citet{2013A&A...551A..43J} and \citet{2012A&A...543A..22W}.
      Red circles are radio/mm interferometric observations from \citet{1984ApJ...287..334C}, \citet{1995A&AS..112..299T}, \citet{2003ApJ...589..386T}, \citet[][]{2005A&A...437..947V} and \citet{2013A&A...551A..43J}. 
      The slopes of the fitted lines are listed in the upper left corner.
      If the emission is from dust, the dust emissivity slope is $\beta=\alpha-2$.}
  \label{fig:cont:spindex}
\end{figure}

In order to explore the change in dust from within the envelope or disc, we derived spectral indices based on dust emission at different scales from single-dish and interferometric observations in the radio/submm regime. 
The former traces cold dust located in the large-scale envelope, whilst the latter trace scales of a few 1000\,au. 
The dust optical depth is assumed to follow a power law $\tau\propto\nu^{\beta}$ in this regime (optically thin).
If it is further assumed that the emission is in the Rayleigh-Jeans regime, the fluxes are then $F_\nu \propto \nu^{\beta+2} \propto \nu^{\alpha}$.
Fig.~\ref{fig:cont:spindex} shows the observations made at different scales and the indices $\alpha$ of the power law fitted to the single dish, mm and radio interferometry data individually.

Dust models with shallower emissivity indices (e.g. WD01, see online Table~\ref{tab:dust}) fit the $\lambda>70$\,\micron\ regime of the SED better than dust with steeper spectral indices (e.g. KMH, OHM92), and allow the fit of visibilities at smaller baselines in the 1.3/1.4\,mm with smaller discs that better reproduce the 1.4\,mm emission. 
This is supported by the emissivity index derived from single dish observations ($\beta=1.2$ from Fig.~\ref{fig:cont:spindex}). 
However, shallower emissivity index models, i.e. models with larger grains (cf. online Table~\ref{tab:dust}), are not good at fitting the mid-IR regime and show an excess in emission along the cavity in the near-IR images because larger grains scatter more photons. 

The dust emissivity index derived from interferometric observations is $\beta=0.2$. 
This shallower emissivity index implies that larger grains are responsible for the mm disc emission, thus it may be evidence of grain growth in the denser regions, i.e. the disc, of the MYSO. 
However, contamination from free-free emission can contribute to this index, hence a higher emissivity index cannot be ruled out.
Although the best-fitting model from the visual inspection step favours a disc with smaller grains, good fits to the 1.3\,mm extended configuration visibilities can be obtained with larger grains (W02-3 grains), albeit reducing the disc mass.
On the whole, both types of grains actually do a good job at reproducing the observations, so there is not strong evidence that grain growth is needed to explain the data.

The spectral index of radio/mm observations and 3.6\,cm observations \citep[e.g.][]{2003ApJ...589..386T,2013A&A...551A..43J} points towards the presence of a jet or wind with $\alpha\sim0.6$. 
Hence, the spectral index from mm interferometry may be steeper if the contribution of the jet/wind emission is subtracted, which will depend on the turnover frequency of the jet emission.

The presence of different dust properties to those used here is also supported by modelling of near-IR polarimetric observations of \citet{2013MNRAS.435.3419S}. 
They found that elongated dust grains rather than spherical ones can reproduce their observations better. 
Therefore, the dust properties may be different depending on the physical conditions at different positions, e.g. dust in the cavity walls is being shocked by the outflows or dust in colder regions may develop ice mantles. 
This may allow dust with $\beta=1.5$ to be present in colder regions of the envelope, thus improving the fit of the submm observations, and dust with $\beta=2.0$ to exist in hotter or shocked regions like the cavity walls, which may improve the mid-IR fit.

\subsubsection{The Ulrich solution limitations}
\label{sect:discussion:ulrich}

The \cite{1976ApJ...210..377U} analytical solution for the accretion problem is able to reproduce some of the observed features. 
However, we could not find one model that can fit all the observations together. 
In particular, the results of the continuum and CH$_3$CN line modelling are contradictory. 
For instance, the stellar mass required by the Ulrich envelope with Keplerian disc velocity model is ${<}15$\,\msun which is much smaller than the one inferred from the source luminosity. 
This problem cannot be solved by changing the inclination angle, as lower inclination angles do not provide good fits to the $J$- and $K$-band observations.
The slower observed rotation than the one suggested by the theoretical model may require the effects of magnetic braking to slow down the rotation of the disc and inner envelope.
These limitations on the physics included in the Ulrich solution do not allow us to obtain better results.

Additionally, the simplicity of the models, where the clumpiness of the cavity or the presence of nearby sources are not taken into account, also limits the quality of the fit. 
The high-resolution 1.3\,mm extended configuration observations in Fig.~\ref{fig:cont:mm:clean}(b) recover more extended emission than the previous 1.4\,mm observations from \citet{2012A&A...543A..22W}, and show that the emission is more clumpy/less axisymmetric than the previous observations.
This has also been observed in other MYSOs, e.g. high-resolution Atacama Large Millimeter/submillimeter Array (ALMA) observations of W33A \citep{2017MNRAS.467L.120M}.
In the case of W33A filaments resembling spiral arms has been interpreted as accretion flows \citep{2018MNRAS.478.2505I}.
The velocity distribution in these accretion flows can be modelled with an Ulrich envelope by selecting some of its streamlines \citep[e.g.][ for low-mass star forming regions]{2012ApJ...748...16T}.
Hence a full Ulrich envelope, rather than selecting some streamlines, is less likely.

Simulations with magnetic fields and radiation transfer show promising results.
The radiation-magnetohydrodynamics (RMHD) simulations of early stages in the formation of high-mass stars calculated by \citet{2011ApJ...742L...9C} show that magnetic fields and radiation can reduce the fragmentation during early stages in the formation of massive stars. 
The collapse of a $10^3$\,\msun\ rotating molecular cloud investigated by \citet{2011ApJ...729...72P} through RMHD simulations also shows that magnetic fields can reduce fragmentation and that they are important in transporting angular momentum at local scales.
Furthermore, the RMHD simulations of \citet{2013ApJ...766...97M}, which explore a larger time evolution of a collapsing core with an initial power law density distribution index of --1.5, show that magnetic fields can remove angular momentum from the infalling material through magnetic braking whilst still forming a massive star (20\,\msun).
Their models can still form an accretion disc but their centrifugal radii are ${<}50$\,au, thus smaller than the one of AFGL~2591.
Larger accretion discs can be obtained when non-ideal MHD effects are taken into account \citep{2018A&A...620A.182K}.
These RMHD models also show that the matter is accreted into the disc from filaments which have collapsed within the envelope at scales of 5000\,au in the mid-plane.
However, these models do not include feedback from an outflow.

The non-magnetic simulations by \citet{2016ApJ...823...28K} produce a Keplerian disc but they focus on 1000\,au scales, thus in the transition from the envelope to disc. 
They also have a higher rotational to kinetic energy ratio than \citet{2013ApJ...766...97M} by a factor of ${\sim}5-30$, and impose an initial solid body rotation.
In the 3-D radiation-hydrodynamic simulations of \citet{2018MNRAS.473.3615M}, Keplerian discs are produced which develop spiral arms and fragment into a multiple stellar system.
The 2-D simulations of \citet{2015ApJ...800...86K,2016ApJ...832...40K,2018A&A...616A.101K} show the effect of radiation on the outflow cavity and its widening from an initially slowly rotating core. 
The cavity distributions in these models resemble the near-IR observations of AFGL~2591. 
However, they do not include magnetic fields.
Newer simulations are including the effects of an increasing number of physical processes, e.g. in the \citet{2018A&A...616A.101K} simulations the effects of photo-ionisation from a potential H\textsc{\lowercase{II}} region are included.
These and future simulations have the potential to describe better the physical properties in MYSOs from high-resolution observations. 

\subsubsection{A recipe for future multi-wavelength modelling}
\label{sec:discussion:constraints}

\begin{figure}
    \resizebox{\columnwidth}{!}{
        \begin{tikzpicture}[node distance=1.5cm, every node/.style={fill=white}, align=left]

            \node (start1) [noderesult, yshift=-0.8cm] {Direct measure:\\Distance (parallax, kinematic otherwise)\\Envelope radius (single dish profiles)\\Envelope mass (single dish emission)\\Disc mass (mm interf. emission)};
            \node (start2) [noderesult, below of=start1, yshift=-0.8cm] {Geometry:\\PA (near-IR, outflow, mm interf. from disc)\\Inclination (near-IR, outflow, mm interf. imaging)\\Cavity shape (near-IR, outflow)\\Disc radius (mm high-res. imaging, mid-IR)\\Disc shape (mid-IR, mm high-res. imaging)};
            \node (start3) [noderesult, below of=start2, yshift=-0.5cm] {Simple modelling:\\Luminosity (SED templates)\\Disc shape (mid-IR visibilities fitting)};

            \node (refine0) [noderesult, below of=start3, yshift=-0.8cm, align=center] {Envelope radius\\(single dish observations)};
            \node (refine1) [noderesult, below of=refine0, xshift=-2.5cm, align=center] {Disc radius - disc mass\\(mm interf. visibilities)};
            \node (refine2) [noderesult, below of=refine0, xshift=2.5cm, align=center] {Disc shape\\(mid-IR high-res.)};
            \node (refine3) [noderesult, below of=refine1, align=center] {Luminosity - envelope mass\\(SED, near-IR)};
            \node (refine4) [noderesult, below of=refine2, align=center] {Inclination - cavity shape\\(SED, near-IR, outflow)};
            \node (refine5) [noderesult, below of=refine3, xshift=2.5cm, align=center] {Cavity density\\(SED, near-IR)};
            \node (refine6) [noderesult, below of=refine5, xshift=-2.5cm, align=center] {(Proto-)Stellar mass\\(High-res. lines)};
            \node (refine7) [noderesult, below of=refine5, xshift=2.5cm, align=center] {Outflow velocity\\(outflow)};

            \node (further1) [noderesult, below of=refine6, yshift=-1.4cm, xshift=2.5cm] {Entrainment parameters\\(near-IR, outflow)};
            \node (further2) [noderesult, below of=further1, xshift=-2.5cm, align=center] {Luminosity - envelope mass\\(SED, near-IR)};
            \node (further3) [noderesult, below of=further1, xshift=2.5cm, align=center] {Inclination - cavity shape\\(SED, near-IR, outflow)};
            \node (further4) [noderesult, below of=further2, xshift=2.5cm, align=center] {Cavity density\\(SED, near-IR)};

 \draw[<->, dashed]      (refine1) -- (refine2);
 \draw[<->, dashed]      (refine3) -- (refine4);
 \draw[<->, dashed]      (refine3) -- (refine5);
 \draw[<->, dashed]      (refine4) -- (refine5);
 \draw[<->, dashed]      (further1) -- (further2);
 \draw[<->, dashed]      (further3) -- (further1);
 \draw[<->, dashed]      (further3) -- (further4);
 \draw[<->, dashed]      (further2) -- (further4);

 \draw [color=blue,thick](-5cm,1cm) rectangle (4.5cm,-6cm);
 \node at (-0.5,1.2) [above=-6.0mm, right=-4.2cm, align=left] {\textsc{Step 1: initial parameters}};
 \draw [color=blue,thick](-5cm,-6.2cm) rectangle (4.5cm,-14.2cm);
 \node at (-0.5,-6.1) [above=-5mm, right=-4.2cm, align=left] {\textsc{Step 2: refining}};
 \draw [color=blue,thick](-5cm,-14.4cm) rectangle (4.5cm,-19.9cm);
 \node at (-0.5,-14.3) [above=-8mm, right=-4.2cm, align=left] {\textsc{Step 3: specific features}\\Same as Step 2 with a modified model\\to explain specific observations e.g.:};
\end{tikzpicture}
}
    \caption{Steps for modelling and bringing up constraints from MYSOs observations.
    Parameters under ``cavity shape'' are opening angle and cavity shape exponent.
    Parameters for ``disc shape'' are scale height and flaring exponent.
    The item ``outflows'' refers to molecular line observations of outflows.
    Simple modelling includes fitting with templates or with models such as 2-D Gaussians.}
    \label{fig:disc:modelling}
\end{figure}
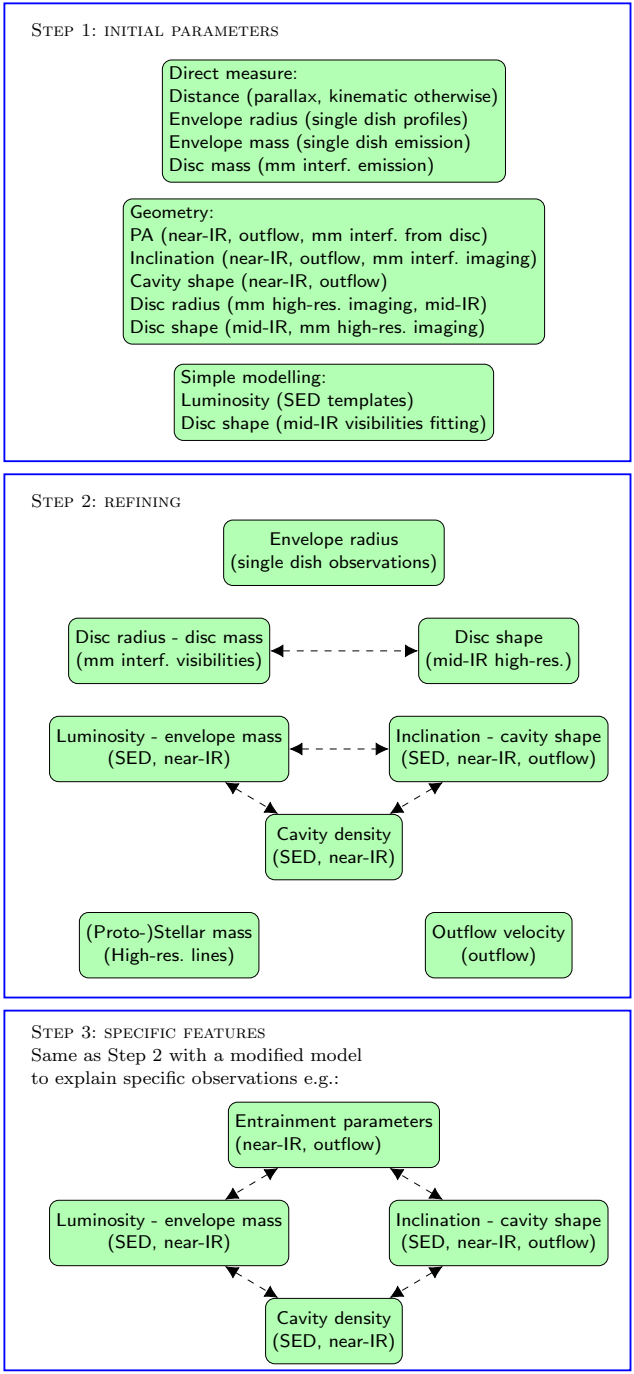

The diagram in Fig.~\ref{fig:disc:modelling} summarises the steps necessary for modelling multi-wavelength data with 2-D radiative transfer models in order to obtain better constraints to the model physical parameters from specific observations.
The modelling is performed in three steps with increasing refinement of key parameters at each step and complexity in the modelling, and each step is divided in a series of sub-steps.
The sub-steps are in sequential order, although some of them can be swapped or skipped depending on the available data.
Here we discuss the main features and caveats of each step.

The aim of the first step is to obtain initial values and constraints for the 2-D model parameters in Table~\ref{tab:cont:best}.
The observations for each parameter in Fig.~\ref{fig:disc:modelling} are in order of preference.
Some of these parameters will then be further constrained in the next steps.
The initial value for the luminosity can be obtained from model templates \citep[e.g.][]{2011A&A...525A.149M}. 
These model templates can account for rotationally flattened envelopes and cavities \citep[e.g.][]{2017A&A...600A..11R}, hence obtaining an initial range of inclination angles and cavity properties can help to constrain the templates.
We have listed the disc shape (disc flaring exponent, scale height) in two sub-steps.
The first one is for cases of discs closer to edge-on using imaging whilst the second relies on the modelling of mid-IR visibilities \citep[e.g.][]{2009A&A...500.1065D}.
Although measuring these disc properties, and to a lesser extent the disc radius, from mid-IR observations is desirable as they are tracing optically thick emission from the surface of the disc, some MYSOs may still only be partially resolved in the mid-IR in high-resolution observations \citep[e.g.][]{2009A&A...494..157D}.
Envelope and disc masses measured from dust emission are obtained by estimating a dust temperature from the SED. 
These are a good approximation in terms of order of magnitude for the next step.
Some deeply embedded MYSOs may be weak in the near-IR, hence to derive the properties of the outflow cavity shape (inclination angle, opening angle, shape exponent) molecular line emission tracing the outflow (e.g. CO) can be used instead.
However, precession of the outflow axis can affect the determination of the cavity geometry, PA and inclination angle.
In such cases, high-resolution mm interferometric observations tracing the disc can provide further constraints.
In general, most geometric properties are degenerate due to projection effects, except at extreme inclination angles.

In Step 2, we have integrated the main relations found in our modelling and added further steps to constrain other parameters.
All the sub-steps in this step constrain the physical parameters through 2-D axisymmetric modelling and are relatively independent from each other.
Some sub-steps iterate between parameter-observations relations in order to optimise the modelling, however $n$-dimensional optimisation, with $n$ being the number of parameters in the sub-step, can also be performed.
Additional parameters constraining the velocity distribution are introduced in this step after modelling the continuum data.
Initial values for these parameters can be derived from the luminosity (stellar mass) or analysis of the spectra (outflows).

The objective of the final step is to model 2-D axisymmetric features of specific observations in order to shrink the area in $\chi^2$ space in the relations of the previous step.
Hence the sub-steps in the previous step must be repeated with the new model and an additional iteration must be performed to constrain the parameters of the newly introduced features (see example in Fig.~\ref{fig:disc:modelling}).
Since the sub-steps are relatively independent, the introduction of other 2-D structures may not affect all the relations.
For instance, introducing entrainment features to better model near-IR observations will likely not affect the envelope radius and disc properties.
On the other hand, introducing a torus to explain mid-IR features \citep[e.g.][]{2011A&A...526L...5D} can affect the disc properties as well as the envelope mass.
Determining the orientation of the 2-D features should be straight forward because the orientation angles (PA and inclination) were constrained in the previous steps.
This step requires that the comparison between model and observations is performed on a pixel-by-pixel basis for images in order to increase the number of data points needed to fit the increasing number of parameters in the model.
Some of the additional model features may be geometrical rather than physical, hence this step may only constrain the physical parameters in the previous one.
Non-axisymmetric features can then be introduced to the models to optimise the relationships in Step 2 (see Section~\ref{sect:discussion:ulrich} for examples).

\section{Conclusions}
\label{sect:conclusions}
We used spatially resolved dust continuum images between 1 to 1400\,\micron\ and the SED to study the dust density and temperature distributions, and methyl cyanide interferometric observations to study the inner envelope velocity distribution of the proto-typical MYSO AFGL~2591. 
We performed a multi-step radiative transfer modelling to constrain the physical model parameters and determined correlations between specific observations and model parameters.
We show for the first time that the \textit{Herschel} 70\,\micron\ emission is extended in the outflow direction and explain its extended nature as the result of dust emission from the outflow cavity, whose walls are heated by the radiation escaping through the less dense cavity medium.
We found that the inner envelope is rotating and this rotation is slower than the disc.

We have attempted to fit a large number of observations with a unified physical model. 
The best-fitting model consists of an Ulrich envelope, a flared disc and bipolar outflow cavities with properties consistent with previous results. 
The presence of a disc is needed in order to fit NOEMA 1.3\,mm extended configuration visibilities. 
We derive a disc mass of $6$~\msun\ and 2200\,au radius. 
These values are relatively independent of the parameters determining the envelope/outflow cavities density distribution.
The 450\,\micron\ and $K$-band speckle interferometry observations were not fitted well by the models.

This model cannot explain the observed CH$_3$CN pv maps, because the model lines are wider due to the high velocity gradients of the Ulrich envelope with Keplerian disc model.
In addition, the model lines are not skewed like the observed ones. 
Accretion flows and magnetic braking may be able to explain why the inner envelope rotation is slower than predicted by the Ulrich envelope.

The physical limitations of the Ulrich solution, together with the simplicity of the models, do not allow an improvement over what we have done in this paper. 
3-D self-consistent theoretical models should be tested in order to improve the fit to the observations. 
Additional improvement to the modelling and high-resolution multi-wavelength observations may also be needed to constrain these parameters.
Future papers including higher-resolution data from CORE and other surveys will allow a more complete study of MYSOs to test these models.

\section*{Acknowledgements}
The authors would like to thank T. Preibisch, J. P. Simpson and K.-S. Wang who kindly provided their data.
F.O. acknowledges the support of CONICYT Becas-Chile 72130407.
A.A., H.B., J.C.M., and C.G. and S.S. acknowledge support from the European Research Council under the European Community's Horizon 2020 framework program (2014-2020) via the ERC Con-solidator Grant ‘From Cloud to Star Formation (CSF)' (project number 648505).
HB also acknowledges funding from the Deutsche Forschungsgemeinschaft (DFG) via the Collaborative Research Center (SFB 881) ‘The Milky Way System' (subproject B1).
A.P. acknowledges financial support from CONACyT and UNAM-PAPIIT IN113119 grant, M\'exico.
DS acknowledges support by the Deutsche Forschungsgemeinschaft through SPP 1833: ``Building a Habitable Earth'' (SE 1962/6-1).
R.K. acknowledges financial support via the Emmy Noether Research Group on Accretion Flows and Feedback in Realistic Models of Massive Star Formation funded by the German Research Foundation (DFG) under grant no. KU 2849/3-1 and KU 2849/3-2.
This research made use of Astropy, a community-developed core Python package for Astronomy (Astropy Collaboration, 2013).
Part of this work is based on archival data, software or online services provided by the ASI Science Data Center (ASDC).

\section*{Data Availability}

The data underlying this article will be shared on reasonable request to the corresponding author.
Parts of these data were derived from: WFCAM Science Archive (\url{http://wsa.roe.ac.uk/}), \textit{Herschel} Science Archive (\url{http://archives.esac.esa.int/hsa/whsa/}), SCUBA Legacy Catalogues (\url{http://www.cadc-ccda.hia-iha.nrc-cnrc.gc.ca/community/scubalegacy/}). 
The $K$-band speckle interferometric data were provided by T. Preibisch by permission.

\Urlmuskip=0mu plus 1mu\relax
\bibliographystyle{mnras}
\bibliography{manuscript}

\appendix

\section{Modelling details}
\label{ap:modelling}

The following sections describe in detail the modelling steps summarised in Fig.~\ref{fig:modelling}.

\subsection{Dust continuum: visual inspection}
\label{ap:modelling:cont:vis}

In order to obtain the dust density and temperature distributions, we used the 3-D Monte Carlo radiative transfer code \textsc{\lowercase{Hyperion}} \citep{2011A&A...536A..79R} to calculate SEDs and images produced by a dust distribution heated by a given heating source. 
We first performed a fitting by varying the model parameters based on $\chi^2$ values and visual inspection.
In this step we explored different density distributions and dust models.
This allowed us to constrain the range of values for some parameters and to identify correlations between parameters and observations.
We then built model grids to constrain parameters strongly related to some of the observations, and to provide fitting uncertainties.

\subsubsection{Density distributions}
\label{ap:modelling:cont:vis:dens}

Table~\ref{tab:densities} presents the different density distribution types considered here, and further details can be found in Appendix~\ref{ap:density}.
For type A, we follow the nomenclature and fiducial values in \citet[][see also Table~\ref{tab:cont:parameters}]{2013A&A...551A..43J} in order to compare the results with their results.
Types B and C are used to study the effects of different types of envelope and disc distributions on the fit. 
All these density distributions have been used and/or proposed to explain the emission of MYSOs.

\begin{table}
 \centering
 \caption{Density distributions.}
 \label{tab:densities}
 \begin{tabular}{ccc}
  \hline
  Type & Density structures$^\rmn{a}$ & $N_{\rm model}^\rmn{b}$ \\
  \hline
  A & Ulrich envelope$^\rmn{c}$ + flared disc & 444\\
  B & Ulrich envelope$^\rmn{c}$ + alpha disc & 131\\
  C & Power law envelope$^\rmn{d}$ + flared disc& 52\\
  \hline
 \end{tabular}
 \begin{minipage}{\columnwidth}
 $^\rmn{a}$ All models include bipolar cavities.\\
 $^\rmn{b}$ Total number of models for each type.\\
 $^\rmn{c}$ \cite{1976ApJ...210..377U}.\\ 
 $^\rmn{d}$ $n\propto r^{-p}$ with $n$ the density and $r$ the distance to the
 source.    
 \end{minipage}
\end{table}

For the density distributions, we used the built-in \textsc{\lowercase{hyperion}} functions. 
The values of the distributions were calculated at each cell of a spherical coordinate grid. 
The densities of all the components are defined between the inner and outer radii, and are defined to model a single source (AFGL~2591 VLA~3).
We fixed the maximum radius of the grid to $3\times10^{5}$\,au, i.e. larger than the envelope radius. 
The optimal inner radius of the grid was chosen by \textsc{\lowercase{hyperion}} based on the inner radius of the density distributions (the dust sublimation radius).
To calculate the total density distribution, \textsc{\lowercase{hyperion}} calculates the density distribution of each component in all cells. 
Then it replaces the cells within the bipolar outflow cavities, which were initially set to the envelope density, by the density of the cavity component, and finally it sums this with the disc component. 
The model grid had 700 cells with a logarithmic distribution in the radial direction and 200 cells with a sinusoidal distribution in the elevation direction ($\theta$ in spherical coordinates), thus improving the number of cells close to the inner radius and the mid-plane. 
For most of the models $5\times10^{6}$ photons were used for the calculations (specific energy and imaging) in order to optimise the computing time, and we set the convergence criteria to that in \citep{2011A&A...536A..79R}.

\subsubsection{Fitting procedure}
\label{ap:modelling:cont:vis:proc}

Table~\ref{tab:cont} lists the data product type (image, intensity radial profile, visibility profile) that we used to compare the observations and the respective synthetic observation in order to extract spatial information.
To obtain each of these data products from the code results and compare them with the observed ones, we binned the images to match the observed pixel size, rotated them to match the cavity position angle \citep[259\degr;][]{2003A&A...412..735P} and convolved them with the respective PSF.
The details about the processing of the model images can be found in Appendix~\ref{ap:dataproc}.

In order to determine a set of models that best fit our data, we first calculated the reduced $\chi^2$, for each of the datasets listed in Table~\ref{tab:cont} and the SED.
The reduced $\chi^2$ were calculated after masking the non-relevant data (see Appendix~\ref{ap:dataproc}).
Similarly to \citet{2005A&A...434..257W}, the rank of the models for each of these reduced $\chi^2$ was calculated, i.e. the rank of the models for each of the types of data. 
Finally, we combined these ranks to obtain an average rank per model. 
In the particular case of the $K$-band and 1.3~mm observations, where we have two measurements for the same wavelength, the average rank of the two observations was calculated at each wavelength before they were combined with the results from the other observations. 
As reference, the best-fitting model from this step is that with the lowest ranking.

\subsubsection{Assumed parameters}
\label{ap:modelling:cont:vis:params}

Most initial parameters were taken from the results of \citet{2013A&A...551A..43J}, where we used average values of their two envelope with disc models, e.g. the disc scale height and density in the cavity. 
We also used initial parameters from \citet{2010A&A...515A..45D} for the MYSO W33A, such as the envelope infall rate. 
The cavity density distribution ($\rho \propto r^{-p_{\rm cav}}$) and shape parameters ($z\propto R^{b_{\rm cav}}$) were taken from observations and theoretical models of outflows \citep[e.g.][]{2003A&A...412..735P,2010ApJ...720.1432B}.
These parameters were varied and new models computed based on the fitting statistics described in Section~\ref{ap:modelling:cont:vis:proc} and a visual inspection of the results. 
Since analyses of the effects of changing each parameter in the model images and SED already exist \citep[e.g.][]{2004A&A...419..203A}, we built our models in this step from this knowledge and the previous studies of this source.

Table~\ref{tab:cont:parameters} summarises the input parameters and the ranges used for the models.
Models with an envelope and a cavity were used to determine the stellar luminosity that best fits the SED peak and shape.
We obtained a value of $L=1.6\times10^5$~\lsun, which is slightly lower than the value in the literature \citep[$L=2\times10^5$\,\lsun;][]{2013ApJS..208...11L}. 
The luminosity was fixed to this value for most models, but we also calculated models with other luminosities in order to improve some fits, e.g. $L=1.4\times10^5$\,\lsun\ for models with WD01 dust (see below). 
The stellar spectrum was assumed to be that of a black body and its radius was calculated self-consistently with the temperature and luminosity. 
Since all the emission is reprocessed by dust and our observations are insensitive to the stellar temperature, we do not expect a large change in our results if a stellar atmosphere model is used instead.

\begin{table*}
 \centering
    \caption{Parameter ranges used in the visual inspection and grid fitting steps.}
\label{tab:cont:parameters}
 \begin{tabular}{lccc}
  \hline
  Parameter &  Visual & \multicolumn{2}{c}{Grid} \\
            &       & Step 1 & Step 2 \\
  \hline
  \multicolumn{4}{c}{Type A}\\
Inclination angle ($i$) & 20--40\degr (5\degr) & 20--40\degr (5\degr) & 20--60\degr$^{a}$ (5\degr)\\
Stellar temperature ($T_\star$/K) & 10000--40000 & \multicolumn{2}{c}{18000} \\
Stellar luminosity ($L_\star/10^5$\,\lsun)     & 1.4--2.0 & 1.6--2.0 (0.2) & 1--2.6 (0.2)\\
Outer radius ($R_\rmn{out}$/au) & 100000--300000$^{b}$ & \multicolumn{2}{c}{195000} \\
Envelope infall rate ($\dot{M}_\rmn{env}/10^{-3}\,\msun\,{\rm yr}^{-1}$) & 0.2--3.0 & 1.3--2.6 & 0.9--1.7 (0.2),\\ 
                                                                         &          &          & 2.0,  2.2--3.0 (0.4)\\
Envelope dust$^c$ & OHM92, KMH, WD01 & \multicolumn{2}{c}{OHM92} \\
Centrifugal radius ($R_\rmn{c}$/au) & 100--1000 & 200--1000 (200), & 2200 \\
                                    &           & 1000--3800 (400) & \\ 
Disc mass ($M_\rmn{d}$/\msun) & 0.25--4 & 0.5, 1, 2--18 (2) & 6 \\
Disc scale height at $r=100$ au ($h_0$/au) & 0.1--11.0 & \multicolumn{2}{c}{0.6} \\
Disc flaring exponent ($\alpha$) & 2.25, 1.88, 2.5 & \multicolumn{2}{c}{2.25} \\
Disc vertical density exponent ($\beta$) & 1.25, 1.13, 1.5 & \multicolumn{2}{c}{1.25} \\
Disc dust$^c$ & OHM92, W02-3 & \multicolumn{2}{c}{OHM92} \\
Cavity half opening angle$^{d}$ ($\theta_{\rm cav}$) & 25--40\degr & 28\degr & 15--50\degr\ (5\degr) \\
Cavity shape exponent ($b_\rmn{cav}$) & 1.5, 2.0 & \multicolumn{2}{c}{2.0} \\
Cavity density exponent ($p_{\rm cav}$) & 0, 1.5, 2.0 & \multicolumn{2}{c}{2.0} \\
Cavity reference density$^{d}$ ($\rho_{\rm cav}/10^{-21}$\,g\,cm$^{-3}$) & $2\times10^{-6}$--21 & \multicolumn{2}{c}{1.4} \\
Cavity dust$^c$ & OHM92, KMH & \multicolumn{2}{c}{KMH} \\
  \multicolumn{4}{c}{Type B}\\
     Stellar Mass ($M_\star/\msun$)                   & 40 & -- & -- \\
     Disc accretion rate ($\dot{M}_\rmn{acc}/\msun\,{\rm yr}^{-1}$) & $1\times10^{-4}$ & -- & -- \\
  \multicolumn{4}{c}{Type C}\\
     Envelope power law index ($p$)                   & 2.0 & -- & -- \\
     Envelope mass ($M_\rmn{env}/10^3\,\msun$)        & 1--5 & -- & --\\
  \hline
 \end{tabular}
    \begin{minipage}{\textwidth}
        \textbf{Notes.} Step 1 refers to the fitting of the 1.3\,mm extended configuration visibilities to constrain the disc mass and radius. 
Step 2 was developed to constrain the envelope infall rate, luminosity, inclination and cavity opening angles (see Section\,\ref{ap:modelling:cont:grid}). 
The numbers in brackets indicate the spacing of the values in the grid.\\
        $^a$ The range depends on the opening angle for grids of luminosity and envelope infall rate, but all cover the 30--40\degr\ range.\\
        $^b$ As an exception, one model with radius 790000\,au was calculated.\\
        $^c$ See Table~\ref{tab:dust}.\\
        $^d$ Defined at a height $z=10^4$~au.
    \end{minipage}
\end{table*}

Based on \citet{2011MNRAS.416..972D} calculations, we estimated a ZAMS stellar mass between $M_\star=30-40$\,\msun\ for a stellar luminosity of $L=1.6\times 10^5$\,\lsun. 
From eq.~\ref{eq:ulrich}, the stellar mass and the infall rate are degenerate parameters for a given envelope density. 
The stellar mass is considered as a free parameter only in models with an alpha disc (type B models) as it determines the accretion luminosity ($L_\rmn{acc}\propto M_\star$). 
The stellar mass can only be constrained by studying the molecular line emission.
Since we perform the line modelling after the continuum fit, we can scale the envelope infall rate to match the envelope mass if needed.

The initial inner radius of the model, $R_{\rm in}$, was calculated using eq.~\ref{eq:ap:rsublimation} with a sublimation temperature of 1500\,K. 
Sublimation of dust was allowed during \textsc{\lowercase{Hyperion}} iterations by decreasing the dust density in those cells exceeding an upper limit dust sublimation temperature of 1600\,K. 
Although this temperature is higher than the one used to estimate the inner radius using the \citet{2004ApJ...617.1177W} method, the range of sublimation radius values are in a similar range (30--40\,au). 
This upper temperature limit was reached by most models in each density component.

In this step, we kept the disc mass $M_\mathrm{disc}\lesssim0.1\,M_\star$ and its radius below 1000\,au, which are close to the values found by \citet{2012A&A...543A..22W}, determined from the total flux of their interferometric observations at 1.4\,mm.
We explored more massive and larger discs during the grid fitting step (see Section~\ref{ap:modelling:cont:grid}). 
For most models, parameters describing the disc flaring and vertical density distribution were fixed to the values in \citet[][$\alpha=2.25$ and $\beta=1.25$]{2013A&A...551A..43J}, and the disc radius was fixed to the centrifugal radius in most cases. 
In the few cases where we varied these parameters, their lower limits were those of a steady Keplerian disc in \citet[][$\alpha=1.88$ and $\beta=1.13$]{2004A&A...419..203A}, and upper limits of $\alpha=2.5$ and $\beta=1.5$. 
The scale height was chosen to produce a physically thin disc. 

For type B models, the $\alpha$-disc parameters were fixed to their flared disc equivalent in type A models, whilst the envelope and cavity parameters were constrained in the same ranges as type A models. 
We expect the accretion rate to be a fraction of the infall rate.
\citet{2010A&A...515A..45D} found that the accretion rate can be three orders of magnitude smaller than the envelope infall rate for a compact (500\,au) and less massive (0.01\,\msun) $\alpha$-disc to fit mid-IR observations of the MYSO W33A.
From eq.~(15) of \citet{2010ApJ...721..478H}, accretion rates slightly higher than $10^{-3}$\,\msun\,yr$^{-1}$, which is comparable with the envelope infall rate, are expected at later phases when the star is settling into the main sequence.
Since we do not have any observation to constrain this parameter \citep[e.g.][]{2010A&A...515A..45D} and assuming that our source is relatively young (e.g. in the swelling phase of \citealp{2010ApJ...721..478H}, which is consistent with the definition of MYSO), the disc accretion rate, which determines the accretion luminosity, was set to $1\times10^{-4}$\,\msun\,yr$^{-1}$ in most models, which is ${\sim}10$~per~cent of the envelope infall rate in the models. 
For a typical disc with an inner radius of 35\,au, outer radius of 100\,au and 1\,\msun, an accretion luminosity of ${\sim}15$\,\lsun\ is obtained from eq.~(6) of \citet{2003ApJ...591.1049W}.
Therefore, the main difference with the flared disc model is the density distribution.
The Disc and cavity parameters in type C models were constrained in the same ranges as type A models.

The observed SED has 41 data points. 
In the images, the number of independent data points can be approximated as the area of the pixels within each observation mask divided by the solid angle of the PSF. 
If the PSF is assumed Gaussian then the solid angle is $\Omega_{\rm PSF} = 2\pi \sigma_{\rm PSF}^2$ with $\sigma_{\rm PSF}$ the standard deviation of the PSF. This gives roughly 160 independent points for the 70\,\micron\ observation and 2700 for each UKIRT masked observation. 
For the other observations the number of fitted points ranges between 7--24, but the number of independent points should be around half of those due to the effect of the beam/PSF convolution. 
Therefore, the number of independent points, and considering that the observations trace different emitting regions, should be enough to constrain the effective 10 free parameters assuming that the data are good probes of these properties.

\subsubsection{Dust models}
\label{ap:modelling:cont:vis:dust}

A number of dust models with different optical properties were used during the modelling.
The purpose of using different dust libraries is two-fold. 
On the one hand, changing the dust properties can improve the fit. 
On the other hand, they can provide information about the dust properties of each density component (e.g. disc, envelope). 
Table~\ref{tab:dust} lists the dust models used for each density component.

For the WD01 dust model, we calculated the dust properties using Mie theory\footnote{\textsc{\lowercase{Hyperion}}'s bhmie code wrapper: \url{https://github.com/hyperion-rt/bhmie}} together with the dust size distribution as calculated by the available online scripts\footnote{\url{http://physics.gmu.edu/~joe/sizedists.html}} and the dust dielectric functions of \cite{1993ApJ...402..441L}, so the dust properties are close to those published by \citet{2001ApJ...548..296W}. 
This dust model was recalculated in order to convert to the appropriate format for \textsc{\lowercase{Hyperion}}.
The dust optical properties of OHM92 dust were imported from the dust files developed by \citet{2010A&A...515A..45D} for the \textsc{\lowercase{HO-CHUNK}} \citep{2013ApJS..207...30W} radiative transfer code, whilst the properties for W02-3 dust were imported from the same code library.
The properties of the KMH dust model are readily available in \textsc{\lowercase{Hyperion}}.

We also used other dust libraries with different far-IR slopes for a few models (e.g. KMH dust with ice mantles), but we did not persist with them as they did not improve the fit. 
We did not include models 1 and 2 from \citet[][]{2002ApJ...564..887W} in order to reduce the parameters in the fit, and we only use model 3 in order to analyse the effect of grain growth and because it is the dust model with the flattest emissivity index.

\begin{table*}
 \centering
 \caption{Dust libraries.}
 \label{tab:dust}
 \begin{tabular}{llccccc}
  \hline
     Dust & Structures & $\beta^\rmn{a}$ & $a_{\rm min}$ & $a_{\rm max}$ & Size dist. & Ref.\\
     & & & (\micron) & (\micron) & & \\
  \hline
     OHM92 & envelope, disc, cavity & 2.0 & 0.005     & 0.25      & MRN+CDE         & (1)\\
     KMH   & envelope, cavity       & 2.0 & 0.0025    & --        & MEM             &(2)\\
     WD01  & envelope               & 1.7 & 0.00035   & ${\sim}1$ & pl + log-normal &(3)\\
     W02-3 & disc                   & 0.4 & --        & 1000      & pl              &(4)\\
  \hline
 \end{tabular}
 \begin{minipage}{\textwidth}
     \textbf{Notes. } MRN is a power law with index --3.5, CDE stands for continuous distribution of ellipsoids, MEM corresponds to maximum entropy method fitted to data and which can be approximated by a power law plus an exponential, and pl stands for power law.\\
     $^\rmn{a}$ Dust emissivity index measured between $100-1000$\,\micron.\\
     \textbf{References.} (1) \cite{1992A&A...261..567O} silicates and MRN amorphous carbon grains \citep{1977ApJ...217..425M}, see also \citet{2010A&A...515A..45D}; 
     (2) \citet{1994ApJ...422..164K}; 
     (3) \citet{2001ApJ...548..296W} for $R_V=5.5$; 
     (4) \citet{2002ApJ...564..887W} model 3.
 \end{minipage}
\end{table*}

\subsection{Dust continuum: grid fitting}
\label{ap:modelling:cont:grid}

To save computing power whilst concentrating on the constrainable parameters, we avoided running a grid consisting of 15 dimensions, and instead we used the parameters of a set of best-fitting models from the visual inspection step to build model grids for combinations of a limited number of parameters in order to fit specific observations.
In this step we only varied the disc radius and mass, the luminosity, the envelope infall rate (i.e. the envelope mass), the cavity opening and inclination angles, in order to further constrain them.
We did not limit the range of the parameters as strongly as in the visual inspection step.
To process the models and the simulated observations we followed the same steps as in the previous section.

We divided the fitting procedure in two steps.
In the first step we constrained the disc mass and radius from the 1.3\,mm extended configuration visibilities by minimising the $\chi^2$ value. 
We then used the SED and UKIRT images to constrain the luminosity, envelope infall rate, and the inclination and cavity opening angles.
A summary of the parameter space explored during each of these two steps can be found in Table~\ref{tab:cont:parameters}.

To obtain a seed model, we determined the most common physical properties of the models in the visual inspection step with a total ranking equal to the best-fitting model ranking plus 100, which gave us a total of 15 different models.
The properties of the seed model determined from these 15 models are listed in Table~\ref{tab:cont:parameters}, these are fixed during all the fitting steps described below.
All these 15 models are type A models (Ulrich envelope and flared disc).
Since the stellar luminosity is less constrained by our observations, we inherit the stellar temperature from the control model in \citet[][18000\,K]{2013A&A...551A..43J}.
We set the disc and envelope dust to the OHM92 model, however good fits may also be obtained with a disc with the W02-3 dust model by scaling the disc mass (disc masses with W02-3 are roughly less than half those with OHM92 dust).

In the visual inspection fitting we found a strong relation between the disc mass and radius with the 1.3\,mm extended configuration visibilities.
The parameters varied during this step are listed under Step 1 in Table~\ref{tab:cont:parameters}.
We did not develop a rectangular grid (evaluating all the combinations of disc mass and radius) in parameter space, but rather focus on a region which minimises the $\chi^2$ values.
This region is closely related to the surface density $\Sigma\propto M_d R_c^{-2}$.
Models with different inclination angles, envelope infall rates and luminosities were used to analyze their influence on the fit of the disc parameters.
We determined the best-fitting disc radius and mass from the model with minimum $\chi^2$.
The fit to the 1.3\,mm visibilities was relatively independent on the luminosity, envelope infall rate and inclination angle, hence the disc properties were fixed during the following steps.

Then, we constrained the stellar luminosity, envelope infall rate, inclination and cavity opening angles. 
The stellar luminosity and envelope mass are strongly determined by the SED and to a lesser extent by the UKIRT observations, whilst the inclination and cavity opening angles can be fitted from the SED and UKIRT observations.
We iterated between grids of luminosity and envelope infall rate at different inclination and fixed cavity opening angles, and grids of inclination and cavity opening angles at fixed luminosity and envelope infall values.
The grids of luminosity and envelope infall rate were calculated at 5 inclination angles.
We observed that the minimum $\chi^2$ for the SED requires larger inclination angles as the opening angles increases, hence the 5 values of the inclination angle varied depending on the opening angle but most models covered the 30--40\degr\ range in steps of 5\degr.
From an initial grid at a seed cavity half opening angle of 28\degr, we used luminosity and envelope infall rate values at the minimum $\chi^2$ for the SED to compute a grid of inclination and cavity opening angles (see Table~\ref{tab:cont:parameters}).
We then repeated the procedure for a different cavity opening angle.
In total we calculated grids of luminosity and envelope infall rate at 5 cavity opening angles (25, 28, 30, 35, 40\degr), and grids of inclination and cavity opening angles at 9 combinations of luminosity (1.6, 1.8, 2.0$\times 10^5\,\lsun$) and envelope infall rate ($1.1, 1.3, 1.5, 1.7, 2.0, 2.2\times10^{-3}$\,\msun\,yr$^{-1}$).
Thus, some regions of the parameter space were covered twice.

We could not find a region in the parameter space that minimised the $\chi^2$ values for the SED and UKIRT observations simultaneously. 
Moreover, we found that due to the probabilistic nature of the radiative transfer code, the values of the $\chi^2$ may vary for different runs of the same model. 
Hence, we can only give a region of the parameter space where the global minimum may be located.
In Table~\ref{tab:cont:chi2std} we list the values of the standard deviation of the $\chi^2$ values for 5 runs of a random model at different inclination angles.
Since these values may depend on model parameters, we determined a conservative overall $\chi^2$ standard deviation in order to define a confidence region in the parameter space where the minimum may be located.
The parameters of the best-fitting model were selected from a model at the intersection of the confidence regions of the observations.
In the inclination and cavity opening angle plane, the best-fitting model is located at position of the minimum $\chi^2$ for the SED and $H$-band observation.
The location of the best-fitting model in the same plane is shifted towards lower values with respect to the location of the minimum $\chi^2$ of the $J$ and $K$-bands.
The position of the best-fitting model in the luminosity and envelope infall rate plane does not coincide with the location of the minimum $\chi^2$ of the observations, but it is located in a compact region of the parameter space with a relatively constant distance to the minimum $\chi^2$ positions for the SED, $H$- and $K$-band observations.

\begin{table}
\centering
\caption{Standard deviation of $\chi^2$ values from a random model run 5 times.}
\label{tab:cont:chi2std}
\begin{tabular}{lcccc}
\hline
Observation & \multicolumn{3}{c}{Inclination angle} & $\sigma_{\chi^2}$ \\
            & $\sigma_{30\degr}$ & $\sigma_{35\degr} $ & $\sigma_{40\degr}$ & \\
\hline
SED             &   0.37  &   0.20  &   0.16    &   0.5 \\
UKIRT $J$       &   2.4   &   2.6   &   2.1     &   3.0 \\
UKIRT $H$       &   4.5   &   2.7   &   1.3     &   5.0 \\
UKIRT $K$       &   19    &   20    &   18      &   20  \\
NOEMA 1.3\,mm ext. &   5.1  &   5.8  &   7.4    &   10  \\
\textit{Herschel} 70\,\micron & 0.5 & 0.6 & 0.7 &   1.0 \\
\hline
\end{tabular}
\end{table}

\subsection{CH$_3$CN molecular line modelling} 
\label{ap:modelling:line:mollie}

\begin{figure}
  \centering
\includegraphics[width=\columnwidth]{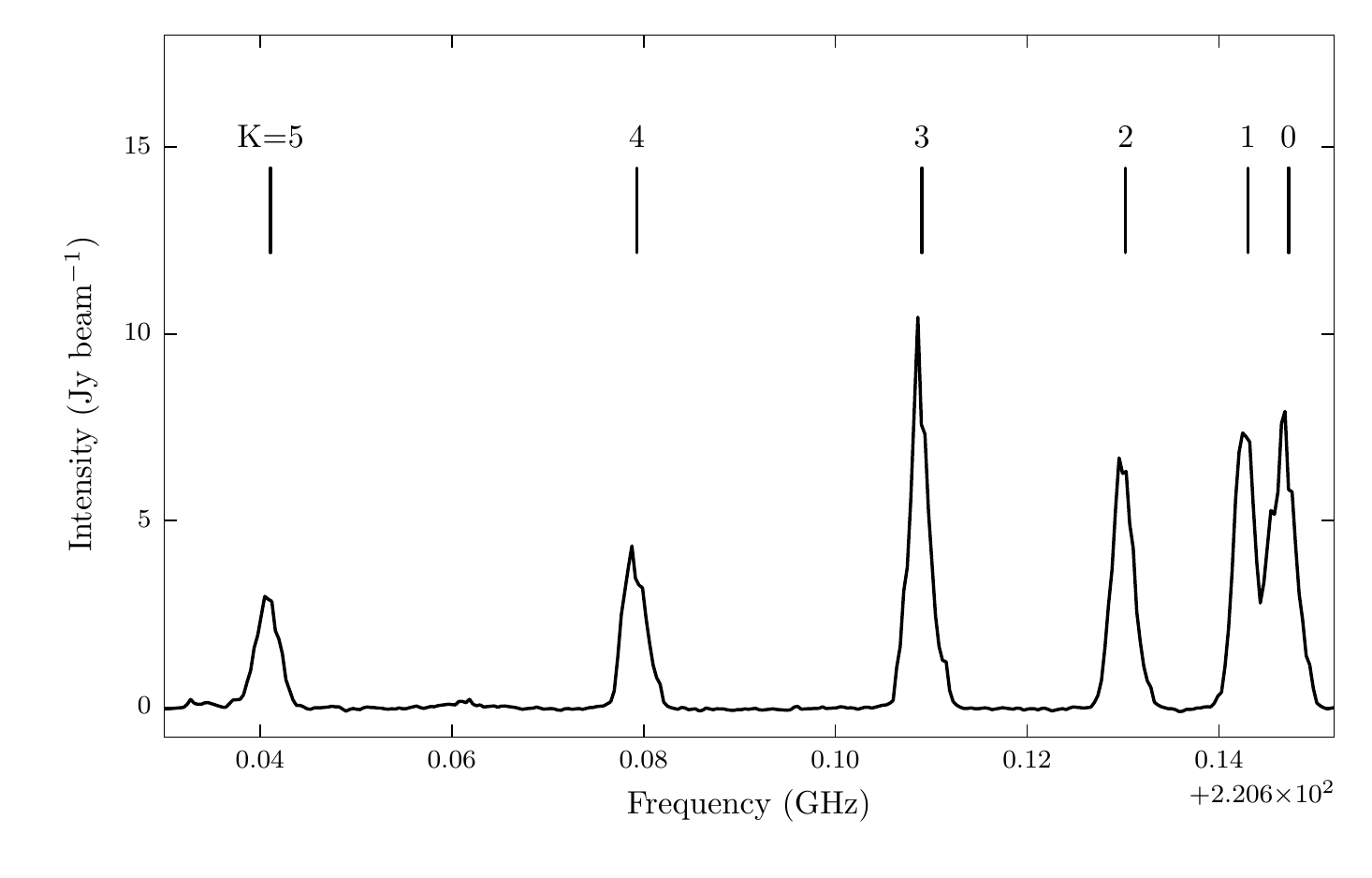}
  \caption{Peak spectrum of the CH$_3$CN $J=12-11$ observations showing the $K$ transitions.}
  \label{fig:line:spectrum}
\end{figure}

Models for the $K=2-5$ lines were calculated using the line radiative transfer code \textsc{\lowercase{Mollie}} \citep{2010ApJ...716.1315K}.
We excluded the blended $K=0$ and 1 transitions (see Fig.~\ref{fig:line:spectrum}) to ease the modelling/analysis and because these should not add any additional information.
The code requires grids describing the density, temperature and velocity distributions of the gas, the abundance of the molecule with respect to the gas density and the width of the line. 
\textsc{\lowercase{Mollie}} calculates the molecular line emission at different velocities, i.e. it produces a data cube, under the LTE approximation for CH$_3$CN. 
Since the radiation field (the gas temperature) is calculated from the continuum modelling and the LTE approximation is used, \textsc{\lowercase{Mollie}} uses the input distributions to calculate the level populations and produce the images.

For the modelling, three nested grids with increasing resolution on the region close to the star were used. 
The grids were defined with $80^3$, $100^3$ and $100\times100\times50$ cells from large to small scale, respectively. 
The former covers a size of $4.0\times10^5$\,au, the second a grid size of $5.0\times10^4$\,au, and the latter a size of 2000\,au in the x-y plane. 
The smallest scale this grid can resolve is 20\,au.

In order to evaluate the density, temperature and velocity in these grids, we first oversample each nested grid by dividing each grid cell into 27 smaller grid cells for the two large-scale grids and 125 smaller grid cells for the grid sampling the smaller scales.
Then the density and velocity distributions are evaluated at each point and the temperature is interpolated. 
Finally, number density weighted averages are taken to obtain the coarser grid values. 
The line width and abundance grids are obtained by evaluating their functions at the positions of the grid. 
We fixed the inner radius of the gas distributions to half the dust sublimation radius. 
We do not expect to constrain this inner radius because the resolution of our observations does not allow it.

We used the Ulrich velocity distribution for the envelope, a Keplerian rotation velocity distribution for the disc and a Hubble expansion law for the outflow.
The equations and details of the velocity distribution can be found in Appendix~\ref{ap:velocity}.

\textsc{\lowercase{MOLLIE}} includes a line width parameter which was set to the sum in quadrature of the thermal speed of sound and a turbulence/non-thermal velocity. 
The turbulent velocity line width $\Delta v_{\rm nth}$ was varied during the modelling (see below).

Following the studies of other molecules \citep[e.g.][]{2015A&A...574A..71K}, the methyl cyanide abundance was defined as a step function of the temperature. 
This approach has also been used for methyl cyanide in other studies \citep[e.g.][]{2015ApJ...813L..19J} and more generally for many species which freeze out \citep[e.g.][]{2004A&A...416..603J}.
The abundance function is then defined as
\begin{equation}\label{eq:abundance}
    X_{\rm CH_3CN}(T) = 
    \begin{cases}
        X_0 & \text{if }\, T<T_0\,,\; \rmn{or} \\
        X_1 & \text{if }\, T>T_0\,. 
    \end{cases}
\end{equation}
This approximation is consistent with the change in abundance due to thermal desorption of molecules from dust grains in the evolution of hot cores when the temperature increases with time \citep[e.g.][]{2004MNRAS.354.1141V}.
We explored several temperature values to match the extension of the zeroth moment map and the ratios between the $K=2-5$ lines with increasing number of temperature steps from 0 to 2. 
It was found that one temperature step is enough to explain the observations and limit the number of parameters to fit. 
The temperature step was fixed to $T_0=100$~K and the abundance below this temperature was fixed to $X_0=5\times10^{-12}$. 
Hence, only $X_1$ was varied during the modelling.

To compare the models with the observations, we obtained pv maps for the $K=2-5$ lines at ${\rm PA}=276\degr$ and $6\degr$, i.e. along and perpendicular to the jet axis, and centred in the peak of the zeroth moment map. 
We use a slit width for the pv maps equal to 3 times the pixel size, i.e. 1.8\,arcsec. 
The pv maps are shown in Fig.~\ref{fig:line:pv}. 
Since the observations are not particularly asymmetric, the two position angles allow us to fit the extension of the emitting region. 
We aligned the spectral axis of the pv maps by matching the centroid of the line at the central position. 
The average shift for all the lines and models was $-0.69\pm0.45$\,km\,s$^{-1}$. 
This value is within the errors in $v_{\rm LSR}$ (see above) and equivalent to a systemic velocity of --6\,km\,s$^{-1}$. 
This shift was applied to all the models prior to the comparison with the observations.

Similar to the continuum modelling visual inspection fitting, the best-fitting model was chosen by averaging the ranking of the $\chi^2$ of the pv maps.
This method was used because both maps are not independent but have a similar number of points.
From the observations we estimated the rms $\sigma_{\rm rms}$ from an emission-free region for each channel and use these values as the errors in the $\chi^2$ calculation. 
We masked points corresponding to the position of NOEMA~2 and emission below $5\sigma_{\rm rms}$.

\subsubsection{Parameter selection}
\label{ap:modelling:line:params}

All the parameters derived from the best-fitting model to the continuum observations were kept unchanged, with the exception of the envelope infall rate. 
We varied key parameters for the velocity distribution in a grid with values listed in Table~\ref{tab:line:grid}.
Changes in centrifugal radius will not necessarily fit the continuum data because these change the density and temperature distributions. 
Therefore, the centrifugal radius was not varied.
The envelope infall rate was scaled to keep the density of the envelope constant (cf. eq.~\ref{eq:ulrich}).

\begin{table}
\centering
\caption{Values of the line modelling parameter grid.}
\label{tab:line:grid}
\begin{tabular}{lc}
\hline
Parameter & Values \\
\hline
    $M_\star^{a}$ & 15, 25, 35\,\msun \\
    Abundance ($X_1$)$^b$ & $(2, 2.5, 3, 3.5, 4, 4.5)\times10^{-8}$\\
    Outflow velocity$^c$ ($v_0$) & 0, 1, 5, 10\,km\,s$^{-1}$ \\
    Turbulent width ($\Delta v_{\rm nth}$) & 0.5, 0.8, 1.0, 1.5, 2.0\,km\,s$^{-1}$\\
\hline
\end{tabular}
\newline
\begin{minipage}{\columnwidth}
	$^a$ we report the results for $M_\star = 35\,\msun$ and use the other masses for comparison.\\
    $^b$ for $T>100$\,K.\\
    $^c$ Defined at $z=10^4$\,au.
\end{minipage}
\end{table}

The range of the abundance values is based on the results of the 1-D line modelling (see Section~\ref{sect:results:line:cassis}), which in turn are consistent with values in the literature for methyl cyanide in high-mass star-forming regions \citep[e.g][]{2014ApJ...788..187H,2015ApJ...813L..19J,2017MNRAS.467.2723P,2018A&A...618A..46A}.

The disc velocity only changed when the stellar mass was varied. 
The direction of the rotation of the envelope and disc was fixed to $\phi_{\rm dir}=1$, i.e. anticlockwise looking down from the blue-shifted cavity axis as suggested by \citet{2012A&A...543A..22W} from their first moment maps of HDO, H$_2$$^{18}$O and SO$_2$ line emission.

The outflow velocities were based on \citet{2013A&A...551A..43J}, who found an average velocity relative to $v_{\rm LSR}$ of $-1.2$\,km\,s$^{-1}$ at a projected outflow length of 47300\,au (i.e. ${\sim}110000$\,au at an inclination angle of 25\degr) from C$^{18}$O line observations.
\citet{2012A&A...543A..22W} found that a wind with a Hubble law constant of 10\,km\,s$^{-1}$\,arcsec$^{-1}$, which is roughly 5 times higher than the highest velocity used here (10\,km\,s$^{-1}$ per 10000\,au). 
Such high-velocity disc wind was not included here because the scale affected by it is not resolved by the methyl cyanide observations. 
In addition, the wind model used by \citet{2012A&A...543A..22W} is equatorial, i.e. without a velocity component in the polar direction, rather than spherical as our model. 
It is worth noticing that the methyl cyanide emission does not show any signatures of outflows, thus it is not expected to constrain this parameter.

\section{Physical distributions}
\label{ap:distributions}

An extensive list of the parameters used in this section can be found in Tables~\ref{tab:cont:parameters} and \ref{tab:line:grid}. 
In what follows, we use the $(r,\,\theta,\,\phi)$ nomenclature for spherical coordinates and $(R,\,z,\,\phi)$ for cylindrical coordinates.

\subsection{Density distribution}
\label{ap:density}

The envelope density distribution of types A and B models (see Table~\ref{tab:densities}) use the \cite{1976ApJ...210..377U} analytical solution for rotating, infalling material (hereafter Ulrich envelope)
\begin{multline}\label{eq:ulrich}
\rho(r, \theta) = \frac{\dot{M}_\rmn{env}}{4\pi(GM_\star R_c^3)^{0.5}} \left( \frac{r}{R_c} \right)^{-1.5} \left( 1 + \frac{\mu}{\mu_0} \right)^{-0.5} \\ 
    \left( \frac{\mu}{\mu_0} + \frac{2 \mu_0^2 R_c}{r} \right)^{-1} \propto \frac{\dot{M}_\rmn{env}}{\sqrt{M_\star}} f(\mathbf{r}, R_\rmn{c})\,,
\end{multline}
where $G$ is the gravitational constant, $f(\mathbf{r}, R_\rmn{c})$ is a function of the position and the centrifugal radius, $\mu=\cos\theta$ and $\mu_0=\cos \theta_0$, with $\theta_0$ the angle of the parabolic orbits of the infalling material, and is calculated by solving
\begin{equation}\label{eq:streamline}
    \mu_0^3 + \mu_0 \left( \frac{r}{R_c} - 1 \right) - \mu \left( \frac{r}{R_c} \right) = 0\,.
\end{equation}
This density distribution behaves like a power law $n(r)\propto r^{-1.5}$ for distances $r$ larger than the centrifugal radius $R_\rmn{c}$, i.e. like free-falling material, while at shorter distances the density is relatively constant. 

Type B models were used to study the MYSO W33A \citep{2010A&A...515A..45D}, and unlike type A they have an alpha disc \citep[e.g.][]{1973A&A....24..337S} with accretion luminosity. 
The disc density distribution of type A models is given by
\begin{equation}
    \rho (R, z) = \rho_0 \left( \frac{R_0}{R} \right)^{\alpha} \exp\left[-0.5 \left(\frac{z}{h(R)}\right)^2 \right]\,,
\end{equation}
with the reference density $\rho_0$ calculated to match the disc mass ($M_d$), a fiducial radius $R_0=100\,{\rm au}$ and the scale height $h=h_0 (R/R_0)^\beta$.
The disc density distribution of type B models differs by a factor $(1-\sqrt{R_\star/R})$, with $R_\star$ the stellar radius, from that of type A models (see \textsc{\lowercase{Hyperion}} documentation\footnote{\url{http://docs.hyperion-rt.org}}). 
The disc radius was set to the centrifugal radius $R_c$.

Type C is one of the proposed solutions to increase the underestimated near and mid-IR fluxes, which results from 1-D spherically symmetric envelopes used to model SEDs and submm intensity radial profiles of MYSOs \citep[e.g.][]{2005A&A...434..257W}. 
The density distribution of this type of models is parametrised as a power law $n=n_0 (r/r_0)^{-p}$ with $r_0=0.5R_\rmn{out}$, $R_\rmn{out}$ the envelope radius and the density reference value $n_0$ is determined by the mass of the envelope $M_\rmn{env}$.

We use the empirical expression found by \citet{2004ApJ...617.1177W} to estimate the dust sublimation radius
\begin{equation}\label{eq:ap:rsublimation}
R_\rmn{in} = R_\star \left( \frac{T_\star}{T_\rmn{sub}} \right)^{2.1} \propto \sqrt{L_\star} \frac{T_\star^{0.1}}{T_\rmn{sub}^{2.1}}
\end{equation} 
where $T_\rmn{sub}$ is the sublimation temperature. 
For a sublimation temperature $T_\rmn{sub}=1500\,{\rm K}$, luminosity $L_\star=1.6\times10^5$\,\lsun\ and stellar temperatures $T_\star=10000-40000$\,K the sublimation radius ranges between 30--40\,au.

\subsection{Velocity distribution}
\label{ap:velocity}

For the envelope, we used the Ulrich envelope velocity distribution. 
The velocity components of the envelope inside the disc radius were set to zero because the Ulrich solution  is undefined at the centrifugal radius resulting in a large increase in velocity. 
This produces a discontinuity in the velocity distribution but at scales which are not resolved by our observations. 
The 3-D components of the envelope velocity distribution are given in spherical coordinates by 
\citep[e.g.][]{2005fost.book.....S}
\begin{equation}
    v_r(r,\theta) = -\left( \frac{GM_\star}{r} \right)^{1/2} \left( 1 + \frac{\mu}{\mu_0} \,,
\right)^{1/2}
\end{equation}
\begin{equation}
    v_\theta(r,\theta) = \left( \frac{GM_\star}{r} \right)^{1/2}
    \frac{\mu_0-\mu}{\sin\theta} \left(
1 + \frac{\mu}{\mu_0} \right)^{1/2} \,,
\end{equation}
\begin{equation}
    v_\phi(r,\theta) = \phi_{\rm dir} \left( \frac{GM_\star}{r} \right)^{1/2} \frac{\sin\theta_0}{\sin\theta}
\left( 1 - \frac{\mu}{\mu_0}\right)^{1/2}\,,
\end{equation}
where $\phi_{\rm dir}$ parametrises the direction of rotation, i.e. has a value of $\pm 1$. 
We use a stellar mass of 35\,\msun\ for the modelling, but as a reference we also calculated models with lower masses.

The outflow velocity was modelled with a Hubble expansion law
\begin{equation}
    v_r(r) = v_0 \frac{r}{r_0}\,,
\end{equation}
where $v_0$ is the velocity at $r_0 = 10^4 \cos^{-1}\theta_{\rm c}$~au with $\theta_{\rm c}$ the half-opening angle of the cavity like in the continuum modelling. 
A similar velocity distribution was used by \citet{2012A&A...543A..22W} to interpret the molecular emission from the PdBI 1.4\,mm observations at a smaller scale, but as a cylindrical radial (i.e. perpendicular to the outflow cavity direction) expansion.
A limit of 12\,km\,s$^{-1}$ was set to the outflow velocity. 
Velocities larger than this limit are achieved in colder less dense gas than the envelope, thus its contribution to the spectra is not important.

For the disc, the azimuthal velocity was parametrised as a Keplerian disc, which is described by \citep[e.g.][]{1976MNRAS.175..613S}
\begin{equation}
    v_\phi(R) = \phi_{\rm dir} \sqrt{\frac{G M_\star}{R}}\,,
\end{equation}
The other two components of the velocity were set to zero in order to limit the number of parameters in the fit and because the disc is not resolved in the observations.

\section{Model image processing}
\label{ap:dataproc}

To obtain the data products in Table~\ref{tab:cont} different processes were used.
For images, we have chosen to compare them on a pixel-by-pixel basis by aligning the emission peak. 
For the UKIRT images, we have used the observed WCS information in the image headers because the peak emission is saturated. 
The images are convolved with the respective PSF, where we have also taken into account the PSF rotations. 
To compare the images, they were also masked to isolate the source and exclude emission which cannot be reproduced by our model because of its symmetry. 
In the $J$, $H$ and $K$ bands this includes features in the outflow cavity (the so-called ``loops''), point-like sources inside and outside the cavity and the saturated central region. 
The diffraction pattern seen in the $K$-band map (see Fig.~\ref{fig:cont:nir}) was not masked.

Monte Carlo noise was present in the images produced by the radiative transfer code because of its probabilistic nature.
This noise affected more strongly the near-IR observations due to scattering. 
In order to reduce this noise in the $JHK$-band images we followed a 2-step process.
First, we identified isolated pixels by approximating pixels with fluxes lower than $10^{-10}$\,Jy\,sr$^{-1}$ to zero and using a median filter of size 2. 
This filters out pixels which are located in the outskirts of the cavity and in other similar regions.
Then, to recover the pixels in the outskirts of the cavity, we calculated the local density of pixels with values larger than zero within $7\times7$\,pixel$^2$ boxes. 
We used the 75th percentile of the local density values of any given image as threshold so pixels in regions of high local density of pixels, like the outflow cavity, are not filtered out. 
The amount of flux loss is ${\sim}1$~per~cent on average.

A similar procedure was followed to obtain the images used for the model radial intensity profiles. 
These images were not rotated since the PSF is symmetric and only radial profiles are compared. 
With these images we follow the same steps utilised to obtain the observed profiles (see Section~\ref{sect:data:multi:submm}). 
In these cases, we have chosen to compare points whose observed intensity is higher than three times the image noise.

In the particular case of the $K$-band speckle interferometric visibilities we Fourier transformed the model image and obtained the visibility radial profile. 
For the mm interferometric data, we used \textsc{\lowercase{CASA}} \citep{2007ASPC..376..127M}  to simulate the observed visibilities from our model images. 
Noise was added to the simulated observations. 
For reproducing the observed PSF properties as close as possible, we used the on-source integration time, antenna configurations and hour angle of the observations. 
The simulated observations have a PSF FWHM of $2\farcs3\times1\farcs8$ with PA$=93\degr$ and $0\farcs46\times0\farcs35$ with PA$=62\degr$ in the compact and extended configurations respectively (cf. Table~\ref{tab:cont}).
We chose to compare points at uv distances larger than $40\,{\rm k}\lambda$ (angular scale of ${\sim}6$\,arcsec) in the compact configuration data because emission not belonging to the core (e.g. background emission from the large scale molecular cloud) may have been included at smaller uv distances. 
Similarly, we limited the comparison of extended configuration data to uv distances between 100 and $400\,{\rm k}\lambda$ (angular scales between ${\sim}2.5-0.6$\,arcsec).
Shorter uv distances ($<100\,{\rm k}\lambda$) are covered by the compact configuration and may be underrepresented in the high-resolution model image, whilst larger uv distances ($>400\,{\rm k}\lambda$) may include substructures that the model cannot reproduce (e.g. clumpy emission, see Fig.~\ref{fig:cont:mm:clean}b). 

\begin{figure*}
  \centering
  \includegraphics[width=\textwidth]{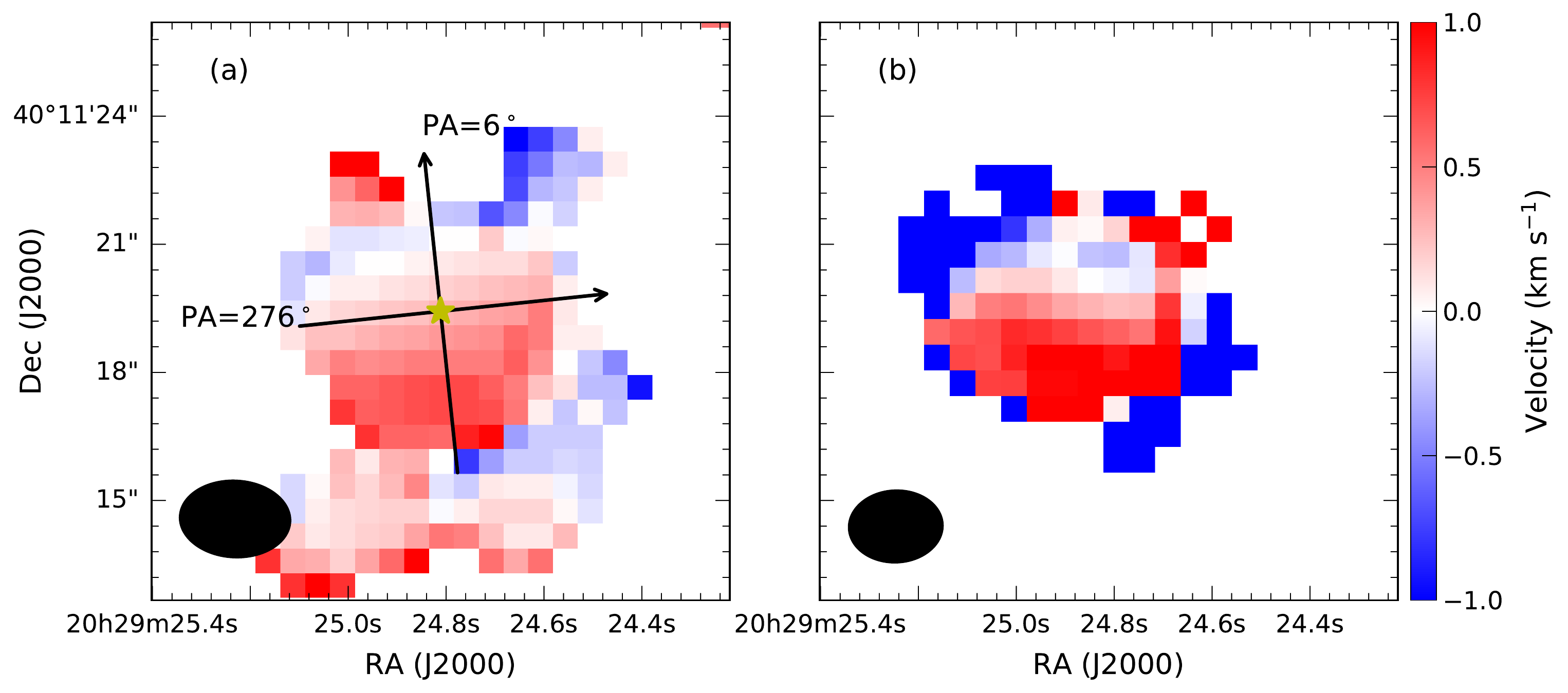} 
  \caption{NOEMA compact configuration first moment map for the observed CH$_3$CN $J=12-11$ $K=3$ line (a) and from the overall best-fitting model (b).
    Black arrows show the direction of the slices in Fig.~\ref{fig:line:pv} with the arrows pointing towards positive offsets. 
    The star marks the position of the radio continuum source.
    The beam is shown in black in the lower left corner.
    }
  \label{fig:line:mom1:k3}
\end{figure*}

The CH$_3$CN model images were rotated to match the jet PA and then synthetic observations were calculated with \textsc{\lowercase{CASA}}. 
The synthetic images included thermal noise which approximately matches that observed. 
The $K=3$ line first moment map in Fig.~\ref{fig:line:mom1:k3}(a) shows that, if the velocity structure was completely due to rotation, the rotation axis will be oriented roughly at a ${\rm PA}=286\degr$. 
However, as pointed out by \citet{2012A&A...543A..22W}, expansion or infalling motions can change the perceived position of the rotation angle. 
The orientation of the disc plane as derived from a Gaussian fit to the 1.3\,mm extended configurations map is $\rmn{PA}=21\degr$, similarly the orientation of the 1.4\,mm continuum emission from \citet{2012A&A...543A..22W} is ${\rm PA}=6\degr$.
These values are nearly perpendicular to the radio jet orientation (${\rm PA}\sim100\degr$ from \citealp{2013A&A...551A..43J}, see Fig.~\ref{fig:line:radio}). 
Therefore, the small scale rotation axis does not have the same orientation as the large scale outflow cavity ($\rmn{PA}=259\degr$). 
Then the orientation of the model rotation axis was set to ${\rm PA}=276\degr$, i.e. perpendicular to the disc plane.

\section{Best-fitting model by types}
\label{ap:bestbytype}

Table~\ref{ap:tab:cont:best} lists the parameters for the continuum best-fitting models from the visual inspection for types B and C, and from the grid fitting for type A.

\begin{table*}
 \centering
    \caption{Parameters of the continuum best-fitting models by density distribution type.}
\label{ap:tab:cont:best}
 \begin{tabular}{lccc}
  \hline
     Parameter & \multicolumn{3}{c}{Type}\\
               & A & B  & C \\
  \hline
     Inclination angle ($i$)                          &  25\degr  & 30\degr & 40\degr \\
     Stellar temperature ($T_\star$/K)                &  18000    & 18000   & 30000   \\
     Stellar luminosity ($L_\star/10^5$\,\lsun)       &  1.6      & 1.6     & 1.4     \\
     Outer radius ($R_\rmn{out}$/au)                  &  195000   & 195000  & 300000  \\
     Envelope infall rate ($\dot{M}_\rmn{env}/10^{-3}\,\msun\,{\rm yr}^{-1}$) & 2.0 & 1.5$^{a}$ & -- \\
     Envelope mass ($M_\rmn{env}/10^3\,\msun$)        & 2.1$^b$   & 1.6$^b$ & 1.0     \\
     Envelope dust                                    & OHM92     & OHM92   & WD01    \\
     Centrifugal radius ($R_\rmn{c}$/au)              & 2200      & 440     & 100     \\
     Disc mass ($M_\rmn{d}$/\msun)                    & 6.0       & 1.0     & 1.2     \\
     Disc scale height at $r=100$ au ($h_0$/au)       & 0.6       & 0.6     & 0.6     \\
     Disc flaring exponent ($\alpha$)                 & 2.25      & 2.25    & 2.25    \\
     Disc vertical density exponent ($\beta$)         & 1.25      & 1.25    & 1.25    \\
     Disc dust                                        & OHM92     & OHM92   & OHM92   \\
     Cavity opening angle$^{c}$ ($2\theta_{\rm cav}$) & 60\degr   & 56\degr & 60\degr \\
     Cavity shape exponent ($b_\rmn{cav}$)            & 2.0       & 2.0     & 2.0     \\
     Cavity density exponent ($p_{\rm cav}$)          & 2.0       & 2.0     & 2.0     \\
  Cavity reference density$^{c}$ ($\rho_{\rm cav}/10^{-21}$\,g\,cm$^{-3}$) & 1.4 & 1.4 & $3.5\times10^{-3}$ \\
     Cavity dust                                      & KMH       & KMH     & KMH     \\
  \hline
 \end{tabular}
    \begin{minipage}{\textwidth}
        $^a$ Scaled to a stellar mass of 35\,\msun.\\
		$^b$ Not an input parameter.\\
        $^c$ Defined at a height $z=10^4$~au.
    \end{minipage}
\end{table*}

\bsp
\label{lastpage}
\end{document}